\documentclass[preprint,12pt,fleqn,number]{elsarticle}

\usepackage[utf8]{inputenc}
\usepackage{makecell}
\usepackage{longtable}
\usepackage{amssymb}
\usepackage{graphicx,amsfonts,graphics}
\usepackage[british]{babel}
\usepackage{url}
\usepackage{amsmath}    
\usepackage{calc}       
\usepackage{colortbl}
\usepackage{bm}
\usepackage{hyperref}
\usepackage{tikz}
\usepackage{afterpage}
\usepackage[normalem]{ulem}
\usepackage{rotating}
\usepackage{tabularray}
\usepackage{arydshln}
\usepackage{booktabs}
\usepackage{subcaption}
\usepackage{multirow}
\usepackage{multibib}
\newcites{main}{References}
\newcites{app}{Appendix References}

\def\x{{\mathbf x}}

\journal{Computers in Biology and Medicine}

\begin{document}

\begin{frontmatter}



\title{ConfLUNet: Multiple sclerosis lesion instance segmentation in presence of confluent lesions} 


\author[1,2]{Maxence Wynen\corref{cor1}}
\author[3,4]{Pedro M. Gordaliza}
\author[1]{Maxime Istasse}
\author[2]{Anna Stölting}
\author[2,5]{Pietro Maggi}
\author[1]{Benoît Macq\corref{eq}}
\author[3,4]{Meritxell Bach Cuadra\corref{eq}}

\affiliation[1]{organization={ICTEAM, Université Catholique de Louvain}, 
                addressline={Louvain-la-Neuve}, 
                country={Belgium}}

\affiliation[2]{organization={Louvain Neuroinflammation Imaging Lab (NIL), Université Catholique de Louvain}, 
                addressline={Brussels}, 
                country={Belgium}}

\affiliation[3]{organization={CIBM Center for Biomedical Imaging}, 
                country={Switzerland}}

\affiliation[4]{organization={Department of Radiology, Lausanne University Hospital and University of Lausanne}, 
                addressline={Lausanne}, 
                country={Switzerland}}

\affiliation[5]{organization={Cliniques Universitaires Saint-Luc, Université Catholique de Louvain}, 
                country={Belgium}}

\cortext[cor1]{Corresponding author.}
\cortext[eq]{The last two authors contributed equally.}


\begin{abstract}
Accurate lesion-level segmentation on MRI is critical for multiple sclerosis (MS) diagnosis, prognosis, and disease monitoring. However, current evaluation practices largely rely on semantic segmentation post-processed with connected components (CC), which cannot separate confluent lesions (aggregates of confluent lesion units, CLUs) due to reliance on spatial connectivity. To address this misalignment with clinical needs, we introduce formal definitions of CLUs and associated CLU-aware detection metrics, and include them in an exhaustive instance segmentation evaluation framework. Within this framework, we systematically evaluate CC and post-processing-based Automated Confluent Splitting (ACLS), the only existing methods for lesion instance segmentation in MS. Our analysis reveals that CC consistently underestimates CLU counts, while ACLS tends to oversplit lesions, leading to overestimated lesion counts and reduced precision. To overcome these limitations, we propose ConfLUNet, the first end-to-end instance segmentation framework for MS lesions. ConfLUNet jointly optimizes lesion detection and delineation from a single FLAIR image. Trained on 50 patients, ConfLUNet significantly outperforms CC and ACLS on the held-out test set (n=13) in instance segmentation (Panoptic Quality: 42.0\% vs. 37.5\%/36.8\%; p = 0.017/0.005) and lesion detection (F1: 67.3\% vs. 61.6\%/59.9\%; p = 0.028/0.013). For CLU detection, ConfLUNet achieves the highest $text{F1}^{\text{CLU}}$ (81.5\%), improving recall over CC (+12.5\%, p = 0.015) and precision over ACLS (+31.2\%, p = 0.003). By combining rigorous definitions, new CLU-aware metrics, a reproducible evaluation framework, and the first dedicated end-to-end model, this work lays the foundation for lesion instance segmentation in MS.

\end{abstract}


\end{frontmatter}



\section{Introduction}
\label{sec:intro}
\begin{figure}[h!]
  \centering
  \includegraphics[width=\textwidth]{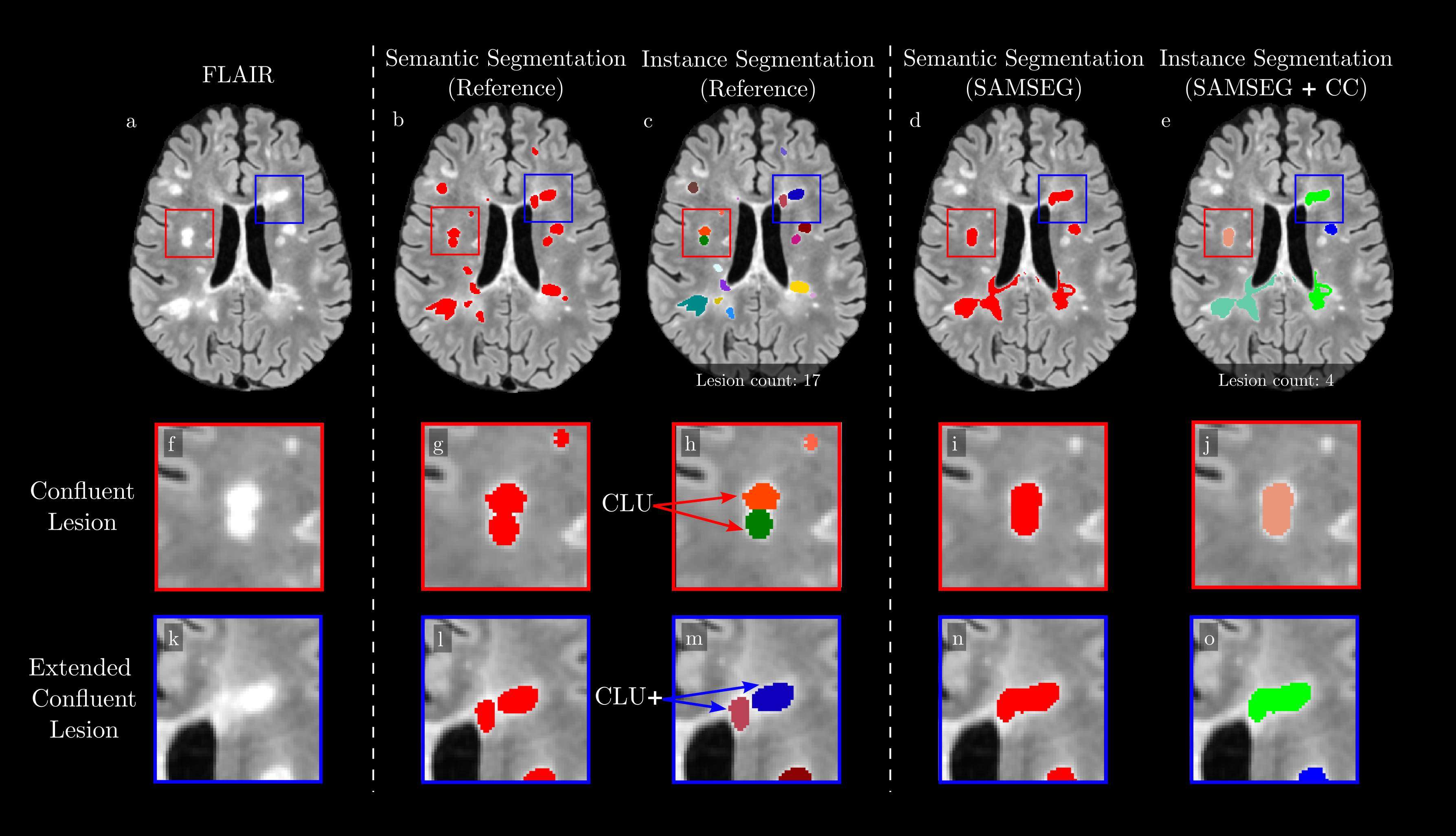}
    \caption{Illustration of key concepts formalized in this study. (a) Axial FLAIR image; (b–c) Reference semantic and instance segmentations; (d–e) SAMSEG's predicted semantic segmentation and corresponding instance segmentation obtained by applying connected components (CC); (f) Close-up of a confluent lesion; (g–h) Reference semantic and instance segmentations of (f); red arrows indicate distinct confluent lesion units (CLUs); (i–j) SAMSEG predictions for (f); CC merges multiple CLUs into a single instance; (k) Close-up of an extended confluent lesion; (l–m) Reference semantic and instance segmentations of (k); red arrows indicate extended CLUs (CLU+), defined as lesions whose connectivity becomes apparent only after semantic mask dilation; (n–o) SAMSEG predictions for (k); CC again merges multiple CLU+ into a single instance.}
    \label{fig:definitions}
\end{figure}

Multiple sclerosis (MS) is a chronic immune-mediated disease and a leading cause of non-traumatic neurological disability in young adults \citepmain{reich_multiple_2018}. MS is characterized by the presence of demyelinated lesions in the central nervous system, mainly located in the white matter (WM) and visible on conventional magnetic resonance imaging (MRI). Both the total lesion count and the lesion load (volume) play crucial roles in the diagnosis \citepmain{thompson_diagnosis_2018}, prognosis \citepmain{on_behalf_of_the_magnims_study_group_magnims_2015}, and monitoring \citepmain{brex_longitudinal_2002,khoury_longitudinal_1994,rudick_significance_2006} of MS. Recently, advanced lesion-level MRI biomarkers, such as the presence of paramagnetic rim or the central vein sign, are being incorporated in clinical guidelines as they enhance diagnostic accuracy \citepmain{montalban_revised_2024,maggi_central_2018}, improve prognostic assessments \citepmain{absinta_association_2019}, and/or uncover underlying pathological mechanisms \citepmain{elliott_slowly_2019}. Automated analyses of these biomarkers necessitate precise localization and delineation of individual lesions \citepmain{la_rosa_cortical_2022,barquero_rimnet_2020,maggi_cvsnet_2020,lou_fully_2021,zhang_qsmrim-net_2022,wynen_longitudinal_2021}, which are generally achieved through lesion instance segmentation masks. Compared to semantic masks (Fig. \ref{fig:definitions}b) which classify voxels as lesion/non-lesion, lesion instance segmentation masks (Fig. \ref{fig:definitions}c) also associate each lesion voxel with a unique lesion identifier (id). 

Naturally, manually annotating 3D lesion instance segmentation masks is even more laborious than creating their semantic counterparts, particularly in cases with many confluent lesions—aggregates of pathologically distinct focal areas of inflammation, arising from the confluence of spatially and usually temporally separated foci of blood-brain barrier disruption, demyelination and axonal damage. \citepmain{zivadinov_effect_2008,lassmann_multiple_2014} (Fig. \ref{fig:definitions}f). To facilitate understanding, we further define the distinct lesions constituting each confluent lesion as \textit{confluent lesion units (CLU)}. Confluent lesions may comprise anywhere from two to dozens of CLUs. These particularities of CLUs render their identification extremely difficult---if not impossible, particularly in late-stage patients (Fig. \ref{fig:unsplittable_lesions})---, especially when using only cross-sectional images. Indeed, regular longitudinal scans with adequate resolution could theoretically help separate spatially confluent lesions based on their temporal appearance, but the frequent acquisition of such scans is very expensive, particularly given that those lesions can form rapidly within the timeframe of weeks, especially in patients with high disease activity \citepmain{harris_serial_1991}. This requirement conflicts with the typical clinical practice of scanning patients only once or twice a year. Although advanced MRI sequences (e.g. EPI, QSM) used to detect paramagnetic rims or perivenular lesions can support lesion instance identification, their high acquisition cost and limited clinical adoption constrain their practical utility.


Despite the increasing interest in lesion-level biomarkers, the vast majority of existing automated approaches for MS lesion instance segmentation do not directly tackle the problem of confluent lesions. Instead, most rely on semantic segmentation followed by connected component (CC) analysis, a simple post-processing step that identifies each connected voxel region in the semantic mask as a separate lesion (Fig. \ref{fig:definitions}d-e). While effective for disconnected lesions, this approach struggles to distinguish CLUs, as their typical spatial connectivity causes CC to consider them as a single instance (Fig. \ref{fig:definitions}e, j, and o), leading to inaccurate lesion counts. To our knowledge, only one alternative, we term \textit{Automated Confluent Lesion Splitting (ACLS)}, attempts to directly address the challenge of lesion splitting \citepmain{dworkin_automated_2018,lou_fully_2021}. However, ACLS, which also operates as a post-processing step on top of semantic segmentation, has not been rigorously validated in the context of instance segmentation. End-to-end instance segmentation could better address CLUs by jointly optimizing detection and segmentation but, to our knowledge, no such methods have yet been proposed.

In parallel, current evaluation strategies often prioritize semantic segmentation metrics, such as the Dice Score Coefficient, which do not fully capture the complexity of instance-level delineation, particularly in the presence of confluent lesions. Our review of recent literature (\ref{app:lit_review}) shows that although 53.5\% of studies since 2014 (61 out of 114) also report lesion detection metrics, nearly all rely on CC to generate both predicted \textit{and} reference lesion instances, thus implicitly ignoring CLUs. This results in a misalignment between current evaluation frameworks and clinical needs for accurate counting and lesion-level delineation. Notably, no prior work has rigorously evaluated methods in a true instance segmentation framework, nor have metrics been developed to specifically assess CLU detection accuracy.

This study makes three primary contributions: (i) we introduce the first end-to-end framework for MS lesion instance segmentation, extending our previous ConfLUNet framework \citepmain{wynen_conflunet_2024}, to jointly optimize lesion detection and delineation in the presence of CLUs; (ii) we propose an evaluation framework tailored to instance segmentation in MS, including new metrics grounded in formal definitions of confluent lesions and CLUs; and (iii) we conduct the first systematic comparison of existing approaches (CC and ACLS) within this framework, and further demonstrate that ConfLUNet achieves superior instance segmentation performance and promising results regarding CLU detection.

\section{Problem Formulation and related works}
\label{sec:definitions}
Building on earlier clinical descriptions, we now provide formal definitions to support the technical formulation of the problem.

Let $I: \Omega \rightarrow \mathbb{R}$ represent a 3D image, where $\Omega = \{(x,y,z) \in \mathbb{N}^3 \mid x < w, y < h, z < d\}$ is the voxel image domain. \textbf{Semantic segmentation masks} $\bm{S}$ classify each voxel in an image (Fig. \ref{fig:definitions}b), in our case, delineating lesion tissue, but do not distinguish between individual lesions (i.e., $\bm{S}: \Omega \rightarrow \{0,1\}$). \textbf{Instance segmentation masks} account for this by extending the semantic mask and assigning a unique identifier to each lesion instance (Figs. \ref{fig:definitions}b and d). Formally, an instance segmentation mask $\bm{M}: \Omega \rightarrow \{0,1,...,K\}$ assigns to each voxel a label, where positive values indicate lesion instances.  The set of lesion instances is defined as $\mathcal{L} = \{L_1, L_2, ..., L_K\}$,  where each $L_k = \{(x,y,z) \in \Omega \mid \bm{M}(x,y,z) = k\}$, is the set of voxel coordinates belonging to the $k$-th lesion. In medical image analysis, the most common method to create an instance segmentation mask from a semantic segmentation mask is via CC analysis (Fig. \ref{fig:definitions}e, Section \ref{sec:instance_seg}). Given a $\bm{S}$, CC analysis identifies disjoint connected regions. Let $\mathcal{B} = \{B_1, B_2, ..., B_J\}$ represent the set of connected components ("blobs") in $ \bm{S} $, where each $B_j \subset \Omega$ is the set of voxel coordinates belonging to the $j$-th component. The instance segmentation mask $ \bm{M}_{\text{\textbf{CC}}} $ obtained by applying CC on $ \bm{S} $ is defined in Table \ref{tab:definitions} (Eq. \ref{eq:cc_instance}). Throughout this work, $\bm{M}_{\text{\textbf{CC}}}$ is always computed with the most conservative structure (6-connectivity). 

\paragraph{Confluent lesions and confluent lesion units} Using $M_{CC}$ and $\mathcal{B}$, we define \textbf{confluent lesions} as connected components from $M_{\text{CC}}$ that overlap with at least two reference lesion instances from the set $\mathcal{L}$ (Fig. \ref{fig:definitions}g). Formally, the set of confluent lesions $\mathcal{B}^{C} \subset \mathcal{B}$ is defined by Eq. \ref{eq:confluent_lesions} (Table \ref{tab:definitions}). Thus, a connected component $B_j$ is confluent if at least two lesions $L_i, L_k$ from $\mathcal{L}$ partially overlap with $B_j$. The individual lesions within a confluent lesion are referred to as \textbf{confluent lesion units (CLUs)} (Fig. \ref{fig:definitions}h). Formally, the set of CLUs $\mathcal{L}^{CLU} \subset\mathcal{L}$ is defined by Eq. \ref{eq:clu} (Table \ref{tab:definitions}). Thus, each CLU $ L_i $ is a distinct lesion instance that overlaps at most partially with a connected component $ B_j $.

\paragraph{Extended definition of confluent lesions}
The standard definition of confluent lesions (Eq. \ref{eq:confluent_lesions}) may prove overly conservative in practice due to two key factors. First, MS lesions on T2-weighted (T2w) images are often surrounded by non-specific hyper-intensities that are difficult to distinguish---even for trained radiologists---and are frequently included in segmentation masks. As a result, automated methods tend to produce larger connected regions than those seen in reference annotations, especially when trained on masks that do not separate these effects (e.g., SAMSEG in Fig.~\ref{fig:definitions}d). Second, manual delineation variability and partial volume effects can fragment visually confluent lesions in the reference masks.

To address these limitations and better reflect clinical confluence, we extend the definition of confluent lesions via morphological dilation. We first obtain $S^+$ by applying a single-iteration binary dilation to the semantic segmentation mask $S$, then extract the set of dilated connected components $\mathcal{B}^+$ via connected component analysis. Substituting $B$ with $B^+$ in Eq.~\ref{eq:confluent_lesions} yields $\mathcal{B}^{C+}$, the set of extended confluent lesions---dilated components containing multiple lesion instances (Fig.~\ref{fig:definitions}l). From $\mathcal{B}^{C+}$, we define the set of extended CLUs $\mathcal{L}^{CLU+}$ as all lesion instances that share a dilated component with at least one other lesion (Eq.~\ref{eq:extended_clu}, Table~\ref{tab:definitions}). For readability, we refer to $\mathcal{L}^{CLU+}$ as CLU+ throughout the manuscript.

\begin{table}[ht]
\centering
\resizebox{\textwidth}{!}{%
\begin{tabular}{@{}m{0.5\textwidth}>{\raggedright\arraybackslash}m{\textwidth}@{}}
\Xhline{4\arrayrulewidth}
\rule{0pt}{3ex}
\textbf{Concept} & \rule{0pt}{3ex} \textbf{Mathematical definition} \\
\Xhline{4\arrayrulewidth}
CC-based Instance Segmentation & 
\begin{minipage}{\linewidth}
\setlength{\mathindent}{0pt}
\begin{equation}
\label{eq:cc_instance}
M_{\text{CC}}(x, y, z) = 
\begin{cases} 
j & \text{if } \exists B_j \in \mathcal{B} \text{ such that } (x, y, z) \in B_j \\
0 & \text{otherwise}.
\end{cases}
\end{equation}
\end{minipage} \\
\addlinespace[0.5em]
Confluent Lesions & 
\begin{minipage}{\linewidth}
\setlength{\mathindent}{0pt}
\begin{equation}
\label{eq:confluent_lesions}
    \begin{aligned}
        \mathcal{B}^{C} = \{B_j \in \mathcal{B} \mid \exists L_i, L_k \in \mathcal{L}, i \neq k,  \text{ such that } & \text{IoU}(L_i, B_j) > 0 \\ 
        \text{ and } & \text{IoU}(L_k, B_j) > 0\}.
    \end{aligned}
\end{equation}
\end{minipage} \\
\addlinespace[0.5em]
Confluent Lesion Units (CLUs) & 
\begin{minipage}{\linewidth}
\setlength{\mathindent}{0pt}
\begin{equation}
\label{eq:clu}
    \begin{aligned}
        \mathcal{L}^{CLU} & = \{k \in \mathcal{L} \mid \exists B_j \in \mathcal{B}^{C} \text{ such that }  \text{IoU}(L_k, B_j) > 0\} \\
                          & = \{k \in \mathcal{L} \mid \exists B_j \in \mathcal{B} \;\;\text{ such that }  0 < \text{IoU}(L_k, B_j) < 1\}.
    \end{aligned}
\end{equation}
\end{minipage} \\
\addlinespace[0.5em]
Extended CLUs (CLU+) & 
\begin{minipage}{\linewidth}
\setlength{\mathindent}{0pt}
\begin{equation}
\label{eq:extended_clu}
\begin{aligned}
\mathcal{L}^{CLU+} 
&= \{L_i \in \mathcal{L} 
 &\mid& \; \exists B_j \in \mathcal{B}^{C+} \text{ such that } \text{IoU}(L_i, B_j) \neq 0 \} \\
&= \{L_i \in \mathcal{L} 
 &\mid& \; \exists B_j \in \mathcal{B}^{+},\; \exists L_k \in \mathcal{L}, L_k \neq L_i \\
&&& \;\;\text{such that } \text{IoU}(L_i, B_j) \neq 0 \\
&&& \;\;\;\;\;\;\;\;\;\;\, \text{and } \text{IoU}(L_k, B_j) \neq 0 \}.
\end{aligned}
\end{equation}
\end{minipage} \\
\Xhline{4\arrayrulewidth}
\end{tabular}%
}
\caption{Summary of mathematical formulations. IoU stands for Intersection over Union.}
\label{tab:definitions}
\end{table}

\subsection{Related works on MS lesion instance segmentation}


\label{sec:instance_seg}
Instance segmentation methods have been widely investigated in computer vision applied to natural images, with extensive reviews highlighting the strengths and limitations of different approaches \citepmain{hafiz_survey_2020,gu_review_2022}. These methods are commonly categorized as either bottom-up (e.g. Panoptic DeepLab \citepmain{cheng_panoptic-deeplab_2020}) or top-down (e.g. Mask-RCNN \citepmain{he_masked_2021}). While end-to-end instance segmentation is common in biomedical imaging (e.g., for nuclei and glands~\citepmain{nasir_nuclei_2023}), radiological imaging typically relies on CC analysis for instance identification and downstream analysis \citepmain{isensee_nnu-net_2021,bilic_liver_2023}. Here, we review CC and ACLS, the only two methods proposed for this task in MS. 


\textit{Connected components} (CC) analysis (Fig.~\ref{fig:definitions}e) identifies lesion instances by grouping adjacent voxels based on a predefined connectivity structure (e.g., 6-, 18-, or 26-connectivity in 3D). While effective for separating spatially disconnected lesions, CC cannot distinguish multiple CLUs within confluent lesions---typically detecting only the instance with the largest overlap and missing the rest. Our previous work~\citepmain{wynen_lesion_2024} showed that even with ideal segmentation, using CC reduces lesion detection sensitivity by 9.3

\textit{Automated confluent lesion splitting} (ACLS), the only method specifically designed to address confluent lesions in MS, was introduced as part of a pipeline for detecting paramagnetic rim lesions~\citepmain{lou_fully_2021}. ACLS operates in two steps and requires a lesion probability map $S_p$, typically generated by a semantic segmentation tool. First, lesion centers are detected; then, each lesion voxel in the binary mask $S$ (obtained by thresholding $S_p$) is assigned to the nearest center, producing an instance map $M$. The key innovation lies in the center detection step, originally proposed by Dworkin et al.~\citepmain{dworkin_automated_2018} to handle lesion counting in presence of confluent regions. It hypothesizes that voxel probabilities peak at the lesion center---where blood-brain barrier breakdown begins---and decrease outward, reflecting radial inflammatory spread. Center voxels are identified as local maxima with negative Hessian eigenvalues in all three spatial directions, then clustered via CC. In the second step, each voxel in $S$ is assigned to its nearest detected center using a nearest-neighbor approach. Despite its innovative approach, ACLS has some limitations. First, its core assumption---that lesion voxel probabilities peak at the center and decrease peripherally---remains unproven. Second, the method does not consider lesion size during clustering. As illustrated in Fig. \ref{fig:acls_issue}, this can lead to incorrect assignments when a small lesion is confluent with a larger one, causing voxels from the larger lesion to be incorrectly attributed to the smaller due to proximity. Finally, ACLS has not been formally validated for lesion instance segmentation. This component was not the primary focus of the original study and has not been evaluated independently. Still, in our previous work~\citepmain{wynen_lesion_2024}, we compared ACLS to CC on a private dataset and found that it achieved higher lesion detection sensitivity (+19.1\% on average across segmentation tools).


\section{Proposed framework}
\begin{figure}[h!]
  \centering
  \includegraphics[width=\textwidth]{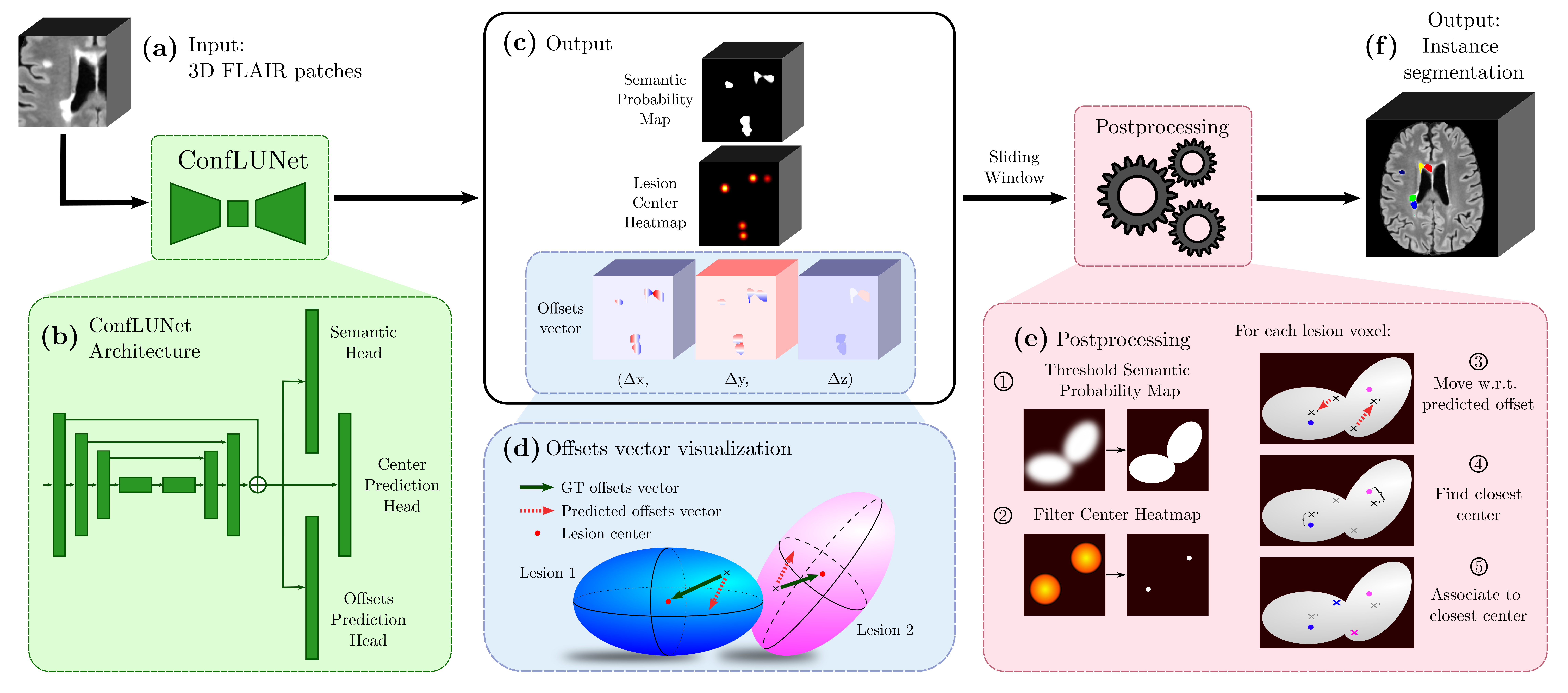}
  \caption{Illustration of ConfLUNet pipeline. 3D FLAIR patches (a) are provided as input to the ConfLUNet network (b), which simultaneously predicts a semantic probability map, a lesion center heatmap, and an offset vector map (c). For each voxel, the offset vector indicates the direction and distance to the center of the lesion to which it belongs (d). After applying a sliding window inference strategy, the outputs are post-processed (e) to generate the final instance segmentation map (f). This post-processing involves binarizing the semantic probability map using a fixed threshold, filtering the center heatmap to identify discrete lesion centers, and assigning each predicted lesion voxel to the center that lies the closest to their initial position shifted by their predicted offset vector (e).}
  \label{fig:conflunet_pipeline}
\end{figure}

ConfLUNet is a 3D instance segmentation framework for MS lesions in MRI, integrating strengths from two established architectures: nnUNet~\citepmain{isensee_nnu-net_2021} and Panoptic DeepLab~\citepmain{cheng_panoptic-deeplab_2020}. It builds upon nnUNet's robust, self-configuring 3D U-Net backbone---including components like data fingerprinting, automatic parameter tuning, and data augmentation---while incorporating key ideas from Panoptic DeepLab for instance-aware representation learning. ConfLUNet's pipeline is described in Fig. \ref{fig:conflunet_pipeline}. Additionally, its source code is open, available on GitHub\footnote{\url{https://github.com/maxencewynen/ConfLUNet/}}, and containerized using Docker\footnote{\url{https://hub.docker.com/repository/docker/petermcgor/conflunet}}.

\label{sec:arch}

\subsection{Architecture}
The resulting architecture (Fig.~\ref{fig:conflunet_pipeline}b) follows a standard encoder-decoder U-Net design with skip connections and three output heads: one for semantic segmentation, one for lesion center detection, and one for predicting offset vectors from voxels positions to their corresponding centers. These additions enable ConfLUNet to capture both the extent and the instance structure of confluent MS lesions.
ConfLUNet's three heads are as follows:
\begin{enumerate}
\item \textbf{Semantic Segmentation:} This head, inherited from nnUNet, outputs a two-channel semantic map (background/lesion). The loss $L_{\rm seg}$ combines focal loss and Dice loss: $L_{\rm seg} = L_{\rm focal} + \gamma L_{\rm dice}$, where $L_{\rm focal}$ mitigates class imbalance by focusing on hard voxels, and $L_{\rm dice}$ measures overlap with reference. We use $\gamma = 0.5$, following prior work~\citepmain{malinin_shifts_2022}.

\item \textbf{Center Prediction:} This head outputs a heatmap where each voxel represents the probability of being a lesion center. The reference is a binary mask of lesion centers of mass $\mathcal{C} = {\mathcal{C}_k = (\mathcal{C}_k^x, \mathcal{C}_k^y, \mathcal{C}_k^z) ; | ; k \in \mathcal{L}}$, convolved with a 3D Gaussian ($\sigma=2$ voxels). The loss $L_{\rm center}$ is the mean squared error between predicted and reference heatmaps.
    
\item \textbf{Offset Prediction:} This head predicts 3D offset vectors from each voxel's position to the center of its lesion instance (Fig.~\ref{fig:conflunet_pipeline}c,d). During training, reference offsets are computed as the difference between each lesion voxel's coordinates $(x, y, z)$ and its corresponding lesion center:
    
$$
    \mathcal{O}(x, y, z) = 
    \begin{cases} 
    (\mathcal{C}_k^x - x, \mathcal{C}_k^y - y, \mathcal{C}_k^z - z) & \text{if } (x, y, z) \in \mathcal{L}_k, \\
    (0, 0, 0) & \text{otherwise},
    \end{cases}
$$
The offset loss $L_{\rm offsets}$ is computed using the L1 norm.
\end{enumerate}

The final instance segmentation mask is obtained by combining semantic, center, and offset outputs during post-processing (Section~\ref{sec:Inference}). 


\subsection{Training}
ConfLUNet is trained using 5-fold cross-validation, with ensemble predictions applied at inference. The total loss combines semantic segmentation ($L_{\text{seg}}$), center prediction ($L_{\text{center}}$), and offset prediction ($L_{\text{offsets}}$) losses: $$L_{\text{total}} = L_{\text{seg}} + \alpha \cdot L_{\text{center}} + \beta \cdot L_{\text{offsets}}.$$
To balance magnitudes, we set $\alpha = 100$ and $\beta = 1$. Optimization is performed with Adam~\citepmain{kingma_adam_2017}, using a constant learning rate of 0.002 and weight decay of $3 \times 10^{-5}$. The final model is trained for 15,000 epochs, while hyperparameter tuning is conducted over 2500 epochs.

We adapt nnUNet’s patch sampling by disabling subject replacement and enforcing a 50\% probability of sampling patches containing lesion voxels. Validation is performed every 25 epochs via full inference on the validation set, using panoptic quality (Sec.~\ref{sec:evaluation}) as the epoch selection metric.

\subsection{Inference}
\label{sec:Inference}
The full inference process is shown in Figure \ref{fig:conflunet_pipeline}e. A sliding window with 50\% overlap and Gaussian importance weighting aggregates the three output patches (segmentation, center and offset maps).  The semantic segmentation mask $\hat{S}$ is obtained by thresholding the softmax output $\hat{S}_p$ at $0.5$. Lesion voxel clustering then starts by detecting predicted lesion centers $\hat{C} = \{\hat{\mathcal{C}}_k: (x_k, y_k, z_k)\}$ via max pooling and filtering of the center heatmap. 

An embedding space is created by adding each voxel’s predicted offset vector $\hat{\mathcal{O}}(x, y, z)$ to its spatial coordinates. Each lesion voxel is then assigned to the nearest predicted center in this space. Specifically, the instance \textit{id} \(\hat{c}(x, y, z)\) of a voxel is the ID of the closest predicted center after applying the offset, expressed as: 
$$\hat{\mathcal{C}}(x, y, z) = \underset{k}{\operatorname{argmin}} \left\| \hat{\mathcal{C}}_k - \left( (x, y, z) + \hat{\mathcal{O}}(x, y, z) \right) \right\|^2,$$ where $\hat{\mathcal{C}}_k$ denotes the coordinates of the $k$-th predicted center. Finally, following the clinical definition of an MS lesion \citepmain{grahl_evidence_2019}, predicted instances smaller than $3mm$ along any axis or under $14mm$\(^3\) in volume are removed. 

\section{Evaluation framework}
\subsection{Materials}
\label{sec:material}
\subsubsection{Dataset and image acquisition details}
This study used a private retrospective cohort of 65 patients recruited between January 2021 and April 2023 at the Saint-Luc University Hospital, Brussels Belgium. Following qualitative evaluation by clinical experts, two patients were excluded due to extreme lesion confluency, rendering their confluent lesions indiscernible. The remaining 63 patients (40 female) were diagnosed with MS per the 2017 McDonald criteria~\citepmain{thompson_diagnosis_2018}, including 45 with relapsing-remitting (RRMS), 11 secondary progressive (SPMS), and 7 primary progressive (PPMS) forms. The cohort’s \textit{mean} (\textit{median}, \textit{[range]}) age is 42.4$\pm$12.1 (42.0, [21.5-67.3]) years old, EDSS 2.7$\pm$2.0 (2.0, [0-7]), and disease duration 7.4$\pm$8.0 years (5.9, [0-33.3]).

The study explores cross-sectional MRI acquisitions conducted at 3T whole-body MR scanner. The protocol (detailed in~\ref{app:mri}) included 3D Fluid Attenuated Inversion Recovery (FLAIR), 3D Magnetization-Prepared Rapid Gradient Echo (MPRAGE) and 3D Echo Planar Imaging (EPI). FLAIR images were registered to the EPI magnitude images ($0.66 \times 0.66 \times 0.7 mm$) for another study, and MPRAGE was aligned to FLAIR. All registrations used ANTS rigid registration (v2.5.0) \citepmain{avants_symmetric_2008}. Brain masks were generated with Synthstrip \citepmain{hoopes_synthstrip_2022} on FLAIR images and applied to all registered scans. Data management and registration were performed using the BIDS Managing and Analysis Tool (BMAT) \citepmain{vanden_bulcke_bmat_2022}.

\subsubsection{Manual segmentation}
Manual instance segmentation was performed by a trained neurobiologist (A.S., 1.5 years’ experience) and reviewed by an experienced neurologist (P.M., >10 years). Annotations were based on FLAIR images, with MPRAGE and EPI phase images consulted for challenging confluent lesions. Only clearly inflammatory WM lesions $\geq$3 mm in diameter \citepmain{grahl_evidence_2019} were included, excluding infratentorial and cortical lesions. All annotations and visualizations were done using ITK-Snap (v3.8.0) \citepmain{yushkevich_itk-snap_2016}.

\begin{figure}[ht!]
  \centering
  \includegraphics[width=\textwidth]{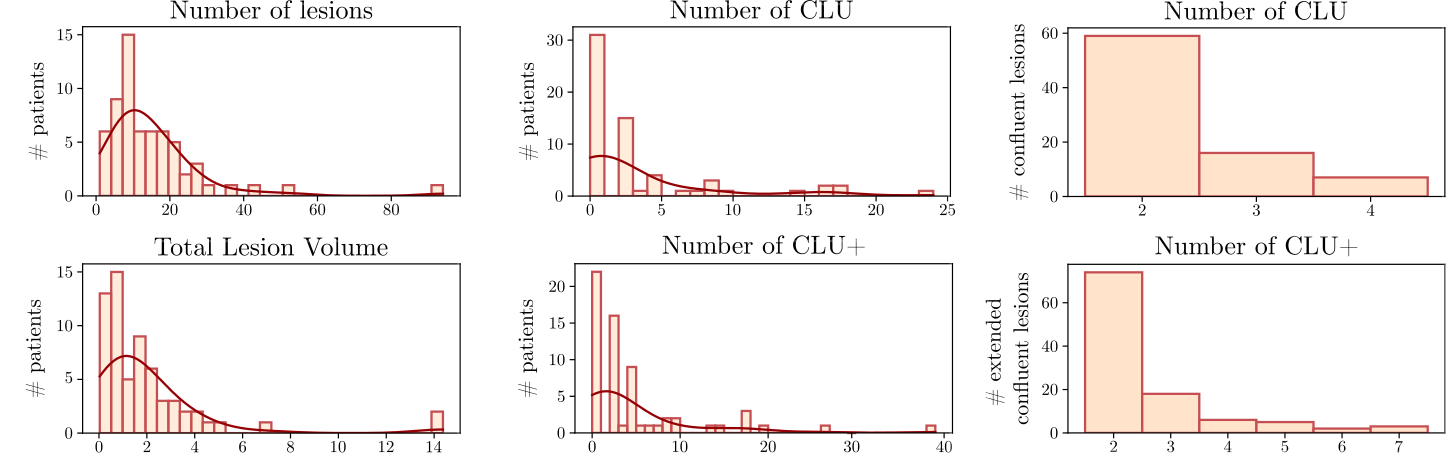}
  \caption{Histograms of lesion and CLU(+) distributions in the dataset.}
  \label{fig:material}
\end{figure}

The dataset included 973 lesions, averaging 15.4 per subject (median: 11; range: 1–94; standard deviation: 13.8). Total lesion volume averaged 2048.8 mm³ per subject (median: 1431.7 mm³; range: 33.0–14368.0 mm³; standard deviation: 2603.7 mm³). According to Section 2 definitions, 32 patients had at least two CLUs and 41 had at least two CLU+. On average, each subject had 3.2 CLUs and 4.5 CLU+, totaling 199 and 285 across the cohort. These distributions are shown in Figure~\ref{fig:material}.

\subsection{Evaluation metrics}
\label{sec:evaluation}

To better reflect clinical relevance, we evaluate MS lesion instance segmentation using established instance segmentation metrics. For completeness, we also report semantic segmentation and detection metrics, following the \textit{Metrics Reloaded} framework \citepmain{maier-hein_metrics_2024}. In addition, we introduce three custom metrics specifically designed to assess CLU detection. All metrics are computed per subject and averaged across the cohort.

\begin{figure}[h!]
  \centering
  \includegraphics[width=\textwidth]{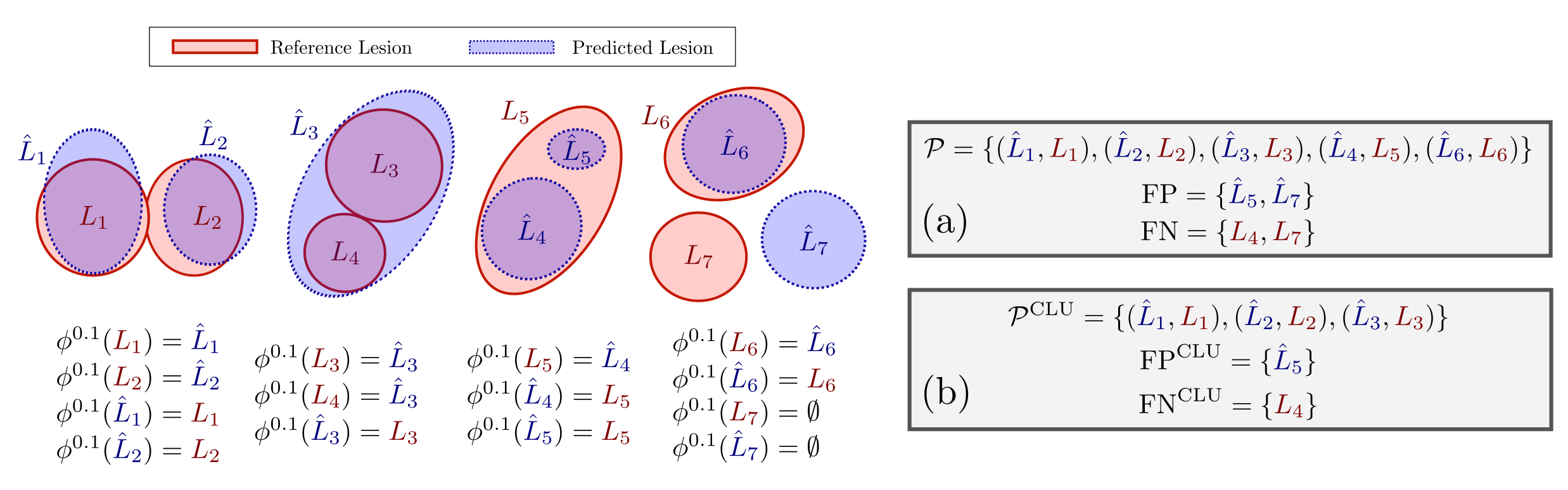}
  \caption{Examples of the mutually consistent lesion matching strategy and of $\mathcal{P}^{\text{CLU}}$ ($=\text{TP}^{\text{CLU}}$), $\text{FP}^{\text{CLU}}$, and $\text{FN}^{\text{CLU}}$.}
  \label{fig:matching_strategy}
\end{figure}

\paragraph{Lesion matching strategy} To compute instance-level metrics, we establish a one-to-one correspondence between predicted and reference lesions (Fig.~\ref{fig:matching_strategy}). Our strategy ensures: (i) each reference lesion is matched to the predicted lesion with the highest overlap (if IoU $\geq \lambda \in ]0,1]$); (ii) each predicted lesion is matched similarly to its best-overlapping reference; and (iii) only mutually consistent (bidirectional) matches are retained.
Formally, we define a matching function $\phi^\lambda: L_1 \rightarrow L_2 \cup {\emptyset}$ that assigns to each instance in $L_1$ its best match in $L_2$ if the maximum IoU exceeds $\lambda$, or $\emptyset$ otherwise:
\begin{equation}
\phi^{\lambda}(l_1) = 
\begin{cases}
\underset{l_2 \in L_2}{\text{argmax}} \, \text{IoU}(l_1, l_2) & \text{if } \text{IoU}(l_1, l_2) \geq \lambda \\
\emptyset & \text{otherwise}
\end{cases}
\end{equation}
While a 0.5 IoU threshold is common in 2D image tasks \citepmain{cheng_panoptic-deeplab_2020}, we use $\lambda = 0.1$ to accommodate the variability in 3D lesion size and shape \citepmain{maier-hein_metrics_2024}.

We define $\mathcal{P}$ as the set of mutually matched pairs between predicted instances $\hat{\mathcal{L}}$ and reference instances $\mathcal{L}$:
\[
\mathcal{P} = \left\{(i, j) \mid j = \phi^{\lambda}(i), i = \phi^{\lambda}(j), ~~  i \in \mathcal{\hat{L}}, j \in \mathcal{L} \right\}
\]
A pair $(i, j)$ is included in $\mathcal{P}$ only if $i$ is the best match for $j$ and vice versa. This strict 1:1 matching avoids cases where a single prediction covers multiple reference lesions, or a reference lesion is split across multiple predictions (Fig.~\ref{fig:matching_strategy}a).


\paragraph{Detection and semantic segmentation accuracy} From \(\mathcal{P}\), we derive the true positive (TP), false negative (FN), and false positives (FP) sets. Then, we compute overall lesion-wise Precision, Recall, and F1 score. 

Additionally, we assess the semantic segmentation performance using classical Dice score coefficient (DSC) and normalized DSC (nDSC) \citepmain{raina_tackling_2023} which reduces the DSC bias due to patients with different lesion loads.

\paragraph{Instance segmentation assessment} Panoptic Quality (PQ), our primary instance segmentation metric, combines Segmentation Quality (SQ) and Recognition Quality (RQ)~\citepmain{cheng_panoptic-deeplab_2020} as $\text{PQ} = \text{SQ} \times \text{RQ}$, where:
\begin{equation}
\text{SQ} = \frac{\underset{(i,j) \in \mathcal{P}}{\sum} \text{IoU}(i,j)}{|\text{TP}|} ;~~~ \text{RQ} = \frac{|\text{TP}|}{|\text{TP}| + \frac{1}{2}|\text{FP}| + \frac{1}{2}|\text{FN}|}.
\end{equation}

To evaluate CLU detection, we define the set of matched pairs $\mathcal{P}^{\text{CLU}}$ between predicted lesions and reference CLU instances:
\[
\mathcal{P}^{\text{CLU}} = \left\{(i, j) \mid j = \phi^{\lambda}(i), i = \phi^{\lambda}(j), ~~~ i \in \mathcal{\hat{L}}, j \in \mathcal{L}^{\text{CLU}}\right\}
\]
where $\mathcal{L}^{\text{CLU}}$ is the set of CLU reference instances (Table \ref{tab:definitions}). Then (Fig. \ref{fig:matching_strategy}b):

\begin{itemize}
    \item $\text{TP}^{\text{CLU}} = \mathcal{P}^{\text{CLU}}$: matched reference CLU considered correctly detected;
    \item $\text{FN}^{\text{CLU}} = \{ j \in \mathcal{L}^{\text{CLU}} \mid \nexists i \in \mathcal{\hat{L}}, (i, j) \in \mathcal{P}^{\text{CLU}}\}$: reference CLU with no matching prediction;
    \item $\text{FP}^{\text{CLU}} = \{i \in \mathcal{\hat{L}} \mid \exists j \in \mathcal{L}, j = \phi^{\lambda}(i) \land i \neq \phi^{\lambda}(j)\}$: predicted lesions that match to a reference lesion ($j = \phi^{\lambda}(i)$) but are \textit{not} the best match for that reference lesion ($i \neq \phi^{\lambda}(j)$)
\end{itemize} 

These sets are used to compute CLU-wise precision ($\text{Precision}^{\text{CLU}}$ sensitive to over-splitting), recall ($\text{Recall}^{\text{CLU}}$, sensitive to under-splitting) and the corresponding F1 score ($\text{F1}^{\text{CLU}}$).
Boundary conditions are handled as follows: if no confluent lesions are present in the reference annotations, $\text{Recall}^{\text{CLU}} = 1$ by definition. Similarly, $\text{Precision}^{\text{CLU}} = 1$ when both true and false positive CLU detections are absent ($\text{TP}^{\text{CLU}} = \text{FP}^{\text{CLU}} = 0$), reflecting that no incorrect predictions were made.

The same logic and evaluation procedure apply to CLU+ lesions, as defined in Eq.~\ref{eq:extended_clu} (Table ~\ref{tab:definitions}), with the corresponding metrics denoted as $\text{Precision}^{\text{CLU+}}$, $\text{Recall}^{\text{CLU+}}$, and $\text{F1}^{\text{CLU+}}$.

\subsection{Experimental setting}

\paragraph{Experiment 1: Instance Segmentation with CC and ACLS}
This experiment formally evaluated the instance segmentation performance of CC and ACLS using 3D FLAIR images from all 63 patients (Section~\ref{sec:material}). Probabilistic semantic segmentation masks ($S_p$) were generated using three established tools: (i) LST-LPA, a logistic regression model in SPM~\citepmain{schmidt_bayesian_2017}; (ii) FLAMeS, a nnUNet model pretrained for MS lesion segmentation~\citepmain{dereskewicz_flames_2025,la_rosa_flames_2023}; and (iii) SAMSEG~\citepmain{cerri_contrast-adaptive_2021}, a probabilistic atlas-based tool. 

All outputs were binarized at 0.5 and post-processed into instance masks using either CC or ACLS. CC used a 6-connectivity structure to avoid over-merging, while ACLS employed 26-connectivity for lesion center detection, which yielded better results. Finally, small predicted lesions were discarded based on earlier described strategy. This yielded six instance segmentation maps per patient, which were evaluated using the metrics described in Section~\ref{sec:evaluation}.

\paragraph{Experiments 2 and 3: Validating ConfLUNet}
To evaluate ConfLUNet, the dataset was split into training (n=50) and test (n=13) sets, ensuring balanced distribution of patients with confluent lesions. A baseline model, matching ConfLUNet’s architecture but excluding the center and offset heads, was trained with only the segmentation loss. Training followed nnUNet's pipeline~\citepmain{isensee_nnu-net_2021}, including a "Poly" learning rate scheduler (initial rate 0.01), patch sampling strategy, and 15,000 training epochs. Validation was performed every 25 epochs, computing PQ on the validation set. To compare ConfLUNet with CC and ACLS fairly, the baseline model’s outputs were post-processed using both CC and ACLS under the same conditions as in Experiment 1. The best-performing epoch was selected based on PQ, ensuring consistency across methods. Experiment 2 reports results from 5-fold cross-validation on the training set for ConfLUNet, U-Net+CC, and U-Net+ACLS, while Experiment 3 evaluates the final models on the held-out test set.


\subsection{Statistical Analyses}
\label{sec:stats}

Statistical analyses were conducted at the patient level using Python's SciPy library \citepmain{noauthor_scipyorg_nodate}. We used the Wilcoxon signed-rank test for paired method comparisons (with Holm-Bonferroni correction for multiple comparisons provided in supplementary materials). Bland-Altman plots were used to assess the agreement between predicted and reference lesion, CLU, and CLU+ counts. Spearman's rank correlation coefficient quantified relationships between means and differences. Statistical significance was set at p < 0.05.

\section{Results}
\subsection{Evaluating CC and ACLS for lesion instance segmentation}
\label{sec:ccvacls}

\begin{figure}[h!]
  \centering
  \includegraphics[width=0.7\textwidth]{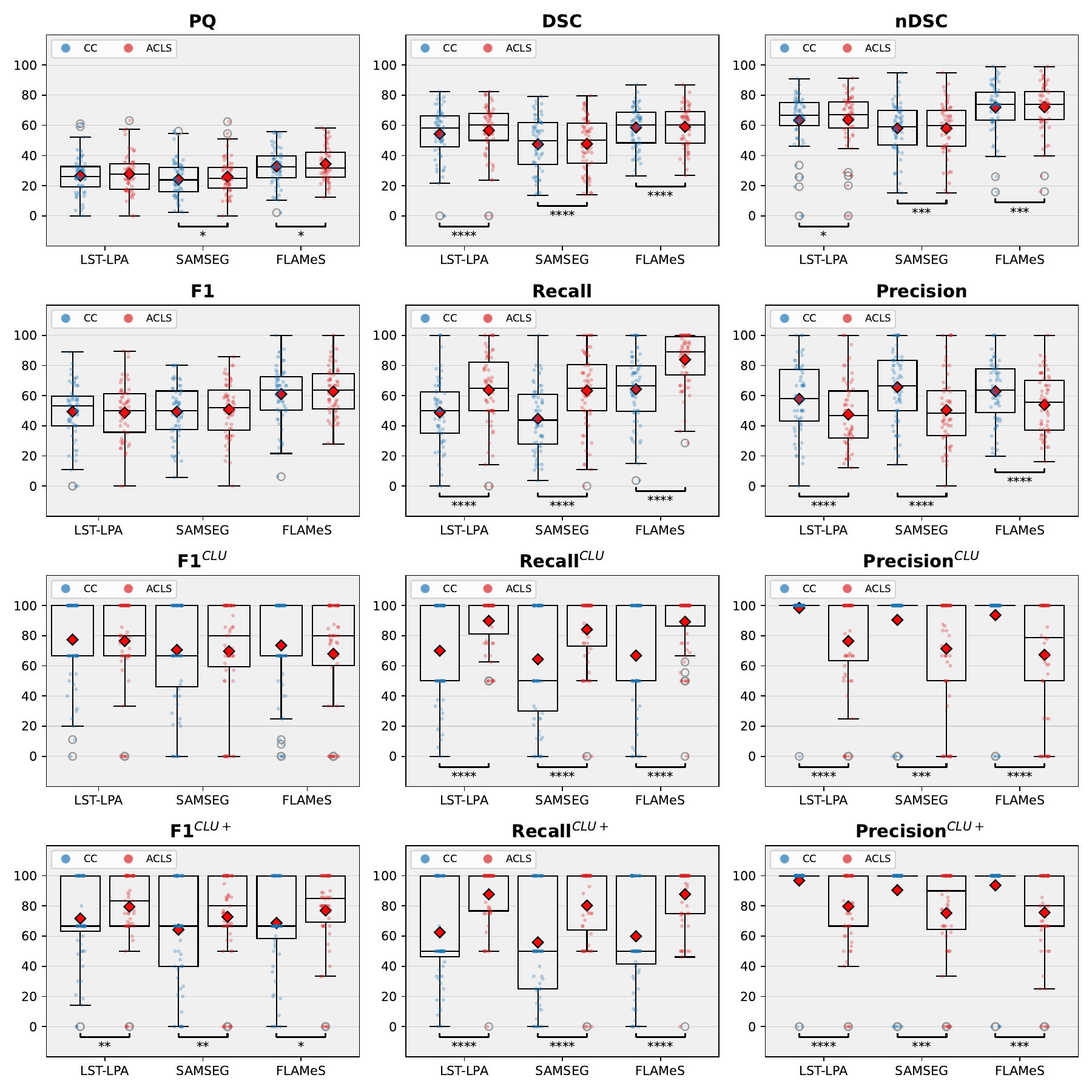}
  \caption{Box plots of the patient-wise metrics obtained for the three lesion segmentation tools, to which we applied either CC or ACLS. P-values are computed with the Wilcoxon signed rank test. ACLS: Automated Confluent Lesion Splitting; CC: Connected Components; DSC: Dice Score Coefficient; nDSC: normalized Dice Score Coefficient; CLU: Confluent Lesion Unit; CLU+: Extended definition of Confluent Lesion Units (Eq.~\ref{eq:extended_clu}).}
  \label{fig:lst_boxplots}
\end{figure}

Our first experiment, which compared the performance of CC and ACLS when applied to outputs from LST-LPA, SAMSEG, and FLAMeS, revealed distinct behavioral patterns between the two methods (Fig.~\ref{fig:lst_boxplots}). ACLS achieved slightly higher instance segmentation performance than CC (PQ: +1.6\% on average), with significant differences for SAMSEG and FLAMeS, but not for LST-LPA. It also outperformed CC in semantic segmentation metrics, with modest yet significant improvements in DSC (+1.0\%) and nDSC (+0.2\%), and a notably higher recall (+17.7\% on average). In contrast, CC yielded significantly higher precision (+11.5\%), but no significant differences were observed in overall lesion detection (F1). For CLU detection, CC consistently achieved significantly higher precision (+22.5\%), while ACLS showed superior recall (+20.7\%). These trends persisted when using the CLU+ definition. Although no significant difference was found in $\text{F1}^{\text{CLU}}$, the use of ACLS instead of CC led to significantly higher $\text{F1}^{\text{CLU+}}$ scores (+8.2\% on average). Nearly all findings were consistent across all three segmentation tools. Interestingly, limiting the analysis to patients with confluent lesions resulted in a substantial drop in CC’s CLU recall (LST-LPA: 70.0\%$\rightarrow$41.0\%; SAMSEG: 64.4\%$\rightarrow$29.8\%; FLAMeS: 66.8\%$\rightarrow$34.7\%), whereas ACLS experienced a more moderate decrease (89.8\%$\rightarrow$80.0\%; 84.1\%$\rightarrow$68.7\%; 89.2\%$\rightarrow$78.8\%).

\begin{figure}[h!]
  \centering
  \includegraphics[width=\textwidth]{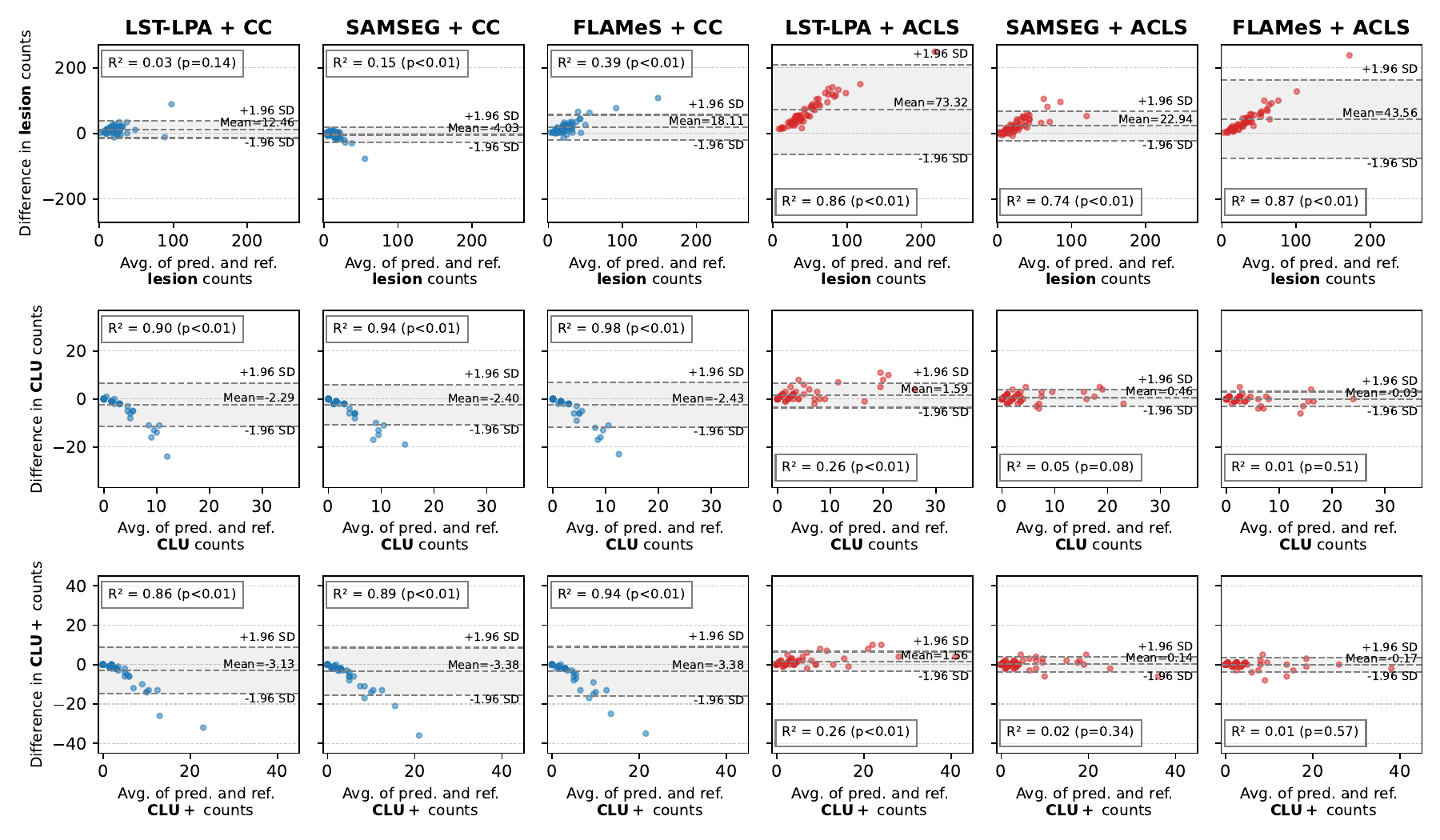}
  \caption{Bland-Altman plots for predicted and reference lesion/CLU/CLU+ counts and across masks provided by lesion segmentation tools to which CC/ACLS was applied. ACLS: Automated Confluent Lesion Splitting; CC: Connected Components; CLU: Confluent Lesion Unit; CLU+: Extended definition of CLUs.}
  \label{fig:ba_lst}
\end{figure}

Bland-Altman plots (Fig. \ref{fig:ba_lst}) for overall lesion count exhibit no clear trend across segmentation tools when using CC. In contrast, ACLS systematically overestimates lesion count, with this overestimation increasing alongside the average of predicted and reference counts. However, ACLS provides more accurate estimates for CLU and CLU+ counts, as indicated by lower mean absolute differences and narrower 95\% limits of agreement. Conversely, CC consistently underestimates these counts, with the difference between predicted and reference CLU count decreasing as their average increases.

\subsection{Cross-validation}
\label{sec:cv}
The results of the 5-fold CV analysis (Table \ref{tab:res_cv_gen}) revealed that ConfLUNet outperformed both CC and ACLS in instance segmentation (+2.3\% and +2.8\%, resp.) and overall lesion detection (F1 +1.4\% and +2.1\%). On the other hand, semantic segmentation-based approaches showed higher DSC (+2.4\%/2.3\% for CC/ACLS) and nDSC (+2.7\%/+2.1\% for CC/ACLS). CC and ACLS demonstrated comparable performance in terms of PQ, DSC, and F1, with their main difference lying in the balance between recall and precision: ACLS favors recall, while CC prioritises precision, following the trend found in Experiment 1 (Section \ref{sec:ccvacls}). ConfLUNet achieves a higher F1 score by striking a compromise between the two, attaining greater recall than CC (though lower than ACLS) and higher precision than ACLS (though lower than CC).

Regarding CLU detection (Table \ref{tab:res_cv_clu}), ConfLUNet achieved the highest F1\textsuperscript{CLU} score, with improvements of +4.1\% and +15.2\% compared to CC and ACLS, respectively, reflecting a strong balance between precision and recall. CC exhibited the highest Precision but the lowest Recall, while ACLS showed the highest Recall and the lowest Precision. These trends persisted upon further examination of CLU+ (eq. (\ref{eq:extended_clu})). All three methods showed moderate gains in precision and slight declines in recall, with corresponding changes in $\text{F1}^{\text{CLU+}}$: ConfLUNet and CC experienced slight decreases (-1.1\% and -1.5\%, respectively), while ACLS improved (+6.2\%). Notably, CC achieved near-perfect precision (99.8\%) under the CLU+ definition.


\begin{table}[h!]
\centering
\setlength{\extrarowheight}{0pt}
\addtolength{\extrarowheight}{\aboverulesep}
\addtolength{\extrarowheight}{\belowrulesep}
\setlength{\aboverulesep}{0pt}
\setlength{\belowrulesep}{0pt}
\resizebox{\textwidth}{!}{
\begin{tabular}{l||c:cc:cccc}
\Xhline{6\arrayrulewidth}
\multicolumn{1}{c||}{Method} & PQ (\%) & DSC (\%) & nDSC (\%) & F1 (\%) & Recall (\%) & Precision (\%) \\ \Xhline{6\arrayrulewidth}
3D U-Net + CC & 47.9 $\pm$ 4.6 & \textbf{70.7 $\pm$ 2.9} & \textbf{77.5 $\pm$ 1.5} & 74.9 $\pm$ 4.4 & 72.3 $\pm$ 7.5 & \textbf{81.9 $\pm$ 3.7}\\
3D U-Net + ACLS & 47.4 $\pm$ 3.2 & 70.6 $\pm$ 2.8 & 76.9 $\pm$ 1.7 & 74.2 $\pm$ 3.1 & \textbf{79.0 $\pm$ 4.2} & 72.6 $\pm$ 3.9 \\
ConfLUNet (ours) & \textbf{50.2 $\pm$ 3.0} & 68.3 $\pm$ 3.2 & 74.8 $\pm$ 3.9 & \textbf{76.3 $\pm$ 3.7} & 75.6 $\pm$ 9.1 & 80.2 $\pm$ 3.3 \\
\Xhline{6\arrayrulewidth}

\end{tabular}
}
\caption{Results of the 5-fold cross-validation comparison between ConfLUNet and 3D U-Net semantic segmentation baseline model + Connected Components (CC) or Automated Confluent Lesion Splitting (ACLS) as post-processing. Variations across folds are indicated. Column groups separated by dashed lines represent the three task evaluations: instance segmentation, semantic segmentation and detection. PQ: Panoptic Quality, DSC: Dice Score, nDSC: normalized dice score.}
\label{tab:res_cv_gen}
\end{table}

\begin{table}[h!]
\centering
\setlength{\extrarowheight}{0pt}
\addtolength{\extrarowheight}{\aboverulesep}
\addtolength{\extrarowheight}{\belowrulesep}
\setlength{\aboverulesep}{0pt}
\setlength{\belowrulesep}{0pt}
\resizebox{\textwidth}{!}{
\begin{tabular}{l||ccc:ccc}
\Xhline{6\arrayrulewidth}
\multicolumn{1}{c||}{Method} & F1$^{\text{CLU}}$ (\%) & Recall$^{\text{CLU}}$ (\%) & Precision$^{\text{CLU}}$ (\%) & F1$^{\text{CLU+}}$ (\%) & Recall$^{\text{CLU+}}$ (\%) & Precision$^{\text{CLU+}}$ (\%) \\ \Xhline{6\arrayrulewidth}
3D U-Net + CC    & 79.9 $\pm$ 5.3 & 72.0 $\pm$ 6.5 & \textbf{97.7 $\pm$ 3.9} & 78.4 $\pm$ 5.8 & 69.3 $\pm$ 7.4 & \textbf{99.8 $\pm$ 0.4} \\
3D U-Net + ACLS  & 68.8 $\pm$ 9.3 & \textbf{88.1 $\pm$ 2.9} & 70.1 $\pm$ 9.9 & 74.6 $\pm$ 8.3 & \textbf{86.8 $\pm$ 3.9} & 75.9 $\pm$ 8.1 \\
ConfLUNet (ours) & \textbf{84.0 $\pm$ 5.3} & 80.0 $\pm$ 8.2 & 92.8 $\pm$ 4.8 & \textbf{82.9 $\pm$ 2.8} & 77.1 $\pm$ 6.2 & 95.5 $\pm$ 4.5 \\
\Xhline{6\arrayrulewidth}

\end{tabular}
}
\caption{Confluent lesion unit (CLU) detection performance resulting from the 5-fold cross-validation comparison between ConfLUNet and 3D U-Net semantic segmentation baseline model + Connected Components (CC) or Automated Confluent Lesion Splitting (ACLS) as post-processing. Variations across folds are indicated.}
\label{tab:res_cv_clu}
\end{table}

\subsection{Performance on held-out test set}
\begin{figure}[h!]
  \centering
  \includegraphics[width=\textwidth]{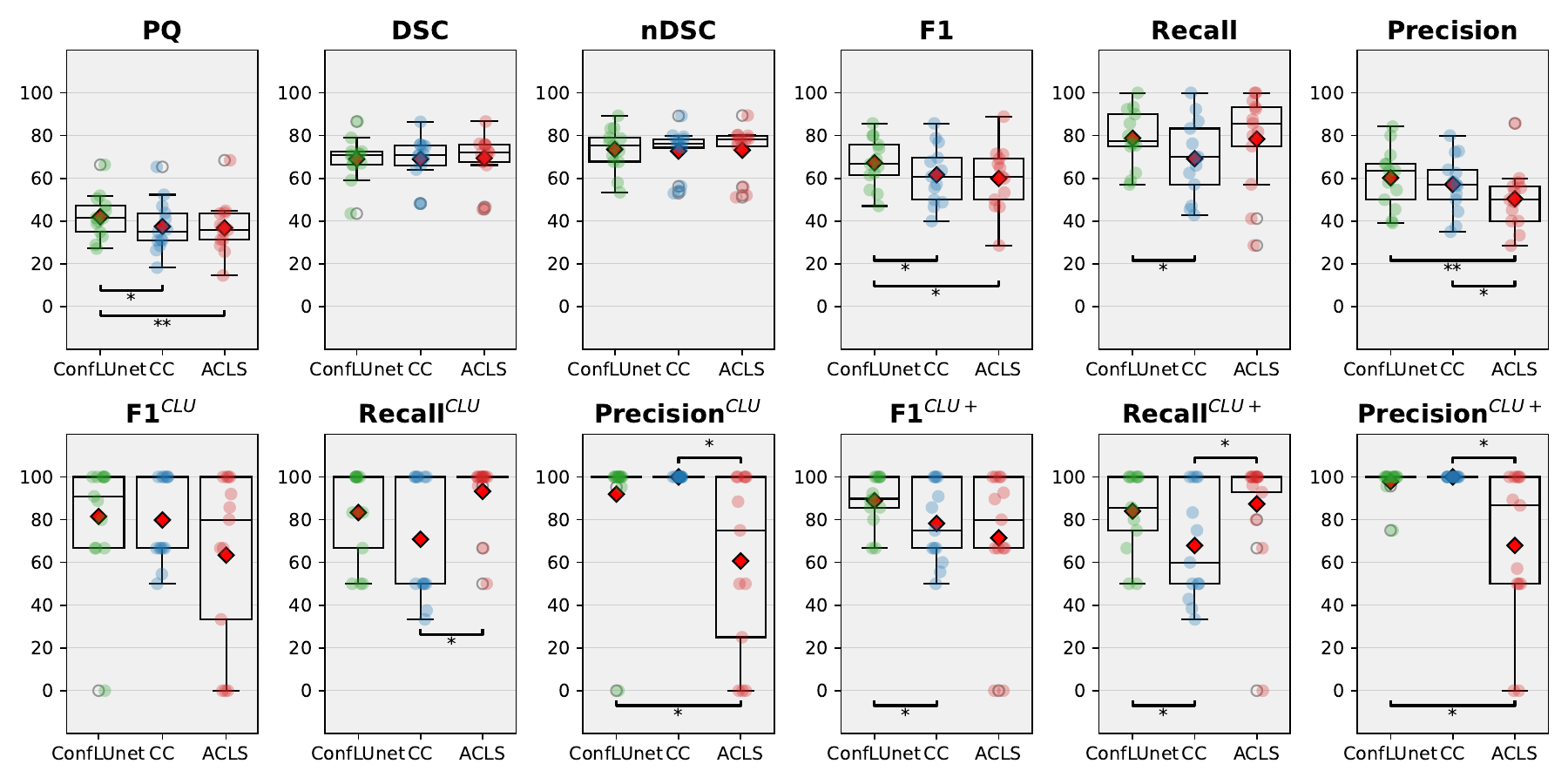}
  \caption{Box plots of the patient-wise general metrics obtained for the three models. The p-values are computed with the Wilcoxon signed rank test. ACLS: Automated Confluent Lesion Splitting; CC: Connected Components; DSC: Dice Score Coefficient; nDSC: normalized Dice Score Coefficient; CLU: Confluent Lesion Unit; CLU+: Extended definition of Confluent Lesion Units (Eq.~\ref{eq:extended_clu}).}
  \label{fig:boxplots}
\end{figure}

Similar to nnUNet, we performed ensemble inference on the independent test set—including 13 patients—using the models trained on each fold. Among these patients, seven presented at least two CLUs, yielding a total of 41 CLUs (range: 0–24 per patient), and ten had at least two CLU+, totaling 66 (range: 0–26 per patient). Table \ref{tab:res_gen} and Fig. \ref{fig:boxplots} highlight ConfLUNet’s improved performance across instance segmentation and general lesion detection metrics on the held-out test set. ConfLUNet significantly outperformed both CC and ACLS in PQ (resp. +4.5\%, $p=0.017$ and +5.2\%, $p=0.005$) and F1 score (resp. +4.5\%, $p=0.028$ and +5.2\%, $p=0.013$). We found no significant difference in semantic segmentation metrics across methods. In terms of detection, ConfLUNet achieved the highest Recall, reaching a significant difference only against CC ($p=0.015$), and the highest Precision, reaching a significant difference only against ACLS ($p=0.003$). Notably, all methods exhibited a general decrease in performance on the test set compared to the 5-fold CV results (Section \ref{sec:cv}).

Regarding confluent lesions (Table~\ref{tab:res_clu}, Fig.~\ref{fig:boxplots}), ConfLUNet achieved the highest $\text{F1}^{\text{CLU}}$ score, surpassing CC by 1.7\% and ACLS by 18.1\%, though without statistical significance. Compared to CC, it showed significantly higher recall (+12.5\%, $p=0.015$) and slightly lower precision (-8.1\%, not significant). Against ACLS, ConfLUNet achieved significantly higher precision (+31.2\%, $p=0.003$), with a non-significant drop in recall (-10.0\%). Under the extended CLU+ definition (Eq.~\ref{eq:extended_clu}), ConfLUNet improved its own $\text{F1}^{\text{CLU}}$ by 7.2\%, while CC declined (-1.6\%) and ACLS gained 6.2\%. The improvement over CC was significant ($p=0.046$), but not over ACLS ($p=0.08$). ConfLUNet also showed significantly higher recall than CC (+16.1\%, $p=0.015$), and significantly higher precision than ACLS (+29.7\%, $p=0.011$), with no significant difference in precision compared to CC. Overall, ConfLUNet achieves a more robust balance between precision and recall, particularly under the clinically motivated CLU+ definition.

When computing metrics across all lesions in the test set—rather than averaging patient-wise—to obtain a global view of CLU detection performance (Table~\ref{tab:res_clu_lesionwise}), the gap between ConfLUNet and CC widens for both $\text{F1}^{\text{CLU}}$ (+25.6\%) and $\text{F1}^{\text{CLU+}}$ (+22.0\%). Interestingly, CC’s recall drops strikingly to 41.5\% for CLU and 50.0\% for CLU+, whereas all metrics are relatively similar to their patient-wise equivalent for ConfLUNet and ACLS. This lesion-wise analysis is valuable, as it gives more weight to patients with a higher number of confluent lesions—typically those with longer disease duration and/or more aggressive progression, for whom accurately separating these confluent lesions is clinically most relevant.

\begin{figure}[h!]
  \centering
  \includegraphics[width=\textwidth]{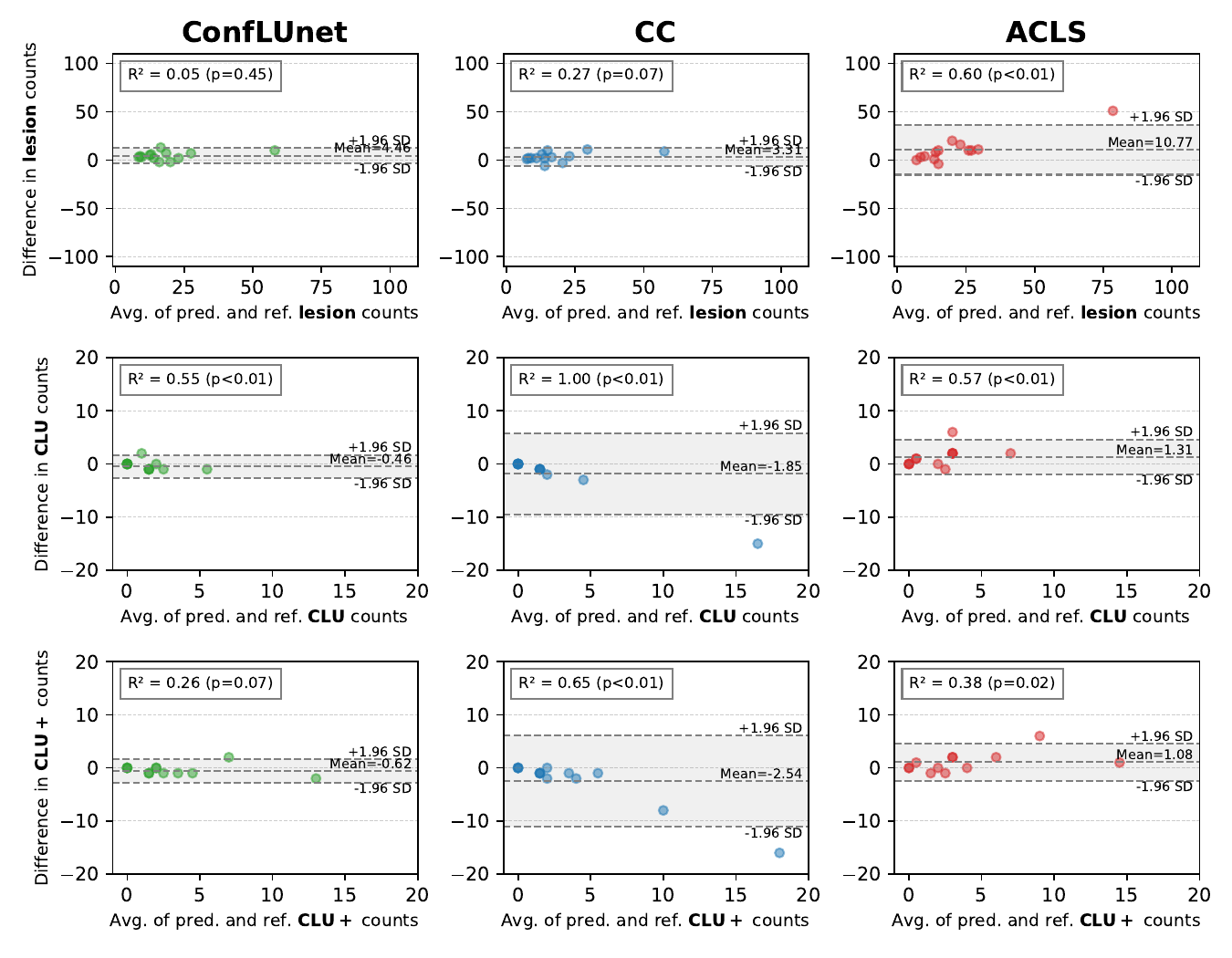}
  \caption{Bland-Altman plot for predicted and reference lesion/CLU/CLU+ counts. Comparison between ConfLUNet and 3D U-Net semantic segmentation baseline model + Connected Components (CC) or Automated Confluent Lesion Splitting (ACLS) as post-processing. CLU: Confluent Lesion Unit; CLU+: Extended definition of CLUs.}
  \label{fig:ba}
\end{figure}

\begin{figure}[h!]
  \centering
  \includegraphics[width=\textwidth]{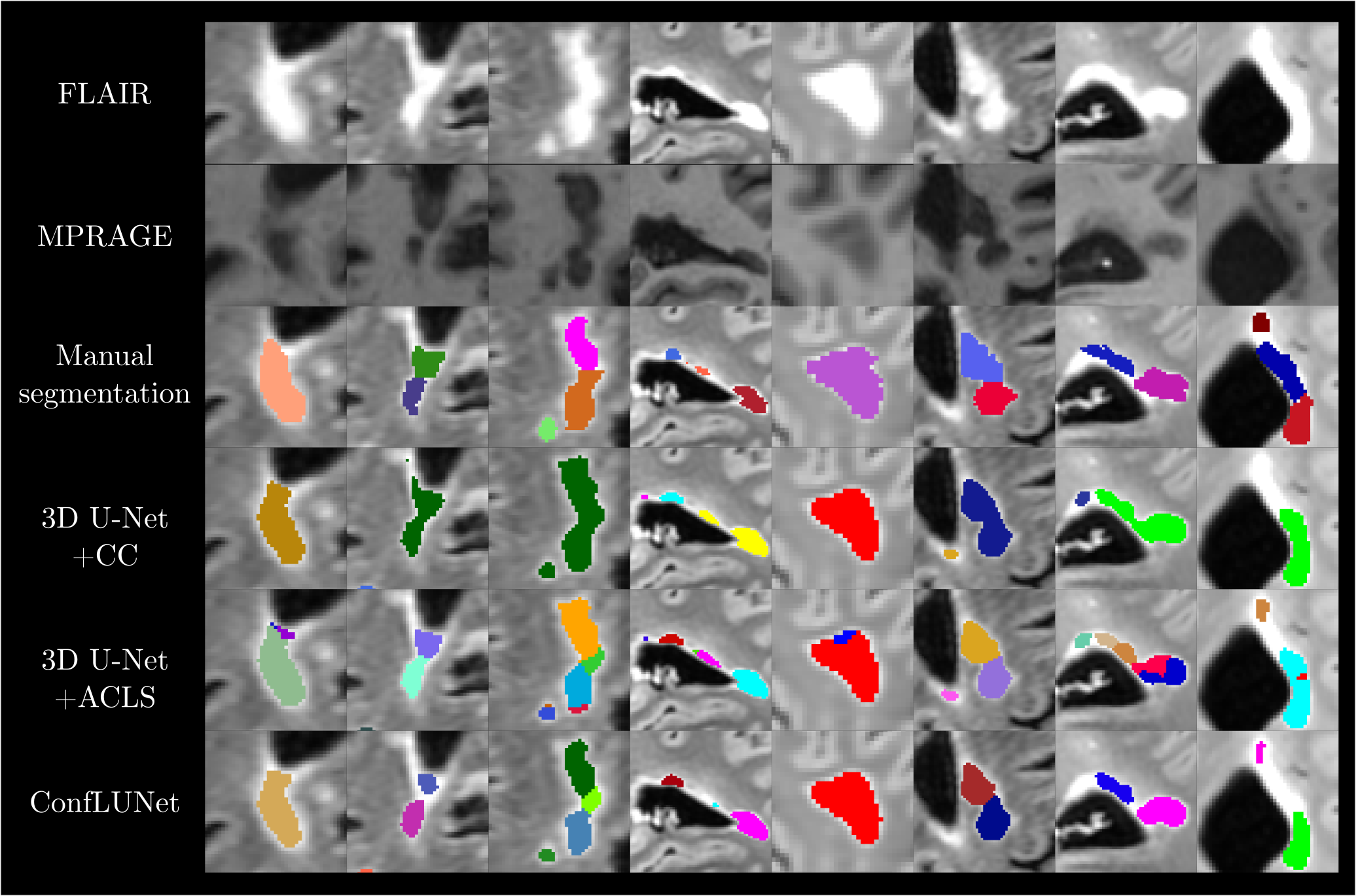}
  \caption{Illustrative examples of predictions made by connected components (CC), automated confluent splitting (ACLS) and ConfLUNet on the test set.}
  \label{fig:qualitative_results}
\end{figure}

Finally, Bland–Altman plots (Fig. \ref{fig:ba}) reveal similar trends as described in Section \ref{sec:ccvacls}, with CC underestimating CLU count and ACLS overestimating overall lesion count. ConfLUNet, however, exhibited a larger bias compared to CC (resp. 4.5 and 3.3), but smaller compared to ACLS (10.8), and weaker trends compared to both CC and ACLS, suggesting greater consistency across lesion load levels. ConfLUNet’s errors in estimating CLU counts were less systematically affected by the reference CLU count, as indicated by the lower coefficient of determination. Illustrative examples of predictions from ConfLUNet, CC and ACLS on the test set are shown in Fig. \ref{fig:qualitative_results}.

\begin{table}[h!]
\centering
\setlength{\extrarowheight}{0pt}
\addtolength{\extrarowheight}{\aboverulesep}
\addtolength{\extrarowheight}{\belowrulesep}
\setlength{\aboverulesep}{0pt}
\setlength{\belowrulesep}{0pt}
\resizebox{\textwidth}{!}{
\begin{tabular}{l||c:cc:cccc}
\Xhline{6\arrayrulewidth}
\multicolumn{1}{c||}{Method} & PQ (\%) & DSC (\%) & nDSC (\%) & F1 (\%) & Recall (\%) & Precision (\%) &  DiC $\searrow$ \\ \Xhline{6\arrayrulewidth}
3D U-Net + CC    & 37.5 $\pm$ 11.9 & 69.0 $\pm$ 10.3 & 72.7 $\pm$ 10.7 & 61.6 $\pm$ 13.2 & 69.2 $\pm$ 17.5 & 57.2 $\pm$ 12.9 & \textbf{4.7 $\pm$ 3.3}  \\
3D U-Net + ACLS  & 36.8 $\pm$ 12.1 & \textbf{69.5 $\pm$ 11.1} & 73.2 $\pm$ 11.6 & 59.9 $\pm$ 14.5 & 78.4 $\pm$ 21.7 & 50.4 $\pm$ 13.8 & 11.4 $\pm$ 12.7  \\
ConfLUNet (ours) & \textbf{42.0 $\pm$ 10.2}  & 69.0 $\pm$ 9.7  & \textbf{73.6 $\pm$ 9.7} & \textbf{67.3 $\pm$ 11.2} & \textbf{78.8 $\pm$ 13.0} & \textbf{60.2 $\pm$ 13.6} & 5.1 $\pm$ 3.3  \\
\Xhline{6\arrayrulewidth}

\end{tabular}
}
\caption{Performance on the test set (average and standard deviation) of ConfLUNet and 3D U-Net semantic segmentation baseline model + Connected Components (CC) or Automated Confluent Lesion Splitting (ACLS) as post-processing. Column groups separated by dashed lines represent the three task evaluations: instance segmentation, semantic segmentation and detection. PQ: Panoptic Quality, DSC: Dice Score, nDSC: normalized dice score, DiC: Absolute Different in Count.}
\label{tab:res_gen}
\end{table}

\begin{table}[h!]
\centering
\setlength{\extrarowheight}{0pt}
\addtolength{\extrarowheight}{\aboverulesep}
\addtolength{\extrarowheight}{\belowrulesep}
\setlength{\aboverulesep}{0pt}
\setlength{\belowrulesep}{0pt}
\resizebox{\textwidth}{!}{
\begin{tabular}{l||ccc:ccc}
\Xhline{6\arrayrulewidth}
\multicolumn{1}{c||}{Method} & F1$^{\text{CLU}}$ (\%) & Recall$^{\text{CLU}}$ (\%) & Precision$^{\text{CLU}}$ (\%) & F1$^{\text{CLU+}}$ (\%) & Recall$^{\text{CLU+}}$ (\%) & Precision$^{\text{CLU+}}$ (\%) \\ \Xhline{6\arrayrulewidth}
3D U-Net + CC    & 79.8 $\pm$ 19.3 & 70.8 $\pm$ 27.4 & \textbf{100.0 $\pm$ 0.0} & 78.2 $\pm$ 17.6 & 67.9 $\pm$ 24.7 & \textbf{100.0 $\pm$ 0.0} \\
3D U-Net + ACLS  & 63.4 $\pm$ 39.2 & \textbf{93.3 $\pm$ 15.3} & 60.7 $\pm$ 40.4 & 71.5 $\pm$ 33.4 & \textbf{87.4 $\pm$ 27.0} & 68.0 $\pm$ 35.0 \\
ConfLUNet (ours) & \textbf{81.5 $\pm$ 27.0} & 83.3 $\pm$ 20.7 & 91.9 $\pm$ 26.6 & \textbf{88.9 $\pm$ 11.0} & 84.0 $\pm$ 17.6 & 97.7 $\pm$ 6.7 \\
\Xhline{6\arrayrulewidth}

\end{tabular}
}
\caption{Performance (average and standard deviation across patients) of confluent lesion unit (CLU) and extended confluent lesions (CLU+) (Eq.~\ref{eq:extended_clu}) detection on the test set. Comparison between ConfLUNet and 3D U-Net semantic segmentation baseline model + Connected Components (CC) or Automated Confluent Lesion Splitting (ACLS) as post-processing. }
\label{tab:res_clu}
\end{table}

\begin{table}[h!]
\centering
\setlength{\extrarowheight}{0pt}
\addtolength{\extrarowheight}{\aboverulesep}
\addtolength{\extrarowheight}{\belowrulesep}
\setlength{\aboverulesep}{0pt}
\setlength{\belowrulesep}{0pt}
\resizebox{\textwidth}{!}{
\begin{tabular}{l||ccc:ccc}
\Xhline{6\arrayrulewidth}
\multicolumn{1}{c||}{Method} & F1$^{\text{CLU}}$ (\%) & Recall$^{\text{CLU}}$ (\%) & Precision$^{\text{CLU}}$ (\%) & F1$^{\text{CLU+}}$ (\%) & Recall$^{\text{CLU+}}$ (\%) & Precision$^{\text{CLU+}}$ (\%) \\ \Xhline{6\arrayrulewidth}
3D U-Net + CC    & 58.6 & 41.5 & \textbf{100.0} & 66.7 & 50.0 & \textbf{100.0} \\
3D U-Net + ACLS  & 76.8 & \textbf{92.7} & 65.5 & 82.2 & \textbf{90.9} & 75.0 \\
ConfLUNet (ours) & \textbf{84.2} & 78.1 & 91.4 &\textbf{88.7} & 83.3 & 94.8 \\
\Xhline{6\arrayrulewidth}

\end{tabular}
}
\caption{Lesion-wise performance of confluent lesion unit (CLU) and extended confluent lesions (CLU+) (Eq.~\ref{eq:extended_clu})) detection on the test set. Comparison between ConfLUNet and 3D U-Net semantic segmentation baseline model + Connected Components (CC) or Automated Confluent Lesion Splitting (ACLS) as post-processing. }
\label{tab:res_clu_lesionwise}
\end{table}

\section{Discussion}

\paragraph{Call for an evaluation framework adapted to lesion instance segmentation in MS} Despite the current need for lesion-level delineation in MS, both for clinical \citepmain{thompson_diagnosis_2018,on_behalf_of_the_magnims_study_group_magnims_2015} and technical \citepmain{zhang_qsmrim-net_2022,wynen_longitudinal_2021,wynen_multi-modal_2024,maggi_b_2023,wynen_machine_2025} applications, current evaluation practices remain poorly aligned with the goals of instance segmentation. Most reported detection metrics are derived from semantic masks processed with CC, which lack the granularity needed to assess individual lesions within confluent regions. Although confluent lesions have been clinically described \citepmain{zivadinov_effect_2008,lassmann_multiple_2014}, no prior work has formally defined them or the distinct units they contain, making their systematic evaluation particularly difficult. To address these limitations, our work introduces formal definitions and proposes CLU-aware metrics grounded in mathematical definitions. By incorporating these metrics alongside instance segmentation metrics, the proposed framework moves toward evaluation practices that are more consistent with clinical and technical needs. In parallel, progress in developing new instance segmentation methods is also hindered by the lack of public benchmarks. This is likely related to the complexity and time required for manual segmentation, which can be infeasible in advanced disease stages or when image quality is insufficient. To drive progress in this area, the community must prioritize the development of reliable public benchmarks and adopt evaluation protocols that reflect the clinical and technical demands of instance-level lesion analysis.

\paragraph{CC underestimates CLU count and lacks sensitivity} The results from Experiments 1, 2 and 3 demonstrate that frequently adopted CC strategy systematically underestimates CLU counts, particularly in cases with higher amounts of confluent lesions. This limitation stems from CC's inherent inability to distinguish spatially connected and/or overlapping lesions, as it can only detect CLUs when lesions are perfectly separated. Whether this spatial separation is correctly done to this purpose is heavily dependent on the predicted lesion semantic segmentation mask. To illustrate this difficulty, none of the three semantic segmentation tools (LST-LPA, SAMSEG, FLAMeS) evaluated in this study produced adequately disconnected instances, as evidenced by CC’s CLU recall dropping below 50\% across all tools—yielding $\text{Recall}^{\text{CLU}}$ scores of 41.2\%, 30.6\%, and 36.2\% for LST-LPA, SAMSEG, and FLAMeS, respectively—when restricted to patients with confluent lesions. While CC achieved extraordinarily high precision in CLU and CLU+ detection across all experiments, this largely reflects its conservative nature, which limits oversplitting and aligns well with how these metrics penalize false positives. Experiment 3 further confirms that these results hold true even when the model was trained on qualitatively conservative annotations. Altogether, these findings suggest that CC, despite its widespread use in current methods and evaluation frameworks (Suppl. \ref{app:lit_review})), is ill-suited for accurate lesion instance segmentation in the presence of confluent lesions and should be reconsidered as a default post-processing choice.

\paragraph{ACLS overestimates lesion count and oversplits lesions} Originally designed to split confluent lesions, our study systematically evaluate ACLS within an instance segmentation framework for the first time, with a particular focus on CLU detection---addressing the original method's limitation of only considering global lesion counts. ACLS consistently and significantly outperforms CC in instance segmentation (PQ: +1.6\%) and shows markedly higher recall for both overall lesion detection (+17.7\%) and CLU-wise detection (Recall$^{\text{CLU}}$: +20.7\%; Recall$^{\text{CLU+}}$: +25.8\%). However, this increased sensitivity comes at a cost: ACLS systematically overestimates total lesion counts (bias: 73.3, Fig.~\ref{fig:ba_lst}) and exhibits low precision across all segmentation tools (47.5\%, 50.4\%, and 53.9\% for LST-LPA, SAMSEG, and FLAMeS, respectively), with approximately half of the predicted lesions being false positives. These findings, corroborated by visual inspection, suggest that ACLS frequently oversplits lesions. This behavior and its excessive sensitivity exhibited provides evidence against ACLS's original assumption, stating local maxima in probability maps correspond to true lesion centers. Interestingly, results from Experiment 1 show virtually the same effects in applying ACLS to either LST-LPA, SAMSEG or FLAMeS, which indicates that ACLS can be applied regardless to segmentation methods with different methodological roots.

This study could not replicate the performances reported in Dworkin et al. \citepmain{dworkin_automated_2018}, especially regarding the correlation between predicted and reference lesion count. Methodological choices might explain these differences, such as chosen connectivity structure or probability threshold---more conservative in Dworkin et al.'s study (0.3), provoking CC to create larger connected regions and thus underperform.

\paragraph{ConfLUNet represents a balanced precision-recall alternative} The proposed ConfLUNet is the first end-to-end framework for instance segmentation of MS lesions with a particular focus on improving detection of CLUs. Our results indicate that ConfLUNet achieves significantly higher instance segmentation performance than both CC and ACLS (resp. +4.5\% and +5.2\% on the test set). Compared to CC, which favors Precision, and ACLS, which prioritizes Recall, ConfLUNet achieves a more balanced lesion-wise and CLU-wise detection performance, combining excellent precision with high sensitivity (F1: +5.7\%/+7.4\%; F1$^{CLU}$: +1.7\%/+18.1\%; F1$^{CLU+}$: +10.7\%/+17.4\%). As further indicated by Fig. \ref{fig:ba}, ConfLUNet mitigates both CC's undercounting bias and ACLS's oversplitting tendency.

\paragraph{Limitations and future work} Several limitations should be noted. Our evaluation used a relatively small single-center dataset with limited cases with confluent lesions (35/63), potentially affecting statistical power and generalizability. Furthermore, among these cases, the maximum number of CLU per confluent lesion was lower ($\leq4$) than the dozens reported by Dworkin and colleagues \citepmain{dworkin_automated_2018}. Future validation should assess performance on cases with higher confluent lesion loads, though we suspect annotation complexity---and thus annotation noise---to increase with CLU density. Finally, the single-rater annotations may introduce bias, and the method's performance on multi-contrast or out-of-domain data remains untested. Despite these limitations, this work remains valuable by being the first to formally assess the shortcomings of existing methods for confluent lesion detection and by offering a comparative evaluation of established approaches (Connected Components and ACLS) against a novel strategy (ConfLUNet). Moreover, we believe that ConfLUNet constitutes a pioneering step toward end-to-end instance segmentation of MS lesions, thereby addressing confluent lesions, and lays the groundwork for broader adoption and future development in this domain. Future work includes annotating larger multi-centric datasets to extend our analysis and assess generalizability of all investigated methods, and the incorporation of test-time training to enhance ConfLUNet's generalizability \citepmain{gerin_exploring_2024}.

\section{Conclusion}
Current evaluation practices for MS lesion segmentation remain misaligned with clinical needs, particularly in their inability to assess individual lesions within confluent regions. To address this gap, we introduced a formal framework for evaluating lesion instance segmentation, and proposed new CLU-aware metrics to assess their detection, grounded in precise definitions of confluent lesions and their constituent units (CLUs). While existing methods like CC and ACLS exhibit opposite failure modes—merging versus oversplitting—our end-to-end approach, ConfLUNet, jointly optimizes detection and delineation, providing a promising direction for more accurate and clinically meaningful lesion-level analysis in MS. Further validation on larger and more heterogeneous multi-centric cohorts is warranted.

\section*{Acknowledgments}
The present research benefited from computational resources made available on Lucia, the Tier-1 supercomputer of the Walloon Region, infrastructure funded by the Walloon Region under the grant agreement n°1910247. The authors thank the study participants; the Radiology departments of the Cliniques universitaires Saint-Luc (UCLouvain, Brussels, Belgium); Colin Vanden Bulcke (Université Catholique de Louvain); Laurence Dricot (Université catholique de Louvain); Serena Borrelli (Université catholique de Louvain); François Guisset (Université catholique de Louvain); Benoît Gérin (Université catholique de Louvain); Karim El Khoury (Université catholique de Louvain) and Maxime Zanella (Université catholique de Louvain) for their precious help.

\section*{Funding}
\noindent\textbf{MW} was funded by the Walloon region under grant No. 2010235 (ARIAC by digitalwallonia4.ai), and by the Swiss Government Excellence Scholarship (No. 2021.0087).  \\
\textbf{PMG} acknowledges the CIBM Center for Biomedical Imaging, a Swiss research center of excellence founded and supported by Lausanne University Hospital (CHUV), University of Lausanne (UNIL), École polytechnique fédérale de Lausanne (EPFL), University of Geneva (UNIGE) and Geneva University Hospitals (HUG).  \\
\textbf{AS} has the financial support of the Fédération Wallonie Bruxelles – FRIA du Fonds de la Recherche Scientifique – FNRS.  \\
\textbf{PM} research activity is supported by the Fondation Charcot Stichting Research Fund 2023, the Fund for Scientific Research (F.S.R, FNRS; grant 40008331), Cliniques universitaires Saint-Luc “Fonds de Recherche Clinique” and Biogen.  \\
\textbf{BM} is funded by Université Catholique de Louvain. \\
\textbf{MBC} acknowledges the CIBM Center for Biomedical Imaging, a Swiss research center of excellence founded and supported by Lausanne University Hospital (CHUV), University of Lausanne (UNIL), École Polytechnique Fédérale de Lausanne (EPFL), University of Geneva (UNIGE) and Geneva University Hospitals (HUG).

\section*{Compliance with ethical standards}
All procedures performed in this study were in accordance with the ethical standards of the Belgian national research committee (n° B4032020000104). Written informed consent was obtained for experimentation with human subjects following the World Medical Association Declaration of Helsinki.

\section*{Author contributions}
\noindent\textbf{MW}: Conceptualization, Data curation, Formal analysis, Investigation, Methodology, Software, Validation, Visualization, Writing – original draft, Writing – review and editing \\
\textbf{PMG}: Formal analysis, Investigation, Methodology, Software, Writing – original draft, Writing – review and editing \\
\textbf{MI}: Conceptualization, Supervision, Investigation, Methodology, Software, Writing – review and editing \\
\textbf{AS}: Data curation, Visualization, Writing – review and editing \\
\textbf{PM}: Conceptualization, Funding acquisition, Resources, Supervision, Project administration, Writing – review and editing \\
\textbf{BM}: Conceptualization, Funding acquisition, Resources, Supervision, Writing – review and editing \\
\textbf{MBC}: Conceptualization, Investigation, Methodology, Supervision, Writing – original draft, Writing – review and editing

\bibliographystylemain{elsarticle-num}
\bibliographymain{references}

\newpage
\appendix

\section{Illustration of unsplittable lesions}
\begin{figure}[h]
  \centering
  \includegraphics[width=\textwidth]{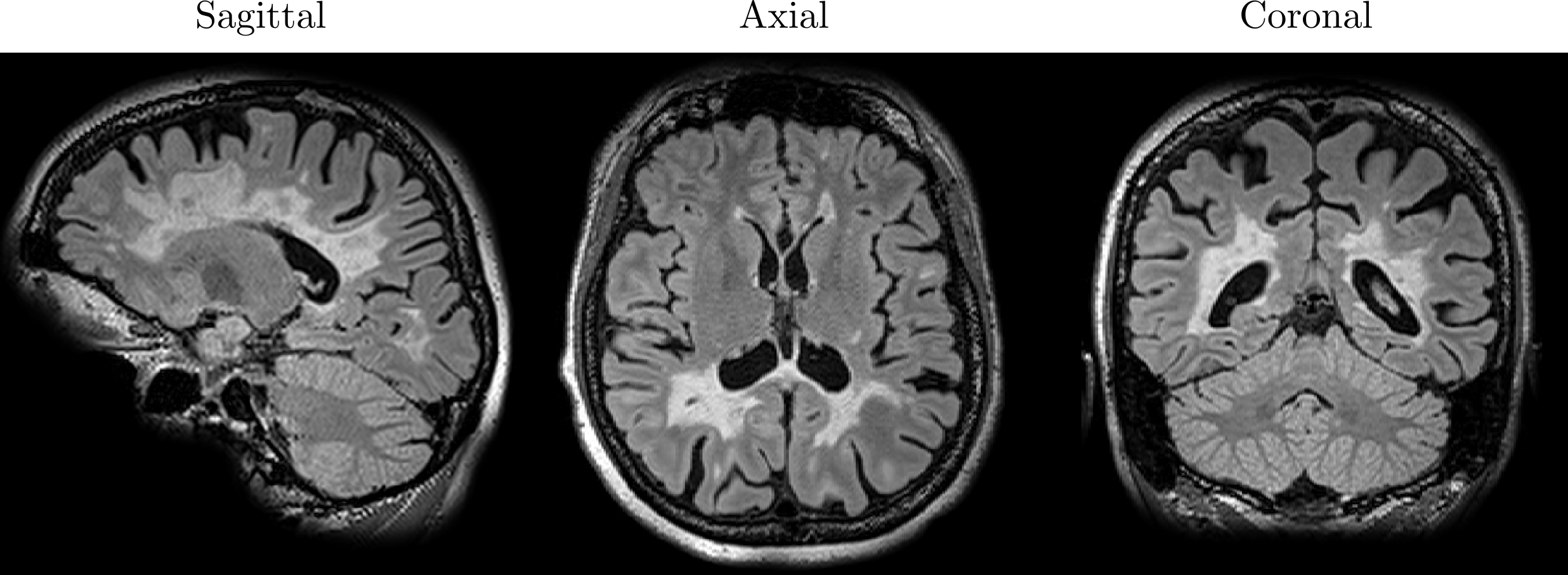}
  \caption{Sagittal, axial, and coronal views of Patient 19 from the UCML public dataset \protect\citepapp{lesjak_novel_2018}. The presence of large confluent lesions creates significant challenges for manual raters, making it extremely difficult, if not impossible, to distinguish individual lesion units.}
  \label{fig:unsplittable_lesions}
\end{figure}

\newpage
\section{MRI protocol}
\label{app:mri}
MRI acquisition was conducted using a 3T whole-body MR scanner (GE SIGNATM Premier research scanner, General Electric, Milwaukee, WI) equip\-ped with a 48-channel head coil. The imaging protocol for MS participants included a sagittal \textbf{3D MPRAGE} sequence with a repetition time (TR) of 2186 ms, echo time (TE) of 3 ms, inversion time (TI) of 900 ms, field of view (FOV) of 256 mm, 156 slices, and a voxel size of 1.0 × 1.0 × 1.0 mm. Additionally, a sagittal \textbf{3D FLAIR} sequence was acquired with a TR of 5000 ms, TE of 105 ms, TI of 1532 ms, FOV of 256 mm, 170 slices, and a voxel size of 1.0 × 1.0 × 1.0 mm. A sagittal high-resolution \textbf{3D EPI} sequence \citepapp{sati_rapid_2014} was also performed, with a TR of 80.2 ms, TE of 35 ms, flip angle of 18°, FOV of 256 mm, 355 slices, and a voxel size of 0.67 × 0.67 × 0.67 mm.

\newpage
\section{Illustration of ACLS's limitation}
\begin{figure}[h]
  \centering
  \includegraphics[width=\textwidth]{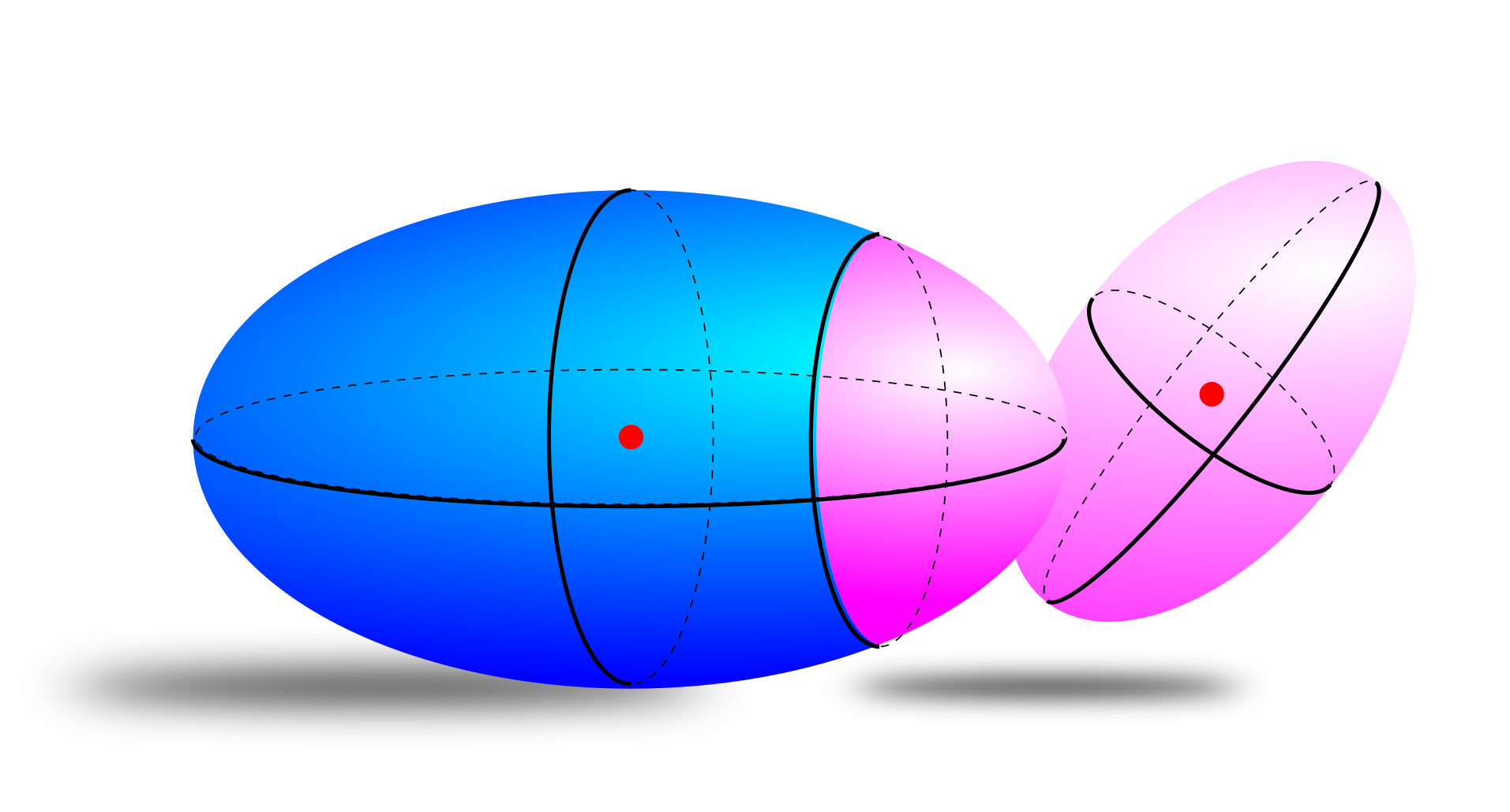}
  \caption{Limitation of nearest-center assignment methods. Even with accurate center detection (dots), assigning voxels to the nearest center can result in incorrect instance delineations. In this example, voxels from a large lesion (blue) are erroneously assigned to a nearby smaller lesion (pink) due to spatial proximity. This illustrates a specific failure mode of nearest-center clustering approaches such as ACLS, which struggle to respect lesion boundaries in cases of asymmetric lesion size or shape.}
  \label{fig:acls_issue}
\end{figure}

\newpage
\section{Test results adjusted for multiple comparisons (Holm-Bonferroni correction)}
\label{app:bonferroni}
\begin{figure}[h!]
  \centering
  \includegraphics[width=\textwidth]{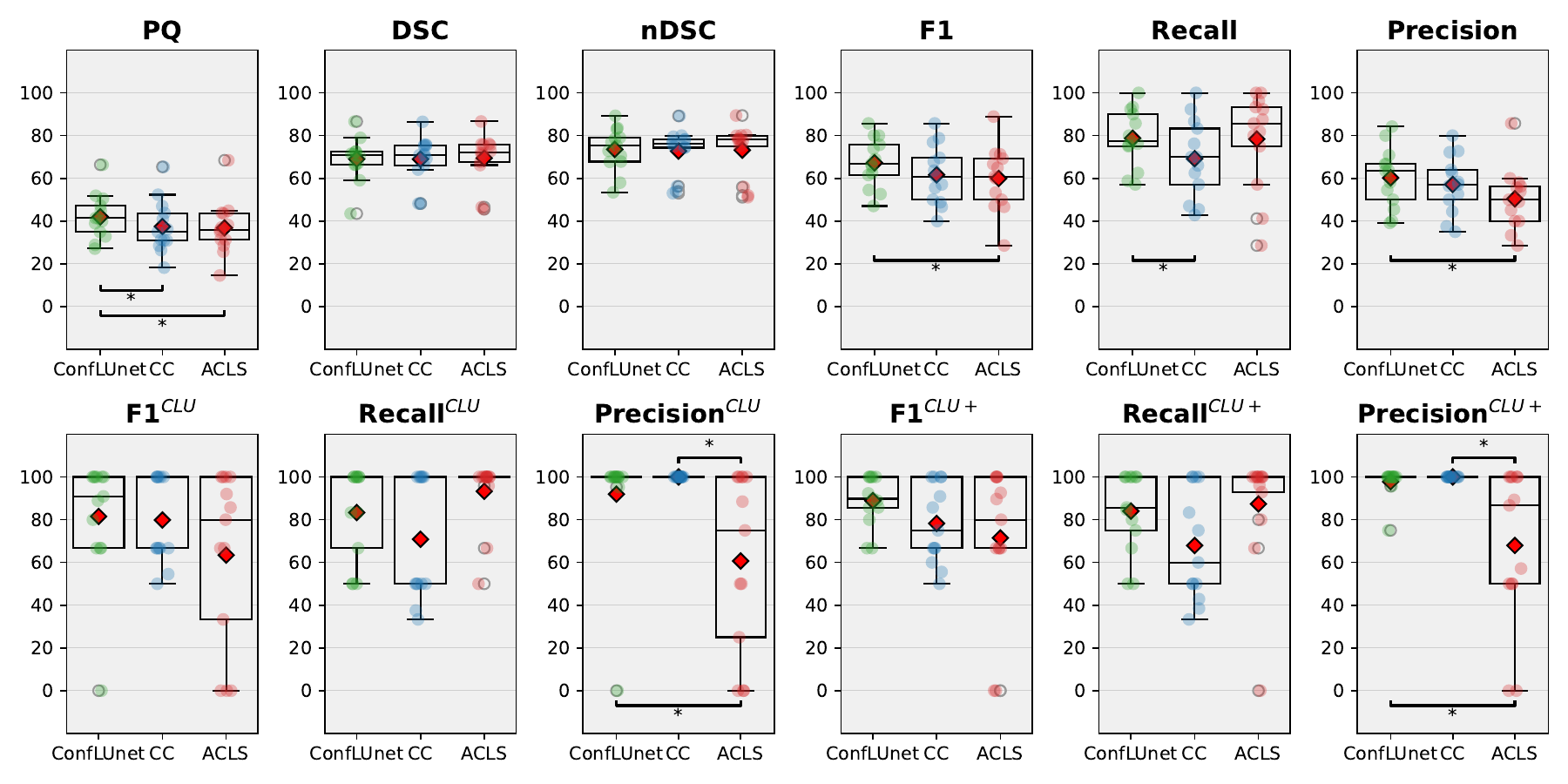}
  \caption{Box plots of the patient-wise metrics obtained for the three models. The p-values are computed with the Wilcoxon signed rank test, adjusted for multiple comparisons with Holm-Bonferroni correction. ACLS: Automated Confluent Lesion Splitting; CC: Connected Components; DSC: Dice Score Coefficient; nDSC: normalized Dice Score Coefficient; CLU: Confluent Lesion Unit; CLU+: Extended definition of Confluent Lesion Units (Eq.~\ref{eq:extended_clu}).}
  \label{fig:boxplots_corrected}
\end{figure}

\newpage
\section{Literature Review}
\label{app:lit_review}
A search on PubMed with the terms "Multiple Sclerosis" and "Lesion Segmentation" (Table \ref{tab:lit_review}) reveals that 114 previous methods have been published since 2014 on this topic (218 papers were excluded due to being reviews, meta-analyses, challenges, data papers, unrelated or unavailable studies, or papers written in non-English language). Of these published methods, 61 are evaluated using detection metrics in addition to semantic segmentation metrics - thereby entering in an instance segmentation evaluation framework - yet all of them specifically only focus on improving semantic segmentation. Of these 61 published methods, 1 used ACLS, 33 used CC, and 27 did not mention how they produced instance segmentation from the semantic segmentation output by the model, leaving the reader to suppose the methods uses CC. This analysis highlights a dissonance in the current literature: while there is a will to improve lesion detection performance, most published methods achieve this by refining semantic segmentation rather than explicitly addressing instance segmentation. This disconnect between evaluation frameworks and methodological design suggests that current evaluation procedures are not completely aligned with clinical needs, rendering existing approaches to be suboptimal for instance-level analysis and underscoring the need for a better defined evaluation approach. Notably, out of the 114 analysed papers, not a single method focuses directly on improving instance segmentation.

\small
\newcolumntype{C}[1]{>{\centering\arraybackslash}p{#1}}
\begin{longtable}{ C{3.5cm} C{0.9cm} C{3cm} C{1.7cm} C{2.4cm} }
\setlength{\extrarowheight}{0pt}
\addtolength{\extrarowheight}{\aboverulesep}
\addtolength{\extrarowheight}{\belowrulesep}
\setlength{\aboverulesep}{0pt}
\setlength{\belowrulesep}{0pt}


\textbf{Authors} & \textbf{Date} & \textbf{Task (evaluation)} & \textbf{Task (Method)} & \textbf{Instantiation Method} \\
\endfirsthead

\textbf{Authors} & \textbf{Date} & \textbf{Task (evaluation)} & \textbf{Task (Method)} & \textbf{Instantiation Method} \\
\endhead
\\ \hline
Cabezas et al. \citepapp{cabezas_boost_2014} & 2014 & SS & SS & None \\ \hline
Gao et al. \citepapp{gao_non-locally_2014} & 2014 & SS & SS & None \\ \hline 
Cabezas et al. \citepapp{cabezas_automatic_2014} & 2014 & SS and Detection & SS & CC \\ \hline
Roy et al. \citepapp{roy_subject_2014} & 2014 & SS & SS & None \\ \hline
Roura et al. \citepapp{roura_toolbox_2015} & 2015 & SS & SS & CC \\ \hline
Jain et al. \citepapp{jain_automatic_2015} & 2015 & SS & SS & None \\ \hline
Valverde et al. \citepapp{valverde_quantifying_2015} & 2015 & SS & SS & None \\ \hline
Jog et al. \citepapp{jog_multi-output_2015} & 2015 & SS & SS & None \\ \hline
Tomas-Fernandez et al. \citepapp{tomas-fernandez_model_2015} & 2015 & SS and Detection & SS & Not mentioned \\ \hline
Roy et al. \citepapp{roy_longitudinal_2015} & 2015 & SS and Detection & SS & Not mentioned \\ \hline
Subbanna et al. \citepapp{subbanna_image_2015} & 2015 & SS and Detection & SS & CC \\ \hline
Guizard et al. \citepapp{guizard_rotation-invariant_2015} & 2015 & SS and Detection & SS & CC \\ \hline
van Opbroek et al. \citepapp{van_opbroek_transfer_2015} & 2015 & SS & SS & None \\ \hline
Gong et al. \citepapp{gong_robust_2015} & 2015 & SS & SS & None \\ \hline
Storelli et al. \citepapp{storelli_semiautomatic_2016} & 2016 & SS & SS & CC \\ \hline
Brosch et al. \citepapp{brosch_deep_2016} & 2016 & SS and Detection & SS & Not mentioned \\ \hline
Freire et al. \citepapp{freire_automatic_2016} & 2016 & SS & SS & CC \\ \hline
Strumia et al. \citepapp{strumia_white_2016} & 2016 & SS and Detection & SS & CC \\ \hline
Jain et al. \citepapp{jain_two_2016} & 2016 & SS and Detection & SS & Not mentioned \\ \hline
Galimzianova et al. \citepapp{galimzianova_stratified_2016} & 2016 & SS and Detection & SS & CC \\ \hline
Lesjak et al. \citepapp{lesjak_validation_2016} & 2016 & SS and Detection & SS & CC \\ \hline
Karimaghaloo et al. \citepapp{karimaghaloo_adaptive_2016} & 2016 & SS and Detection & SS & CC \\ \hline
Mechrez et al. \citepapp{mechrez_patch-based_2016} & 2016 & SS & SS & None \\ \hline
Meier et al. \citepapp{meier_dual-sensitivity_2018} & 2017 & SS & SS & CC \\ \hline
Valverde et al. \citepapp{valverde_improving_2017} & 2017 & SS and Detection & SS & Not mentioned \\ \hline
Dong et al. \citepapp{dong_multiple_2017} & 2017 & SS & SS & Not Mentioned \\ \hline
Roy et al. \citepapp{roy_effective_2017} & 2017 & SS & SS & CC \\ \hline
Zhao et al. \citepapp{zhao_energy_2017} & 2017 & SS & SS & CC \\ \hline
Ghribi et al. \citepapp{ghribi_advanced_2017} & 2017 & SS and Detection & SS & Not mentioned \\ \hline
Khastavaneh et al. \citepapp{khastavaneh_neural_2017} & 2017 & SS & SS & None \\ \hline
Salem et al. \citepapp{salem_supervised_2018} & 2017 & SS & SS & None \\ \hline
Zhao et al. \citepapp{zhao_level_2018} & 2018 & SS & SS & None \\ \hline
Fleishman et al. \citepapp{fleishman_joint_2018} & 2018 & SS & SS & None \\ \hline
Galimzianova et al. \citepapp{galimzianova_locally_2018} & 2018 & SS & SS & None \\ \hline
da Silva Senra Filho et al. \citepapp{da_silva_senra_filho_hybrid_2018} & 2018 & SS & SS & None \\ \hline
Valcarcel et al. \citepapp{valcarcel_mimosa_2018} & 2018 & SS & SS & None \\ \hline
Fartaria et al. \citepapp{fartaria_partial_2018} & 2018 & SS and Detection & SS & CC \\ \hline
Oguz et al. \citepapp{oguz_dice_2018} & 2018 & SS and Detection & SS & CC \\ \hline
Zhang et al. \citepapp{zhang_multiple_2019} & 2019 & SS and Detection & SS & Not mentioned \\ \hline
Fartaria et al. \citepapp{fartaria_automated_2019} & 2019 & SS and Detection & SS & CC \\ \hline
Aslani et al. \citepapp{aslani_multi-branch_2019} & 2019 & SS and Detection & SS & Not mentioned \\ \hline
Valverde et al. \citepapp{valverde_one-shot_2019} & 2019 & SS and Detection & SS & Not mentioned \\ \hline
McKinley et al. \citepapp{mckinley_automatic_2020} & 2019 & SS and Detection & SS & CC \\ \hline
Hashemi et al. \citepapp{hashemi_asymmetric_2019} & 2019 & SS and Detection & SS & Not mentioned \\ \hline
HosseiniPanah et al. \citepapp{hosseinipanah_multiple_2019} & 2019 & SS & SS & None \\ \hline
Fartaria et al. \citepapp{fartaria_longitudinal_2019} & 2019 & SS and Detection & SS & Not mentioned \\ \hline
Weeda et al. \citepapp{weeda_comparing_2019} & 2019 & SS & SS & CC \\ \hline
Köhler et al. \citepapp{kohler_exploring_2019} & 2019 & SS and Detection & SS & CC \\ \hline
Narayana et al. \citepapp{narayana_deep-learning-based_2020} & 2019 & SS & SS & None \\ \hline
Le et al. \citepapp{le_flair2_2019} & 2019 & SS & SS & None \\ \hline
Schmidt et al. \citepapp{schmidt_automated_2019} & 2019 & SS and Detection & SS & Not mentioned \\ \hline
Rachmadi et al. \citepapp{rachmadi_limited_2020} & 2019 & SS & SS & CC \\ \hline
Ge et al. \citepapp{ge_brain_2019} & 2019 & SS & SS & None \\ \hline 
Gessert et al. \citepapp{gessert_multiple_2020} & 2020 & SS and Detection & SS & CC \\ \hline
La Rosa et al. \citepapp{la_rosa_multiple_2020} & 2020 & SS and Detection & SS & CC \\ \hline
Gabr et al. \citepapp{gabr_brain_2020} & 2020 & SS and Detection & SS & Not mentioned \\ \hline
Essa et al. \citepapp{essa_neuro-fuzzy_2020} & 2020 & SS and Detection & SS & Not mentioned \\ \hline
Valcarcel et al. \citepapp{valcarcel_tapas_2020} & 2020 & SS & SS & None \\ \hline
Krüger et al. \citepapp{kruger_fully_2020} & 2020 & SS and Detection & SS & CC \\ \hline
Salem et al. \citepapp{salem_fully_2020} & 2020 & SS and Detection & SS & CC \\ \hline
McKinley et al. \citepapp{mckinley_simultaneous_2021} & 2021 & SS & SS & None \\ \hline
Alijamaat et al. \citepapp{alijamaat_multiple_2021} & 2021 & SS and Detection & SS & Not mentioned \\ \hline
Fenneteau et al. \citepapp{fenneteau_investigating_2021} & 2021 & SS & SS & None \\ \hline
Ansari et al. \citepapp{ansari_multiple_2021} & 2021 & SS and Detection & SS & Not mentioned \\ \hline
Cerri et al. \citepapp{cerri_contrast-adaptive_2021} & 2021 & SS & SS & None \\ \hline
Zhang et al. \citepapp{zhang_all-net_2021} & 2021 & SS and Detection & SS & CC \\ \hline
Billot et al. \citepapp{billot_joint_2021} & 2021 & SS & SS & Not mentioned \\ \hline 
Priya et al. \citepapp{krishna_priya_improved_2021} & 2021 & SS & SS & Not Mentioned \\ \hline
Rakić et al. \citepapp{rakic_icobrain_2021} & 2021 & SS and Detection & SS & CC \\ \hline
Bonanno et al. \citepapp{bonanno_multiple_2021} & 2021 & SS & SS & None \\ \hline
Lou et al. \citepapp{lou_fully_2021} & 2021 & SS and Detection & SS & ACLS \\ \hline
Chen et al. \citepapp{chen_mtans_2021} & 2021 & SS and Detection & SS & CC \\ \hline
Gros et al. \citepapp{gros_softseg_2021} & 2021 & SS & SS & None \\ \hline
Ding et al. \citepapp{ding_improved_2020} & 2021 & SS & SS & None \\ \hline
Gordon et al. \citepapp{gordon_atlas_2021} & 2021 & SS & SS & None \\ \hline
Mehta et al. \citepapp{mehta_propagating_2022} & 2021 & SS and Detection & SS & CC \\ \hline
Basaran et al. \citepapp{basaran_new_2022} & 2022 & SS and Detection & SS & CC \\ \hline
Ashtari et al. \citepapp{ashtari_new_2022} & 2022 & SS and Detection & SS & CC \\ \hline
Papadopoulos et al. \citepapp{papadopoulos_white_2022} & 2022 & SS & SS & None \\ \hline
Krishnamoorthy et al. \citepapp{krishnamoorthy_framework_2022} & 2022 & SS & SS & None \\ \hline
Chen et al. \citepapp{chen_deep_2022} & 2022 & SS and Detection & SS & CC \\ \hline
Kamraoui et al. \citepapp{kamraoui_deeplesionbrain_2022} & 2022 & SS and Detection & SS & Not mentioned \\ \hline
Zelilidou et al. \citepapp{zelilidou_segmentation_2022} & 2022 & SS & SS & None \\ \hline
Krishnan et al. \citepapp{krishnan_joint_2022} & 2022 & SS and Detection & SS & CC \\ \hline
SadeghiBakhi et al. \citepapp{sadeghibakhi_multiple_2022} & 2022 & SS and Detection & SS & CC \\ \hline
Hashemi et al. \citepapp{hashemi_delve_2022} & 2022 & SS & SS & None \\ \hline
Tran et al. \citepapp{tran_automatic_2022} & 2022 & SS and Detection & SS & CC \\ \hline
Zhang et al. \citepapp{zhang_deep_2022} & 2022 & SS & SS & None \\ \hline
de Oliveira et al. \citepapp{de_oliveira_lesion_2022} & 2022 & SS & SS & None \\ \hline
Hindsholm et al. \citepapp{hindsholm_assessment_2022} & 2022 & SS and Detection & SS & Not mentioned \\ \hline
Yamamoto et al. \citepapp{yamamoto_validation_2022} & 2022 & SS and Detection & SS & CC \\ \hline
Hitziger et al. \citepapp{hitziger_triplanar_2022} & 2022 & SS and Detection & SS & CC \\ \hline
Sarica et al. \citepapp{sarica_new_2022} & 2022 & SS and Detection & SS & CC \\ \hline
Andresen et al. \citepapp{andresen_image_2022} & 2022 & SS and Detection & SS & Not mentioned \\ \hline
Kamraoui et al. \citepapp{kamraoui_longitudinal_2022} & 2022 & SS and Detection & SS & CC \\ \hline
Rondinella et al. \citepapp{rondinella_boosting_2023} & 2023 & SS & SS & None \\ \hline
Gentile et al. \citepapp{gentile_bianca-ms_2023} & 2023 & SS and Detection & SS & CC \\ \hline
Krishnan et al. \citepapp{krishnan_multi-arm_2023} & 2023 & SS and Detection & SS & CC \\ \hline
Cerri et al. \citepapp{cerri_open-source_2023} & 2023 & SS & SS & None \\ \hline
Sarica et al. \citepapp{sarica_dense_2023} & 2023 & SS and Detection & SS & Not mentioned \\ \hline
Raab et al. \citepapp{raab_investigation_2023} & 2023 & SS and Detection & SS & Not mentioned \\ \hline
Zhang et al. \citepapp{zhang_learning_2023} & 2023 & SS & SS & None \\ \hline
Wang et al. \citepapp{wang_energy_2023} & 2023 & SS & SS & None \\ \hline
Hindsholm et al. \citepapp{hindsholm_scanner_2023} & 2023 & SS and Detection & SS & Not mentioned \\ \hline
Donnay et al. \citepapp{donnay_pseudo-label_2023} & 2023 & SS & SS & None \\ \hline
Wahlig et al. \citepapp{wahlig_3d_2023} & 2023 & SS and Detection & SS & Not mentioned \\ \hline
Ramona Todea et al. \citepapp{todea_multicenter_2023} & 2023 & SS and Detection & SS & CC \\ \hline
Uwaeze et al. \citepapp{uwaeze_automatic_2024} & 2024 & SS & SS & None \\ \hline
De Rosa et al. \citepapp{de_rosa_consensus_2024} & 2024 & SS and Detection & SS & Not mentioned \\ \hline
Bai et al. \citepapp{bai_improving_2024} & 2024 & SS and Detection & SS & Not mentioned \\ \hline
Huang et al. \citepapp{huang_joint_2024} & 2024 & SS & SS & None \\ \hline
Wiltgen et al. \citepapp{wiltgen_lst-ai_2024} & 2024 & SS and Detection & SS & Not mentioned \\ \hline
Gaj et al. \citepapp{gaj_subject-based_2025} & 2025 & SS & SS & None \\ \hline
Placidi et al. \citepapp{placidi_context-dependent_2025} & 2025 & SS and Detection & SS & Not mentioned \\ \hline
\Xhline{3\arrayrulewidth}

\caption{Literature review. ACLS: Automated confluent lesion splitting; CC: Connected components; SS: Semantic segmentation.}
\label{tab:lit_review}
\end{longtable}

\bibliographystyleapp{elsarticle-num}
\bibliographyapp{references}


\begin{thebibliography}{100}
\expandafter\ifx\csname url\endcsname\relax
  \def\url#1{\texttt{#1}}\fi
\expandafter\ifx\csname urlprefix\endcsname\relax\def\urlprefix{URL }\fi
\expandafter\ifx\csname href\endcsname\relax
  \def\href#1#2{#2} \def\path#1{#1}\fi

\bibitem{lesjak_novel_2018}
v.~Lesjak, A.~Galimzianova, A.~Koren, M.~Lukin, F.~Pernu\v{s}, B.~Likar, v.~\v{S}piclin, \href{https://doi.org/10.1007/s12021-017-9348-7}{A {Novel} {Public} {MR} {Image} {Dataset} of {Multiple} {Sclerosis} {Patients} {With} {Lesion} {Segmentations} {Based} on {Multi}-rater {Consensus}}, Neuroinformatics 16~(1) (2018) 51--63.
\newblock \href {https://doi.org/10.1007/s12021-017-9348-7} {\path{doi:10.1007/s12021-017-9348-7}}.
\newline\urlprefix\url{https://doi.org/10.1007/s12021-017-9348-7}

\bibitem{sati_rapid_2014}
P.~Sati, D.~M. Thomasson, N.~Li, D.~L. Pham, N.~M. Biassou, D.~S. Reich, J.~A. Butman, \href{https://www.ncbi.nlm.nih.gov/pmc/articles/PMC4167170/}{Rapid, high-resolution, whole-brain, susceptibility-based {MRI} of multiple sclerosis}, Multiple sclerosis (Houndmills, Basingstoke, England) 20~(11) (2014) 1464--1470.
\newblock \href {https://doi.org/10.1177/1352458514525868} {\path{doi:10.1177/1352458514525868}}.
\newline\urlprefix\url{https://www.ncbi.nlm.nih.gov/pmc/articles/PMC4167170/}

\bibitem{cabezas_boost_2014}
M.~Cabezas, A.~Oliver, S.~Valverde, B.~Beltran, J.~Freixenet, J.~C. Vilanova, L.~Rami\'o-Torrent\a`, A.~Rovira, X.~Llad\'o, \href{https://www.sciencedirect.com/science/article/pii/S0165027014003161}{{BOOST}: {A} supervised approach for multiple sclerosis lesion segmentation}, Journal of Neuroscience Methods 237 (2014) 108--117.
\newblock \href {https://doi.org/10.1016/j.jneumeth.2014.08.024} {\path{doi:10.1016/j.jneumeth.2014.08.024}}.
\newline\urlprefix\url{https://www.sciencedirect.com/science/article/pii/S0165027014003161}

\bibitem{gao_non-locally_2014}
J.~Gao, C.~Li, C.~Feng, M.~Xie, Y.~Yin, C.~Davatzikos, \href{https://www.sciencedirect.com/science/article/pii/S0730725X14000885}{Non-locally regularized segmentation of multiple sclerosis lesion from multi-channel {MRI} data}, Magnetic Resonance Imaging 32~(8) (2014) 1058--1066.
\newblock \href {https://doi.org/10.1016/j.mri.2014.03.006} {\path{doi:10.1016/j.mri.2014.03.006}}.
\newline\urlprefix\url{https://www.sciencedirect.com/science/article/pii/S0730725X14000885}

\bibitem{cabezas_automatic_2014}
M.~Cabezas, A.~Oliver, E.~Roura, J.~Freixenet, J.~C. Vilanova, L.~Rami\'o-Torrent\a`, A.~Rovira, X.~Llad\'o, Automatic multiple sclerosis lesion detection in brain {MRI} by {FLAIR} thresholding, Computer Methods and Programs in Biomedicine 115~(3) (2014) 147--161.
\newblock \href {https://doi.org/10.1016/j.cmpb.2014.04.006} {\path{doi:10.1016/j.cmpb.2014.04.006}}.

\bibitem{roy_subject_2014}
S.~Roy, A.~Carass, J.~L. Prince, D.~L. Pham, Subject {Specific} {Sparse} {Dictionary} {Learning} for {Atlas} based {Brain} {MRI} {Segmentation}, Machine learning in medical imaging. MLMI (Workshop) 8679 (2014) 248--255.
\newblock \href {https://doi.org/10.1007/978-3-319-10581-9_31} {\path{doi:10.1007/978-3-319-10581-9_31}}.

\bibitem{roura_toolbox_2015}
E.~Roura, A.~Oliver, M.~Cabezas, S.~Valverde, D.~Pareto, J.~C. Vilanova, L.~Rami\'o-Torrent\a`, A.~Rovira, X.~Llad\'o, \href{https://doi.org/10.1007/s00234-015-1552-2}{A toolbox for multiple sclerosis lesion segmentation}, Neuroradiology 57~(10) (2015) 1031--1043.
\newblock \href {https://doi.org/10.1007/s00234-015-1552-2} {\path{doi:10.1007/s00234-015-1552-2}}.
\newline\urlprefix\url{https://doi.org/10.1007/s00234-015-1552-2}

\bibitem{jain_automatic_2015}
S.~Jain, D.~M. Sima, A.~Ribbens, M.~Cambron, A.~Maertens, W.~Van~Hecke, J.~De~Mey, F.~Barkhof, M.~D. Steenwijk, M.~Daams, F.~Maes, S.~Van~Huffel, H.~Vrenken, D.~Smeets, Automatic segmentation and volumetry of multiple sclerosis brain lesions from {MR} images, NeuroImage. Clinical 8 (2015) 367--375.
\newblock \href {https://doi.org/10.1016/j.nicl.2015.05.003} {\path{doi:10.1016/j.nicl.2015.05.003}}.

\bibitem{valverde_quantifying_2015}
S.~Valverde, A.~Oliver, E.~Roura, D.~Pareto, J.~C. Vilanova, L.~Rami\'o-Torrent\a`, J.~Sastre-Garriga, X.~Montalban, A.~Rovira, X.~Llad\'o, \href{https://www.sciencedirect.com/science/article/pii/S2213158215300127}{Quantifying brain tissue volume in multiple sclerosis with automated lesion segmentation and filling}, NeuroImage: Clinical 9 (2015) 640--647.
\newblock \href {https://doi.org/10.1016/j.nicl.2015.10.012} {\path{doi:10.1016/j.nicl.2015.10.012}}.
\newline\urlprefix\url{https://www.sciencedirect.com/science/article/pii/S2213158215300127}

\bibitem{jog_multi-output_2015}
A.~Jog, A.~Carass, D.~L. Pham, J.~L. Prince, \href{https://www.ncbi.nlm.nih.gov/pmc/articles/PMC5041594/}{Multi-{Output} {Decision} {Trees} for {Lesion} {Segmentation} in {Multiple} {Sclerosis}}, Proceedings of SPIE--the International Society for Optical Engineering 9413 (2015) 94131C.
\newblock \href {https://doi.org/10.1117/12.2082157} {\path{doi:10.1117/12.2082157}}.
\newline\urlprefix\url{https://www.ncbi.nlm.nih.gov/pmc/articles/PMC5041594/}

\bibitem{tomas-fernandez_model_2015}
X.~Tomas-Fernandez, S.~K. Warfield, \href{https://ieeexplore.ieee.org/document/7014271}{A {Model} of {Population} and {Subject} ({MOPS}) {Intensities} {With} {Application} to {Multiple} {Sclerosis} {Lesion} {Segmentation}}, IEEE Transactions on Medical Imaging 34~(6) (2015) 1349--1361.
\newblock \href {https://doi.org/10.1109/TMI.2015.2393853} {\path{doi:10.1109/TMI.2015.2393853}}.
\newline\urlprefix\url{https://ieeexplore.ieee.org/document/7014271}

\bibitem{roy_longitudinal_2015}
S.~Roy, A.~Carass, J.~L. Prince, D.~L. Pham, \href{https://www.ncbi.nlm.nih.gov/pmc/articles/PMC4996470/}{Longitudinal {Patch}-{Based} {Segmentation} of {Multiple} {Sclerosis} {White} {Matter} {Lesions}}, Machine learning in medical imaging. MLMI (Workshop), author 9352 (2015) 194--202.
\newblock \href {https://doi.org/10.1007/978-3-319-24888-2_24} {\path{doi:10.1007/978-3-319-24888-2_24}}.
\newline\urlprefix\url{https://www.ncbi.nlm.nih.gov/pmc/articles/PMC4996470/}

\bibitem{subbanna_image_2015}
N.~Subbanna, D.~Precup, D.~Arnold, T.~Arbel, {IMaGe}: {Iterative} {Multilevel} {Probabilistic} {Graphical} {Model} for {Detection} and {Segmentation} of {Multiple} {Sclerosis} {Lesions} in {Brain} {MRI}, in: S.~Ourselin, D.~C. Alexander, C.-F. Westin, M.~J. Cardoso (Eds.), Information {Processing} in {Medical} {Imaging}, Springer International Publishing, Cham, 2015, pp. 514--526.
\newblock \href {https://doi.org/10.1007/978-3-319-19992-4_40} {\path{doi:10.1007/978-3-319-19992-4_40}}.

\bibitem{guizard_rotation-invariant_2015}
N.~Guizard, P.~Coupé, V.~S. Fonov, J.~V. Manj\'on, D.~L. Arnold, D.~L. Collins, \href{https://www.sciencedirect.com/science/article/pii/S2213158215000935}{Rotation-invariant multi-contrast non-local means for {MS} lesion segmentation}, NeuroImage: Clinical 8 (2015) 376--389.
\newblock \href {https://doi.org/10.1016/j.nicl.2015.05.001} {\path{doi:10.1016/j.nicl.2015.05.001}}.
\newline\urlprefix\url{https://www.sciencedirect.com/science/article/pii/S2213158215000935}

\bibitem{van_opbroek_transfer_2015}
A.~van Opbroek, M.~A. Ikram, M.~W. Vernooij, M.~de~Bruijne, Transfer learning improves supervised image segmentation across imaging protocols, IEEE transactions on medical imaging 34~(5) (2015) 1018--1030.
\newblock \href {https://doi.org/10.1109/TMI.2014.2366792} {\path{doi:10.1109/TMI.2014.2366792}}.

\bibitem{gong_robust_2015}
Z.~Gong, D.~Zhao, C.~Li, W.~Tan, C.~Davatzikos, A {Robust} {Energy} {Minimization} {Algorithm} for {MS}-{Lesion} {Segmentation}, Advances in visual computing: ... international symposium, ISVC ...: proceedings. International Symposium on Visual Computing 9474 (2015) 521--530.
\newblock \href {https://doi.org/10.1007/978-3-319-27857-5_47} {\path{doi:10.1007/978-3-319-27857-5_47}}.

\bibitem{storelli_semiautomatic_2016}
L.~Storelli, E.~Pagani, M.~A. Rocca, M.~A. Horsfield, A.~Gallo, A.~Bisecco, M.~Battaglini, N.~De~Stefano, H.~Vrenken, D.~L. Thomas, L.~Mancini, S.~Ropele, C.~Enzinger, P.~Preziosa, M.~Filippi, A {Semiautomatic} {Method} for {Multiple} {Sclerosis} {Lesion} {Segmentation} on {Dual}-{Echo} {MR} {Imaging}: {Application} in a {Multicenter} {Context}, AJNR. American journal of neuroradiology 37~(11) (2016) 2043--2049.
\newblock \href {https://doi.org/10.3174/ajnr.A4874} {\path{doi:10.3174/ajnr.A4874}}.

\bibitem{brosch_deep_2016}
T.~Brosch, L.~Y.~W. Tang, Y.~Yoo, D.~K.~B. Li, A.~Traboulsee, R.~Tam, \href{https://ieeexplore.ieee.org/document/7404285}{Deep {3D} {Convolutional} {Encoder} {Networks} {With} {Shortcuts} for {Multiscale} {Feature} {Integration} {Applied} to {Multiple} {Sclerosis} {Lesion} {Segmentation}}, IEEE Transactions on Medical Imaging 35~(5) (2016) 1229--1239.
\newblock \href {https://doi.org/10.1109/TMI.2016.2528821} {\path{doi:10.1109/TMI.2016.2528821}}.
\newline\urlprefix\url{https://ieeexplore.ieee.org/document/7404285}

\bibitem{freire_automatic_2016}
P.~G.~L. Freire, R.~J. Ferrari, \href{https://www.sciencedirect.com/science/article/pii/S0010482516300798}{Automatic iterative segmentation of multiple sclerosis lesions using {Student}'s \textit{t} mixture models and probabilistic anatomical atlases in {FLAIR} images}, Computers in Biology and Medicine 73 (2016) 10--23.
\newblock \href {https://doi.org/10.1016/j.compbiomed.2016.03.025} {\path{doi:10.1016/j.compbiomed.2016.03.025}}.
\newline\urlprefix\url{https://www.sciencedirect.com/science/article/pii/S0010482516300798}

\bibitem{strumia_white_2016}
M.~Strumia, F.~R. Schmidt, C.~Anastasopoulos, C.~Granziera, G.~Krueger, T.~Brox, White {Matter} {MS}-{Lesion} {Segmentation} {Using} a {Geometric} {Brain} {Model}, IEEE transactions on medical imaging 35~(7) (2016) 1636--1646.
\newblock \href {https://doi.org/10.1109/TMI.2016.2522178} {\path{doi:10.1109/TMI.2016.2522178}}.

\bibitem{jain_two_2016}
S.~Jain, A.~Ribbens, D.~M. Sima, M.~Cambron, J.~De~Keyser, C.~Wang, M.~H. Barnett, S.~Van~Huffel, F.~Maes, D.~Smeets, \href{https://www.frontiersin.org/journals/neuroscience/articles/10.3389/fnins.2016.00576/full}{Two {Time} {Point} {MS} {Lesion} {Segmentation} in {Brain} {MRI}: {An} {Expectation}-{Maximization} {Framework}}, Frontiers in Neuroscience 10, publisher: Frontiers (Dec. 2016).
\newblock \href {https://doi.org/10.3389/fnins.2016.00576} {\path{doi:10.3389/fnins.2016.00576}}.
\newline\urlprefix\url{https://www.frontiersin.org/journals/neuroscience/articles/10.3389/fnins.2016.00576/full}

\bibitem{galimzianova_stratified_2016}
A.~Galimzianova, F.~Pernu\v{s}, B.~Likar, v.~\v{S}piclin, \href{https://www.sciencedirect.com/science/article/pii/S1053811915008629}{Stratified mixture modeling for segmentation of white-matter lesions in brain {MR} images}, NeuroImage 124 (2016) 1031--1043.
\newblock \href {https://doi.org/10.1016/j.neuroimage.2015.09.047} {\path{doi:10.1016/j.neuroimage.2015.09.047}}.
\newline\urlprefix\url{https://www.sciencedirect.com/science/article/pii/S1053811915008629}

\bibitem{lesjak_validation_2016}
v.~Lesjak, F.~Pernu\v{s}, B.~Likar, v.~\v{S}piclin, \href{https://doi.org/10.1007/s12021-016-9301-1}{Validation of {White}-{Matter} {Lesion} {Change} {Detection} {Methods} on a {Novel} {Publicly} {Available} {MRI} {Image} {Database}}, Neuroinformatics 14~(4) (2016) 403--420.
\newblock \href {https://doi.org/10.1007/s12021-016-9301-1} {\path{doi:10.1007/s12021-016-9301-1}}.
\newline\urlprefix\url{https://doi.org/10.1007/s12021-016-9301-1}

\bibitem{karimaghaloo_adaptive_2016}
Z.~Karimaghaloo, D.~L. Arnold, T.~Arbel, Adaptive multi-level conditional random fields for detection and segmentation of small enhanced pathology in medical images, Medical Image Analysis 27 (2016) 17--30.
\newblock \href {https://doi.org/10.1016/j.media.2015.06.004} {\path{doi:10.1016/j.media.2015.06.004}}.

\bibitem{mechrez_patch-based_2016}
R.~Mechrez, J.~Goldberger, H.~Greenspan, Patch-{Based} {Segmentation} with {Spatial} {Consistency}: {Application} to {MS} {Lesions} in {Brain} {MRI}, International Journal of Biomedical Imaging 2016 (2016) 7952541.
\newblock \href {https://doi.org/10.1155/2016/7952541} {\path{doi:10.1155/2016/7952541}}.

\bibitem{meier_dual-sensitivity_2018}
D.~S. Meier, C.~R.~G. Guttmann, S.~Tummala, N.~Moscufo, M.~Cavallari, S.~Tauhid, R.~Bakshi, H.~L. Weiner, Dual-{Sensitivity} {Multiple} {Sclerosis} {Lesion} and {CSF} {Segmentation} for {Multichannel} {3T} {Brain} {MRI}, Journal of Neuroimaging: Official Journal of the American Society of Neuroimaging 28~(1) (2018) 36--47.
\newblock \href {https://doi.org/10.1111/jon.12491} {\path{doi:10.1111/jon.12491}}.

\bibitem{valverde_improving_2017}
S.~Valverde, M.~Cabezas, E.~Roura, S.~González-Vill\a`, D.~Pareto, J.~C. Vilanova, L.~Rami\'o-Torrent\a`, A.~Rovira, A.~Oliver, X.~Llad\'o, \href{https://www.sciencedirect.com/science/article/pii/S1053811917303270}{Improving automated multiple sclerosis lesion segmentation with a cascaded {3D} convolutional neural network approach}, NeuroImage 155 (2017) 159--168.
\newblock \href {https://doi.org/10.1016/j.neuroimage.2017.04.034} {\path{doi:10.1016/j.neuroimage.2017.04.034}}.
\newline\urlprefix\url{https://www.sciencedirect.com/science/article/pii/S1053811917303270}

\bibitem{dong_multiple_2017}
M.~Dong, I.~Oguz, N.~Subbana, P.~Calabresi, R.~T. Shinohara, P.~Yushkevich, \href{https://www.ncbi.nlm.nih.gov/pmc/articles/PMC5918408/}{Multiple {Sclerosis} {Lesion} {Segmentation} {Using} {Joint} {Label} {Fusion}}, Patch-based techniques in medical imaging : third International Workshop, Patch-MI 2017, held in conjunction with MICCAI 2017, Quebec City, QC, Canada, September 14, 2017, Proceedings. Patch-MI (Workshop) (3rd : 2017 : Quebec, Quebec) 10530 (2017) 138--145.
\newblock \href {https://doi.org/10.1007/978-3-319-67434-6_16} {\path{doi:10.1007/978-3-319-67434-6_16}}.
\newline\urlprefix\url{https://www.ncbi.nlm.nih.gov/pmc/articles/PMC5918408/}

\bibitem{roy_effective_2017}
S.~Roy, D.~Bhattacharyya, S.~K. Bandyopadhyay, T.-H. Kim, \href{https://www.sciencedirect.com/science/article/pii/S016926071631450X}{An effective method for computerized prediction and segmentation of multiple sclerosis lesions in brain {MRI}}, Computer Methods and Programs in Biomedicine 140 (2017) 307--320.
\newblock \href {https://doi.org/10.1016/j.cmpb.2017.01.003} {\path{doi:10.1016/j.cmpb.2017.01.003}}.
\newline\urlprefix\url{https://www.sciencedirect.com/science/article/pii/S016926071631450X}

\bibitem{zhao_energy_2017}
Y.~Zhao, S.~Guo, M.~Luo, Y.~Liu, M.~Bilello, C.~Li, \href{https://www.sciencedirect.com/science/article/pii/S0730725X16300157}{An energy minimization method for {MS} lesion segmentation from {T1}-w and {FLAIR} images}, Magnetic Resonance Imaging 39 (2017) 1--6.
\newblock \href {https://doi.org/10.1016/j.mri.2016.04.003} {\path{doi:10.1016/j.mri.2016.04.003}}.
\newline\urlprefix\url{https://www.sciencedirect.com/science/article/pii/S0730725X16300157}

\bibitem{ghribi_advanced_2017}
O.~Ghribi, L.~Sellami, M.~Ben~Slima, A.~Ben~Hamida, C.~Mhiri, K.~B. Mahfoudh, An {Advanced} {MRI} {Multi}-{Modalities} {Segmentation} {Methodology} {Dedicated} to {Multiple} {Sclerosis} {Lesions} {Exploration} and {Differentiation}, IEEE transactions on nanobioscience 16~(8) (2017) 656--665.
\newblock \href {https://doi.org/10.1109/TNB.2017.2763246} {\path{doi:10.1109/TNB.2017.2763246}}.

\bibitem{khastavaneh_neural_2017}
H.~Khastavaneh, H.~Ebrahimpour-Komleh, Neural {Network}-{Based} {Learning} {Kernel} for {Automatic} {Segmentation} of {Multiple} {Sclerosis} {Lesions} on {Magnetic} {Resonance} {Images}, Journal of Biomedical Physics \& Engineering 7~(2) (2017) 155--162.

\bibitem{salem_supervised_2018}
M.~Salem, M.~Cabezas, S.~Valverde, D.~Pareto, A.~Oliver, J.~Salvi, A.~Rovira, X.~Llad\'o, A supervised framework with intensity subtraction and deformation field features for the detection of new {T2}-w lesions in multiple sclerosis, NeuroImage. Clinical 17 (2018) 607--615.
\newblock \href {https://doi.org/10.1016/j.nicl.2017.11.015} {\path{doi:10.1016/j.nicl.2017.11.015}}.

\bibitem{zhao_level_2018}
Y.~Zhao, S.~Guo, M.~Luo, X.~Shi, M.~Bilello, S.~Zhang, C.~Li, \href{https://www.sciencedirect.com/science/article/pii/S0730725X17300528}{A level set method for multiple sclerosis lesion segmentation}, Magnetic Resonance Imaging 49 (2018) 94--100.
\newblock \href {https://doi.org/10.1016/j.mri.2017.03.002} {\path{doi:10.1016/j.mri.2017.03.002}}.
\newline\urlprefix\url{https://www.sciencedirect.com/science/article/pii/S0730725X17300528}

\bibitem{fleishman_joint_2018}
G.~M. Fleishman, A.~Valcarcel, D.~L. Pham, S.~Roy, P.~A. Calabresi, P.~Yushkevich, R.~T. Shinohara, I.~Oguz, Joint {Intensity} {Fusion} {Image} {Synthesis} {Applied} to {Multiple} {Sclerosis} {Lesion} {Segmentation}, Brainlesion: Glioma, Multiple Sclerosis, Stroke and Traumatic Brain Injuries. BrainLes (Workshop) 10670 (2018) 43--54.

\bibitem{galimzianova_locally_2018}
A.~Galimzianova, v.~Lesjak, D.~L. Rubin, B.~Likar, F.~Pernu\v{s}, v.~\v{S}piclin, Locally adaptive magnetic resonance intensity models for unsupervised segmentation of multiple sclerosis lesions, Journal of Medical Imaging (Bellingham, Wash.) 5~(1) (2018) 011007.
\newblock \href {https://doi.org/10.1117/1.JMI.5.1.011007} {\path{doi:10.1117/1.JMI.5.1.011007}}.

\bibitem{da_silva_senra_filho_hybrid_2018}
A.~C. da~Silva Senra~Filho, \href{https://doi.org/10.1007/s11517-017-1747-2}{A hybrid approach based on logistic classification and iterative contrast enhancement algorithm for hyperintense multiple sclerosis lesion segmentation}, Medical \& Biological Engineering \& Computing 56~(6) (2018) 1063--1076.
\newblock \href {https://doi.org/10.1007/s11517-017-1747-2} {\path{doi:10.1007/s11517-017-1747-2}}.
\newline\urlprefix\url{https://doi.org/10.1007/s11517-017-1747-2}

\bibitem{valcarcel_mimosa_2018}
A.~M. Valcarcel, K.~A. Linn, S.~N. Vandekar, T.~D. Satterthwaite, J.~Muschelli, P.~A. Calabresi, D.~L. Pham, M.~L. Martin, R.~T. Shinohara, \href{https://onlinelibrary.wiley.com/doi/abs/10.1111/jon.12506}{{MIMoSA}: {An} {Automated} {Method} for {Intermodal} {Segmentation} {Analysis} of {Multiple} {Sclerosis} {Brain} {Lesions}}, Journal of Neuroimaging 28~(4) (2018) 389--398, \_eprint: https://onlinelibrary.wiley.com/doi/pdf/10.1111/jon.12506.
\newblock \href {https://doi.org/10.1111/jon.12506} {\path{doi:10.1111/jon.12506}}.
\newline\urlprefix\url{https://onlinelibrary.wiley.com/doi/abs/10.1111/jon.12506}

\bibitem{fartaria_partial_2018}
M.~J. Fartaria, A.~Todea, T.~Kober, K.~O'brien, G.~Krueger, R.~Meuli, C.~Granziera, A.~Roche, M.~Bach~Cuadra, Partial volume-aware assessment of multiple sclerosis lesions, NeuroImage. Clinical 18 (2018) 245--253.
\newblock \href {https://doi.org/10.1016/j.nicl.2018.01.011} {\path{doi:10.1016/j.nicl.2018.01.011}}.

\bibitem{oguz_dice_2018}
I.~Oguz, A.~Carass, D.~L. Pham, S.~Roy, N.~Subbana, P.~A. Calabresi, P.~A. Yushkevich, R.~T. Shinohara, J.~L. Prince, Dice {Overlap} {Measures} for {Objects} of {Unknown} {Number}: {Application} to {Lesion} {Segmentation}, Brainlesion: Glioma, Multiple Sclerosis, Stroke and Traumatic Brain Injuries. BrainLes (Workshop) 10670 (2018) 3--14.
\newblock \href {https://doi.org/10.1007/978-3-319-75238-9_1} {\path{doi:10.1007/978-3-319-75238-9_1}}.

\bibitem{zhang_multiple_2019}
H.~Zhang, A.~M. Valcarcel, R.~Bakshi, R.~Chu, F.~Bagnato, R.~T. Shinohara, K.~Hett, I.~Oguz, Multiple {Sclerosis} {Lesion} {Segmentation} with {Tiramisu} and 2.{5D} {Stacked} {Slices}, Medical image computing and computer-assisted intervention: MICCAI ... International Conference on Medical Image Computing and Computer-Assisted Intervention 11766 (2019) 338--346.
\newblock \href {https://doi.org/10.1007/978-3-030-32248-9_38} {\path{doi:10.1007/978-3-030-32248-9_38}}.

\bibitem{fartaria_automated_2019}
M.~J. Fartaria, P.~Sati, A.~Todea, E.-W. Radue, R.~Rahmanzadeh, K.~O’Brien, D.~S. Reich, M.~B. Cuadra, T.~Kober, C.~Granziera, \href{https://www.ncbi.nlm.nih.gov/pmc/articles/PMC6499666/}{Automated detection and segmentation of multiple sclerosis lesions using ultra-high-field {MP2RAGE}}, Investigative radiology 54~(6) (2019) 356--364.
\newblock \href {https://doi.org/10.1097/RLI.0000000000000551} {\path{doi:10.1097/RLI.0000000000000551}}.
\newline\urlprefix\url{https://www.ncbi.nlm.nih.gov/pmc/articles/PMC6499666/}

\bibitem{aslani_multi-branch_2019}
S.~Aslani, M.~Dayan, L.~Storelli, M.~Filippi, V.~Murino, M.~A. Rocca, D.~Sona, \href{https://www.sciencedirect.com/science/article/pii/S105381191930268X}{Multi-branch convolutional neural network for multiple sclerosis lesion segmentation}, NeuroImage 196 (2019) 1--15.
\newblock \href {https://doi.org/10.1016/j.neuroimage.2019.03.068} {\path{doi:10.1016/j.neuroimage.2019.03.068}}.
\newline\urlprefix\url{https://www.sciencedirect.com/science/article/pii/S105381191930268X}

\bibitem{valverde_one-shot_2019}
S.~Valverde, M.~Salem, M.~Cabezas, D.~Pareto, J.~C. Vilanova, L.~Rami\'o-Torrent\a`, A.~Rovira, J.~Salvi, A.~Oliver, X.~Llad\'o, \href{https://www.sciencedirect.com/science/article/pii/S2213158218303863}{One-shot domain adaptation in multiple sclerosis lesion segmentation using convolutional neural networks}, NeuroImage: Clinical 21 (2019) 101638.
\newblock \href {https://doi.org/10.1016/j.nicl.2018.101638} {\path{doi:10.1016/j.nicl.2018.101638}}.
\newline\urlprefix\url{https://www.sciencedirect.com/science/article/pii/S2213158218303863}

\bibitem{mckinley_automatic_2020}
R.~McKinley, R.~Wepfer, L.~Grunder, F.~Aschwanden, T.~Fischer, C.~Friedli, R.~Muri, C.~Rummel, R.~Verma, C.~Weisstanner, B.~Wiestler, C.~Berger, P.~Eichinger, M.~Muhlau, M.~Reyes, A.~Salmen, A.~Chan, R.~Wiest, F.~Wagner, \href{https://www.sciencedirect.com/science/article/pii/S2213158219304516}{Automatic detection of lesion load change in {Multiple} {Sclerosis} using convolutional neural networks with segmentation confidence}, NeuroImage: Clinical 25 (2020) 102104.
\newblock \href {https://doi.org/10.1016/j.nicl.2019.102104} {\path{doi:10.1016/j.nicl.2019.102104}}.
\newline\urlprefix\url{https://www.sciencedirect.com/science/article/pii/S2213158219304516}

\bibitem{hashemi_asymmetric_2019}
S.~R. Hashemi, S.~S. Mohseni~Salehi, D.~Erdogmus, S.~P. Prabhu, S.~K. Warfield, A.~Gholipour, \href{https://ieeexplore.ieee.org/document/8573779}{Asymmetric {Loss} {Functions} and {Deep} {Densely}-{Connected} {Networks} for {Highly}-{Imbalanced} {Medical} {Image} {Segmentation}: {Application} to {Multiple} {Sclerosis} {Lesion} {Detection}}, IEEE Access 7 (2019) 1721--1735.
\newblock \href {https://doi.org/10.1109/ACCESS.2018.2886371} {\path{doi:10.1109/ACCESS.2018.2886371}}.
\newline\urlprefix\url{https://ieeexplore.ieee.org/document/8573779}

\bibitem{hosseinipanah_multiple_2019}
S.~HosseiniPanah, A.~Zamani, F.~Emadi, F.~HamtaeiPour, \href{https://www.ncbi.nlm.nih.gov/pmc/articles/PMC6943841/}{Multiple {Sclerosis} {Lesions} {Segmentation} in {Magnetic} {Resonance} {Imaging} using {Ensemble} {Support} {Vector} {Machine} ({ESVM})}, Journal of Biomedical Physics \& Engineering 9~(6) (2019) 699--710.
\newblock \href {https://doi.org/10.31661/jbpe.v0i0.986} {\path{doi:10.31661/jbpe.v0i0.986}}.
\newline\urlprefix\url{https://www.ncbi.nlm.nih.gov/pmc/articles/PMC6943841/}

\bibitem{fartaria_longitudinal_2019}
M.~J. Fartaria, T.~Kober, C.~Granziera, M.~Bach~Cuadra, Longitudinal analysis of white matter and cortical lesions in multiple sclerosis, NeuroImage. Clinical 23 (2019) 101938.
\newblock \href {https://doi.org/10.1016/j.nicl.2019.101938} {\path{doi:10.1016/j.nicl.2019.101938}}.

\bibitem{weeda_comparing_2019}
M.~M. Weeda, I.~Brouwer, M.~L. de~Vos, M.~S. de~Vries, F.~Barkhof, P.~J.~W. Pouwels, H.~Vrenken, \href{https://www.sciencedirect.com/science/article/pii/S2213158219304218}{Comparing lesion segmentation methods in multiple sclerosis: {Input} from one manually delineated subject is sufficient for accurate lesion segmentation}, NeuroImage: Clinical 24 (2019) 102074.
\newblock \href {https://doi.org/10.1016/j.nicl.2019.102074} {\path{doi:10.1016/j.nicl.2019.102074}}.
\newline\urlprefix\url{https://www.sciencedirect.com/science/article/pii/S2213158219304218}

\bibitem{kohler_exploring_2019}
C.~Köhler, H.~Wahl, T.~Ziemssen, J.~Linn, H.~H. Kitzler, Exploring individual multiple sclerosis lesion volume change over time: {Development} of an algorithm for the analyses of longitudinal quantitative {MRI} measures, NeuroImage. Clinical 21 (2019) 101623.
\newblock \href {https://doi.org/10.1016/j.nicl.2018.101623} {\path{doi:10.1016/j.nicl.2018.101623}}.

\bibitem{narayana_deep-learning-based_2020}
P.~A. Narayana, I.~Coronado, S.~J. Sujit, J.~S. Wolinsky, F.~D. Lublin, R.~E. Gabr, \href{https://onlinelibrary.wiley.com/doi/abs/10.1002/jmri.26959}{Deep-{Learning}-{Based} {Neural} {Tissue} {Segmentation} of {MRI} in {Multiple} {Sclerosis}: {Effect} of {Training} {Set} {Size}}, Journal of Magnetic Resonance Imaging 51~(5) (2020) 1487--1496, \_eprint: https://onlinelibrary.wiley.com/doi/pdf/10.1002/jmri.26959.
\newblock \href {https://doi.org/10.1002/jmri.26959} {\path{doi:10.1002/jmri.26959}}.
\newline\urlprefix\url{https://onlinelibrary.wiley.com/doi/abs/10.1002/jmri.26959}

\bibitem{le_flair2_2019}
M.~Le, L.~Y.~W. Tang, E.~Hernández-Torres, M.~Jarrett, T.~Brosch, L.~Metz, D.~K.~B. Li, A.~Traboulsee, R.~C. Tam, A.~Rauscher, V.~Wiggermann, {FLAIR2} improves {LesionTOADS} automatic segmentation of multiple sclerosis lesions in non-homogenized, multi-center, {2D} clinical magnetic resonance images, NeuroImage. Clinical 23 (2019) 101918.
\newblock \href {https://doi.org/10.1016/j.nicl.2019.101918} {\path{doi:10.1016/j.nicl.2019.101918}}.

\bibitem{schmidt_automated_2019}
P.~Schmidt, V.~Pongratz, P.~Küster, D.~Meier, J.~Wuerfel, C.~Lukas, B.~Bellenberg, F.~Zipp, S.~Groppa, P.~G. Sämann, F.~Weber, C.~Gaser, T.~Franke, M.~Bussas, J.~Kirschke, C.~Zimmer, B.~Hemmer, M.~Mühlau, Automated segmentation of changes in {FLAIR}-hyperintense white matter lesions in multiple sclerosis on serial magnetic resonance imaging, NeuroImage. Clinical 23 (2019) 101849.
\newblock \href {https://doi.org/10.1016/j.nicl.2019.101849} {\path{doi:10.1016/j.nicl.2019.101849}}.

\bibitem{rachmadi_limited_2020}
M.~F. Rachmadi, M.~d.~C. Valdés-Hernández, H.~Li, R.~Guerrero, R.~Meijboom, S.~Wiseman, A.~Waldman, J.~Zhang, D.~Rueckert, J.~Wardlaw, T.~Komura, \href{https://www.sciencedirect.com/science/article/pii/S0895611119301004}{Limited {One}-time {Sampling} {Irregularity} {Map} ({LOTS}-{IM}) for {Automatic} {Unsupervised} {Assessment} of {White} {Matter} {Hyperintensities} and {Multiple} {Sclerosis} {Lesions} in {Structural} {Brain} {Magnetic} {Resonance} {Images}}, Computerized Medical Imaging and Graphics 79 (2020) 101685.
\newblock \href {https://doi.org/10.1016/j.compmedimag.2019.101685} {\path{doi:10.1016/j.compmedimag.2019.101685}}.
\newline\urlprefix\url{https://www.sciencedirect.com/science/article/pii/S0895611119301004}

\bibitem{ge_brain_2019}
T.~Ge, N.~Mu, T.~Zhan, Z.~Chen, W.~Gao, S.~Mu, \href{https://onlinelibrary.wiley.com/doi/abs/10.1155/2019/9378014}{Brain {Lesion} {Segmentation} {Based} on {Joint} {Constraints} of {Low}-{Rank} {Representation} and {Sparse} {Representation}}, Computational Intelligence and Neuroscience 2019~(1) (2019) 9378014, \_eprint: https://onlinelibrary.wiley.com/doi/pdf/10.1155/2019/9378014.
\newblock \href {https://doi.org/10.1155/2019/9378014} {\path{doi:10.1155/2019/9378014}}.
\newline\urlprefix\url{https://onlinelibrary.wiley.com/doi/abs/10.1155/2019/9378014}

\bibitem{gessert_multiple_2020}
N.~Gessert, J.~Krüger, R.~Opfer, A.-C. Ostwaldt, P.~Manogaran, H.~H. Kitzler, S.~Schippling, A.~Schlaefer, \href{https://www.sciencedirect.com/science/article/pii/S0895611120300732}{Multiple sclerosis lesion activity segmentation with attention-guided two-path {CNNs}}, Computerized Medical Imaging and Graphics 84 (2020) 101772.
\newblock \href {https://doi.org/10.1016/j.compmedimag.2020.101772} {\path{doi:10.1016/j.compmedimag.2020.101772}}.
\newline\urlprefix\url{https://www.sciencedirect.com/science/article/pii/S0895611120300732}

\bibitem{la_rosa_multiple_2020}
F.~La~Rosa, A.~Abdulkadir, M.~J. Fartaria, R.~Rahmanzadeh, P.-J. Lu, R.~Galbusera, M.~Barakovic, J.-P. Thiran, C.~Granziera, M.~B. Cuadra, \href{https://www.ncbi.nlm.nih.gov/pmc/articles/PMC7358270/}{Multiple sclerosis cortical and {WM} lesion segmentation at {3T} {MRI}: a deep learning method based on {FLAIR} and {MP2RAGE}}, NeuroImage : Clinical 27 (Jun. 2020).
\newblock \href {https://doi.org/10.1016/j.nicl.2020.102335} {\path{doi:10.1016/j.nicl.2020.102335}}.
\newline\urlprefix\url{https://www.ncbi.nlm.nih.gov/pmc/articles/PMC7358270/}

\bibitem{gabr_brain_2020}
R.~E. Gabr, I.~Coronado, M.~Robinson, S.~J. Sujit, S.~Datta, X.~Sun, W.~J. Allen, F.~D. Lublin, J.~S. Wolinsky, P.~A. Narayana, Brain and lesion segmentation in multiple sclerosis using fully convolutional neural networks: {A} large-scale study, Multiple Sclerosis (Houndmills, Basingstoke, England) 26~(10) (2020) 1217--1226.
\newblock \href {https://doi.org/10.1177/1352458519856843} {\path{doi:10.1177/1352458519856843}}.

\bibitem{essa_neuro-fuzzy_2020}
E.~Essa, D.~Aldesouky, S.~E. Hussein, M.~Z. Rashad, \href{https://doi.org/10.1007/s11517-020-02225-6}{Neuro-fuzzy patch-wise {R}-{CNN} for multiple sclerosis segmentation}, Medical \& Biological Engineering \& Computing 58~(9) (2020) 2161--2175.
\newblock \href {https://doi.org/10.1007/s11517-020-02225-6} {\path{doi:10.1007/s11517-020-02225-6}}.
\newline\urlprefix\url{https://doi.org/10.1007/s11517-020-02225-6}

\bibitem{valcarcel_tapas_2020}
A.~M. Valcarcel, J.~Muschelli, D.~L. Pham, M.~L. Martin, P.~Yushkevich, R.~Brandstadter, K.~R. Patterson, M.~K. Schindler, P.~A. Calabresi, R.~Bakshi, R.~T. Shinohara, {TAPAS}: {A} {Thresholding} {Approach} for {Probability} {Map} {Automatic} {Segmentation} in {Multiple} {Sclerosis}, NeuroImage. Clinical 27 (2020) 102256.
\newblock \href {https://doi.org/10.1016/j.nicl.2020.102256} {\path{doi:10.1016/j.nicl.2020.102256}}.

\bibitem{kruger_fully_2020}
J.~Krüger, R.~Opfer, N.~Gessert, A.-C. Ostwaldt, P.~Manogaran, H.~H. Kitzler, A.~Schlaefer, S.~Schippling, \href{https://www.sciencedirect.com/science/article/pii/S2213158220302825}{Fully automated longitudinal segmentation of new or enlarged multiple sclerosis lesions using {3D} convolutional neural networks}, NeuroImage: Clinical 28 (2020) 102445.
\newblock \href {https://doi.org/10.1016/j.nicl.2020.102445} {\path{doi:10.1016/j.nicl.2020.102445}}.
\newline\urlprefix\url{https://www.sciencedirect.com/science/article/pii/S2213158220302825}

\bibitem{salem_fully_2020}
M.~Salem, S.~Valverde, M.~Cabezas, D.~Pareto, A.~Oliver, J.~Salvi, A.~Rovira, X.~Llad\'o, \href{https://www.sciencedirect.com/science/article/pii/S2213158219304954}{A fully convolutional neural network for new {T2}-w lesion detection in multiple sclerosis}, NeuroImage: Clinical 25 (2020) 102149.
\newblock \href {https://doi.org/10.1016/j.nicl.2019.102149} {\path{doi:10.1016/j.nicl.2019.102149}}.
\newline\urlprefix\url{https://www.sciencedirect.com/science/article/pii/S2213158219304954}

\bibitem{mckinley_simultaneous_2021}
R.~McKinley, R.~Wepfer, F.~Aschwanden, L.~Grunder, R.~Muri, C.~Rummel, R.~Verma, C.~Weisstanner, M.~Reyes, A.~Salmen, A.~Chan, F.~Wagner, R.~Wiest, \href{https://www.nature.com/articles/s41598-020-79925-4}{Simultaneous lesion and brain segmentation in multiple sclerosis using deep neural networks}, Scientific Reports 11~(1) (2021) 1087, publisher: Nature Publishing Group.
\newblock \href {https://doi.org/10.1038/s41598-020-79925-4} {\path{doi:10.1038/s41598-020-79925-4}}.
\newline\urlprefix\url{https://www.nature.com/articles/s41598-020-79925-4}

\bibitem{alijamaat_multiple_2021}
A.~Alijamaat, A.~NikravanShalmani, P.~Bayat, \href{https://doi.org/10.1007/s11548-021-02327-y}{Multiple sclerosis lesion segmentation from brain {MRI} using {U}-{Net} based on wavelet pooling}, International Journal of Computer Assisted Radiology and Surgery 16~(9) (2021) 1459--1467.
\newblock \href {https://doi.org/10.1007/s11548-021-02327-y} {\path{doi:10.1007/s11548-021-02327-y}}.
\newline\urlprefix\url{https://doi.org/10.1007/s11548-021-02327-y}

\bibitem{fenneteau_investigating_2021}
A.~Fenneteau, P.~Bourdon, D.~Helbert, C.~Fernandez-Maloigne, C.~Habas, R.~Guillevin, Investigating efficient {CNN} architecture for multiple sclerosis lesion segmentation, Journal of Medical Imaging (Bellingham, Wash.) 8~(1) (2021) 014504.
\newblock \href {https://doi.org/10.1117/1.JMI.8.1.014504} {\path{doi:10.1117/1.JMI.8.1.014504}}.

\bibitem{ansari_multiple_2021}
S.~U. Ansari, K.~Javed, S.~M. Qaisar, R.~Jillani, U.~Haider, \href{https://onlinelibrary.wiley.com/doi/abs/10.1155/2021/4138137}{Multiple {Sclerosis} {Lesion} {Segmentation} in {Brain} {MRI} {Using} {Inception} {Modules} {Embedded} in a {Convolutional} {Neural} {Network}}, Journal of Healthcare Engineering 2021~(1) (2021) 4138137, \_eprint: https://onlinelibrary.wiley.com/doi/pdf/10.1155/2021/4138137.
\newblock \href {https://doi.org/10.1155/2021/4138137} {\path{doi:10.1155/2021/4138137}}.
\newline\urlprefix\url{https://onlinelibrary.wiley.com/doi/abs/10.1155/2021/4138137}

\bibitem{cerri_contrast-adaptive_2021}
S.~Cerri, O.~Puonti, D.~S. Meier, J.~Wuerfel, M.~Mühlau, H.~R. Siebner, K.~Van~Leemput, \href{https://www.sciencedirect.com/science/article/pii/S1053811920309563}{A contrast-adaptive method for simultaneous whole-brain and lesion segmentation in multiple sclerosis}, NeuroImage 225 (2021) 117471.
\newblock \href {https://doi.org/10.1016/j.neuroimage.2020.117471} {\path{doi:10.1016/j.neuroimage.2020.117471}}.
\newline\urlprefix\url{https://www.sciencedirect.com/science/article/pii/S1053811920309563}

\bibitem{zhang_all-net_2021}
H.~Zhang, J.~Zhang, C.~Li, E.~M. Sweeney, P.~Spincemaille, T.~D. Nguyen, S.~A. Gauthier, Y.~Wang, M.~Marcille, \href{https://www.sciencedirect.com/science/article/pii/S2213158221002989}{{ALL}-{Net}: {Anatomical} information lesion-wise loss function integrated into neural network for multiple sclerosis lesion segmentation}, NeuroImage: Clinical 32 (2021) 102854.
\newblock \href {https://doi.org/10.1016/j.nicl.2021.102854} {\path{doi:10.1016/j.nicl.2021.102854}}.
\newline\urlprefix\url{https://www.sciencedirect.com/science/article/pii/S2213158221002989}

\bibitem{billot_joint_2021}
B.~Billot, S.~Cerri, K.~V. Leemput, A.~V. Dalca, J.~E. Iglesias, \href{https://ieeexplore.ieee.org/document/9434127}{Joint {Segmentation} {Of} {Multiple} {Sclerosis} {Lesions} {And} {Brain} {Anatomy} {In} {MRI} {Scans} {Of} {Any} {Contrast} {And} {Resolution} {With} {CNNs}}, in: 2021 {IEEE} 18th {International} {Symposium} on {Biomedical} {Imaging} ({ISBI}), 2021, pp. 1971--1974, iSSN: 1945-8452.
\newblock \href {https://doi.org/10.1109/ISBI48211.2021.9434127} {\path{doi:10.1109/ISBI48211.2021.9434127}}.
\newline\urlprefix\url{https://ieeexplore.ieee.org/document/9434127}

\bibitem{krishna_priya_improved_2021}
R.~Krishna~Priya, S.~Chacko, \href{https://onlinelibrary.wiley.com/doi/abs/10.1002/cnm.3506}{Improved particle swarm optimized deep convolutional neural network with super-pixel clustering for multiple sclerosis lesion segmentation in brain {MRI} imaging}, International Journal for Numerical Methods in Biomedical Engineering 37~(9) (2021) e3506, \_eprint: https://onlinelibrary.wiley.com/doi/pdf/10.1002/cnm.3506.
\newblock \href {https://doi.org/10.1002/cnm.3506} {\path{doi:10.1002/cnm.3506}}.
\newline\urlprefix\url{https://onlinelibrary.wiley.com/doi/abs/10.1002/cnm.3506}

\bibitem{rakic_icobrain_2021}
M.~Rakić, S.~Vercruyssen, S.~Van~Eyndhoven, E.~de~la Rosa, S.~Jain, S.~Van~Huffel, F.~Maes, D.~Smeets, D.~M. Sima, \href{https://www.sciencedirect.com/science/article/pii/S2213158221001510}{icobrain ms 5.1: {Combining} unsupervised and supervised approaches for improving the detection of multiple sclerosis lesions}, NeuroImage: Clinical 31 (2021) 102707.
\newblock \href {https://doi.org/10.1016/j.nicl.2021.102707} {\path{doi:10.1016/j.nicl.2021.102707}}.
\newline\urlprefix\url{https://www.sciencedirect.com/science/article/pii/S2213158221001510}

\bibitem{bonanno_multiple_2021}
L.~Bonanno, N.~Mammone, S.~De~Salvo, A.~Bramanti, C.~Rifici, E.~Sessa, P.~Bramanti, S.~Marino, R.~Ciurleo, \href{https://www.sciencedirect.com/science/article/pii/S0899707120304290}{Multiple {Sclerosis} lesions detection by a hybrid {Watershed}-{Clustering} algorithm}, Clinical Imaging 72 (2021) 162--167.
\newblock \href {https://doi.org/10.1016/j.clinimag.2020.11.006} {\path{doi:10.1016/j.clinimag.2020.11.006}}.
\newline\urlprefix\url{https://www.sciencedirect.com/science/article/pii/S0899707120304290}

\bibitem{lou_fully_2021}
C.~Lou, P.~Sati, M.~Absinta, K.~Clark, J.~D. Dworkin, A.~M. Valcarcel, M.~K. Schindler, D.~S. Reich, E.~M. Sweeney, R.~T. Shinohara, Fully automated detection of paramagnetic rims in multiple sclerosis lesions on {3T} susceptibility-based {MR} imaging, NeuroImage. Clinical 32 (2021) 102796.
\newblock \href {https://doi.org/10.1016/j.nicl.2021.102796} {\path{doi:10.1016/j.nicl.2021.102796}}.

\bibitem{chen_mtans_2021}
G.~Chen, J.~Ru, Y.~Zhou, I.~Rekik, Z.~Pan, X.~Liu, Y.~Lin, B.~Lu, J.~Shi, {MTANS}: {Multi}-{Scale} {Mean} {Teacher} {Combined} {Adversarial} {Network} with {Shape}-{Aware} {Embedding} for {Semi}-{Supervised} {Brain} {Lesion} {Segmentation}, NeuroImage 244 (2021) 118568.
\newblock \href {https://doi.org/10.1016/j.neuroimage.2021.118568} {\path{doi:10.1016/j.neuroimage.2021.118568}}.

\bibitem{gros_softseg_2021}
C.~Gros, A.~Lemay, J.~Cohen-Adad, \href{https://www.sciencedirect.com/science/article/pii/S1361841521000840}{{SoftSeg}: {Advantages} of soft versus binary training for image segmentation}, Medical Image Analysis 71 (2021) 102038.
\newblock \href {https://doi.org/10.1016/j.media.2021.102038} {\path{doi:10.1016/j.media.2021.102038}}.
\newline\urlprefix\url{https://www.sciencedirect.com/science/article/pii/S1361841521000840}

\bibitem{ding_improved_2020}
T.~Ding, A.~D. Cohen, E.~E. O'Connor, H.~T. Karim, A.~Crainiceanu, J.~Muschelli, O.~Lopez, W.~E. Klunk, H.~J. Aizenstein, R.~Krafty, C.~M. Crainiceanu, D.~L. Tudorascu, An improved algorithm of white matter hyperintensity detection in elderly adults, NeuroImage. Clinical 25 (2020) 102151.
\newblock \href {https://doi.org/10.1016/j.nicl.2019.102151} {\path{doi:10.1016/j.nicl.2019.102151}}.

\bibitem{gordon_atlas_2021}
S.~Gordon, B.~Kodner, T.~Goldfryd, M.~Sidorov, J.~Goldberger, T.~R. Raviv, An atlas of classifiers-a machine learning paradigm for brain {MRI} segmentation, Medical \& Biological Engineering \& Computing 59~(9) (2021) 1833--1849.
\newblock \href {https://doi.org/10.1007/s11517-021-02414-x} {\path{doi:10.1007/s11517-021-02414-x}}.

\bibitem{mehta_propagating_2022}
R.~Mehta, T.~Christinck, T.~Nair, A.~Bussy, S.~Premasiri, M.~Costantino, M.~M. Chakravarthy, D.~L. Arnold, Y.~Gal, T.~Arbel, Propagating {Uncertainty} {Across} {Cascaded} {Medical} {Imaging} {Tasks} for {Improved} {Deep} {Learning} {Inference}, IEEE transactions on medical imaging 41~(2) (2022) 360--373.
\newblock \href {https://doi.org/10.1109/TMI.2021.3114097} {\path{doi:10.1109/TMI.2021.3114097}}.

\bibitem{basaran_new_2022}
B.~D. Basaran, P.~M. Matthews, W.~Bai, \href{https://www.frontiersin.org/journals/neuroscience/articles/10.3389/fnins.2022.1007453/full}{New lesion segmentation for multiple sclerosis brain images with imaging and lesion-aware augmentation}, Frontiers in Neuroscience 16, publisher: Frontiers (Oct. 2022).
\newblock \href {https://doi.org/10.3389/fnins.2022.1007453} {\path{doi:10.3389/fnins.2022.1007453}}.
\newline\urlprefix\url{https://www.frontiersin.org/journals/neuroscience/articles/10.3389/fnins.2022.1007453/full}

\bibitem{ashtari_new_2022}
P.~Ashtari, B.~Barile, S.~Van~Huffel, D.~Sappey-Marinier, \href{https://www.frontiersin.org/journals/neuroscience/articles/10.3389/fnins.2022.975862/full}{New multiple sclerosis lesion segmentation and detection using pre-activation {U}-{Net}}, Frontiers in Neuroscience 16, publisher: Frontiers (Oct. 2022).
\newblock \href {https://doi.org/10.3389/fnins.2022.975862} {\path{doi:10.3389/fnins.2022.975862}}.
\newline\urlprefix\url{https://www.frontiersin.org/journals/neuroscience/articles/10.3389/fnins.2022.975862/full}

\bibitem{papadopoulos_white_2022}
T.~G. Papadopoulos, E.~E. Tripoliti, D.~Plati, S.~Zelilidou, K.~Vlachos, S.~Konitsiotis, D.~I. Fotiadis, \href{https://ieeexplore.ieee.org/document/9871401}{White {Matter} {Lesion} {Segmentation} for {Multiple} {Sclerosis} {Patients} implementing deep learning}, in: 2022 44th {Annual} {International} {Conference} of the {IEEE} {Engineering} in {Medicine} \& {Biology} {Society} ({EMBC}), 2022, pp. 3818--3821, iSSN: 2694-0604.
\newblock \href {https://doi.org/10.1109/EMBC48229.2022.9871401} {\path{doi:10.1109/EMBC48229.2022.9871401}}.
\newline\urlprefix\url{https://ieeexplore.ieee.org/document/9871401}

\bibitem{krishnamoorthy_framework_2022}
S.~Krishnamoorthy, Y.~Zhang, S.~Kadry, W.~Yu, \href{https://onlinelibrary.wiley.com/doi/abs/10.1155/2022/4928096}{Framework to {Segment} and {Evaluate} {Multiple} {Sclerosis} {Lesion} in {MRI} {Slices} {Using} {VGG}-{UNet}}, Computational Intelligence and Neuroscience 2022~(1) (2022) 4928096, \_eprint: https://onlinelibrary.wiley.com/doi/pdf/10.1155/2022/4928096.
\newblock \href {https://doi.org/10.1155/2022/4928096} {\path{doi:10.1155/2022/4928096}}.
\newline\urlprefix\url{https://onlinelibrary.wiley.com/doi/abs/10.1155/2022/4928096}

\bibitem{chen_deep_2022}
Z.~Chen, X.~Wang, J.~Huang, J.~Lu, J.~Zheng, Deep {Attention} and {Graphical} {Neural} {Network} for {Multiple} {Sclerosis} {Lesion} {Segmentation} {From} {MR} {Imaging} {Sequences}, IEEE journal of biomedical and health informatics 26~(3) (2022) 1196--1207.
\newblock \href {https://doi.org/10.1109/JBHI.2021.3109119} {\path{doi:10.1109/JBHI.2021.3109119}}.

\bibitem{kamraoui_deeplesionbrain_2022}
R.~A. Kamraoui, V.-T. Ta, T.~Tourdias, B.~Mansencal, J.~V. Manjon, P.~Coupé, \href{https://www.sciencedirect.com/science/article/pii/S1361841521003571}{{DeepLesionBrain}: {Towards} a broader deep-learning generalization for multiple sclerosis lesion segmentation}, Medical Image Analysis 76 (2022) 102312.
\newblock \href {https://doi.org/10.1016/j.media.2021.102312} {\path{doi:10.1016/j.media.2021.102312}}.
\newline\urlprefix\url{https://www.sciencedirect.com/science/article/pii/S1361841521003571}

\bibitem{zelilidou_segmentation_2022}
S.~P. Zelilidou, E.~E. Tripoliti, K.~I. Vlachos, S.~Konitsiotis, D.~I. Fotiadis, Segmentation and volume quantification of {MR} {Images} for the detection and monitoring multiple sclerosis progression, Annual International Conference of the IEEE Engineering in Medicine and Biology Society. IEEE Engineering in Medicine and Biology Society. Annual International Conference 2022 (2022) 4745--4748.
\newblock \href {https://doi.org/10.1109/EMBC48229.2022.9871533} {\path{doi:10.1109/EMBC48229.2022.9871533}}.

\bibitem{krishnan_joint_2022}
A.~P. Krishnan, Z.~Song, D.~Clayton, L.~Gaetano, X.~Jia, A.~de~Crespigny, T.~Bengtsson, R.~A.~D. Carano, \href{https://pubs.rsna.org/doi/10.1148/radiol.211528}{Joint {MRI} {T1} {Unenhancing} and {Contrast}-enhancing {Multiple} {Sclerosis} {Lesion} {Segmentation} with {Deep} {Learning} in {OPERA} {Trials}}, Radiology 302~(3) (2022) 662--673, publisher: Radiological Society of North America.
\newblock \href {https://doi.org/10.1148/radiol.211528} {\path{doi:10.1148/radiol.211528}}.
\newline\urlprefix\url{https://pubs.rsna.org/doi/10.1148/radiol.211528}

\bibitem{sadeghibakhi_multiple_2022}
M.~Sadeghibakhi, H.~Pourreza, H.~Mahyar, \href{https://ieeexplore.ieee.org/document/9766141}{Multiple {Sclerosis} {Lesions} {Segmentation} {Using} {Attention}-{Based} {CNNs} in {FLAIR} {Images}}, IEEE Journal of Translational Engineering in Health and Medicine 10 (2022) 1--11.
\newblock \href {https://doi.org/10.1109/JTEHM.2022.3172025} {\path{doi:10.1109/JTEHM.2022.3172025}}.
\newline\urlprefix\url{https://ieeexplore.ieee.org/document/9766141}

\bibitem{hashemi_delve_2022}
M.~Hashemi, M.~Akhbari, C.~Jutten, Delve into {Multiple} {Sclerosis} ({MS}) lesion exploration: {A} modified attention {U}-{Net} for {MS} lesion segmentation in {Brain} {MRI}, Computers in Biology and Medicine 145 (2022) 105402.
\newblock \href {https://doi.org/10.1016/j.compbiomed.2022.105402} {\path{doi:10.1016/j.compbiomed.2022.105402}}.

\bibitem{tran_automatic_2022}
P.~Tran, U.~Thoprakarn, E.~Gourieux, C.~L. dos Santos, E.~Cavedo, N.~Guizard, F.~Cotton, P.~Krolak-Salmon, C.~Delmaire, D.~Heidelberg, N.~Pyatigorskaya, S.~Ströer, D.~Dormont, J.-B. Martini, M.~Chupin, \href{https://www.sciencedirect.com/science/article/pii/S2213158222000055}{Automatic segmentation of white matter hyperintensities: validation and comparison with state-of-the-art methods on both {Multiple} {Sclerosis} and elderly subjects}, NeuroImage: Clinical 33 (2022) 102940.
\newblock \href {https://doi.org/10.1016/j.nicl.2022.102940} {\path{doi:10.1016/j.nicl.2022.102940}}.
\newline\urlprefix\url{https://www.sciencedirect.com/science/article/pii/S2213158222000055}

\bibitem{zhang_deep_2022}
Y.~Zhang, Y.~Duan, X.~Wang, Z.~Zhuo, S.~Haller, F.~Barkhof, Y.~Liu, \href{https://doi.org/10.1007/s00234-021-02820-w}{A deep learning algorithm for white matter hyperintensity lesion detection and segmentation}, Neuroradiology 64~(4) (2022) 727--734.
\newblock \href {https://doi.org/10.1007/s00234-021-02820-w} {\path{doi:10.1007/s00234-021-02820-w}}.
\newline\urlprefix\url{https://doi.org/10.1007/s00234-021-02820-w}

\bibitem{de_oliveira_lesion_2022}
M.~de~Oliveira, M.~Piacenti-Silva, F.~C.~G. da~Rocha, J.~M. Santos, J.~D.~S. Cardoso, P.~N. Lisboa-Filho, Lesion {Volume} {Quantification} {Using} {Two} {Convolutional} {Neural} {Networks} in {MRIs} of {Multiple} {Sclerosis} {Patients}, Diagnostics (Basel, Switzerland) 12~(2) (2022) 230.
\newblock \href {https://doi.org/10.3390/diagnostics12020230} {\path{doi:10.3390/diagnostics12020230}}.

\bibitem{hindsholm_assessment_2022}
A.~M. Hindsholm, S.~P. Cramer, H.~J. Simonsen, J.~L. Frederiksen, F.~Andersen, L.~Højgaard, C.~N. Ladefoged, U.~Lindberg, Assessment of {Artificial} {Intelligence} {Automatic} {Multiple} {Sclerosis} {Lesion} {Delineation} {Tool} for {Clinical} {Use}, Clinical Neuroradiology 32~(3) (2022) 643--653.
\newblock \href {https://doi.org/10.1007/s00062-021-01089-z} {\path{doi:10.1007/s00062-021-01089-z}}.

\bibitem{yamamoto_validation_2022}
T.~Yamamoto, C.~Lacheret, H.~Fukutomi, R.~A. Kamraoui, L.~Denat, B.~Zhang, V.~Prevost, L.~Zhang, A.~Ruet, B.~Triaire, V.~Dousset, P.~Coupé, T.~Tourdias, Validation of a {Denoising} {Method} {Using} {Deep} {Learning}-{Based} {Reconstruction} to {Quantify} {Multiple} {Sclerosis} {Lesion} {Load} on {Fast} {FLAIR} {Imaging}, AJNR. American journal of neuroradiology 43~(8) (2022) 1099--1106.
\newblock \href {https://doi.org/10.3174/ajnr.A7589} {\path{doi:10.3174/ajnr.A7589}}.

\bibitem{hitziger_triplanar_2022}
S.~Hitziger, W.~X. Ling, T.~Fritz, T.~D'Albis, A.~Lemke, J.~Grilo, \href{https://www.frontiersin.org/journals/neuroscience/articles/10.3389/fnins.2022.964250/full}{Triplanar {U}-{Net} with lesion-wise voting for the segmentation of new lesions on longitudinal {MRI} studies}, Frontiers in Neuroscience 16, publisher: Frontiers (Aug. 2022).
\newblock \href {https://doi.org/10.3389/fnins.2022.964250} {\path{doi:10.3389/fnins.2022.964250}}.
\newline\urlprefix\url{https://www.frontiersin.org/journals/neuroscience/articles/10.3389/fnins.2022.964250/full}

\bibitem{sarica_new_2022}
B.~Sarica, D.~Z. Seker, \href{https://www.frontiersin.org/journals/neuroscience/articles/10.3389/fnins.2022.912000/full}{New {MS} lesion segmentation with deep residual attention gate {U}-{Net} utilizing {2D} slices of {3D} {MR} images}, Frontiers in Neuroscience 16, publisher: Frontiers (Jul. 2022).
\newblock \href {https://doi.org/10.3389/fnins.2022.912000} {\path{doi:10.3389/fnins.2022.912000}}.
\newline\urlprefix\url{https://www.frontiersin.org/journals/neuroscience/articles/10.3389/fnins.2022.912000/full}

\bibitem{andresen_image_2022}
J.~Andresen, H.~Uzunova, J.~Ehrhardt, T.~Kepp, H.~Handels, \href{https://www.frontiersin.org/journals/neuroscience/articles/10.3389/fnins.2022.981523/full}{Image registration and appearance adaptation in non-correspondent image regions for new {MS} lesions detection}, Frontiers in Neuroscience 16, publisher: Frontiers (Sep. 2022).
\newblock \href {https://doi.org/10.3389/fnins.2022.981523} {\path{doi:10.3389/fnins.2022.981523}}.
\newline\urlprefix\url{https://www.frontiersin.org/journals/neuroscience/articles/10.3389/fnins.2022.981523/full}

\bibitem{kamraoui_longitudinal_2022}
R.~A. Kamraoui, B.~Mansencal, J.~V. Manjon, P.~Coupé, \href{https://www.frontiersin.org/journals/neuroimaging/articles/10.3389/fnimg.2022.948235/full}{Longitudinal detection of new {MS} lesions using deep learning}, Frontiers in Neuroimaging 1, publisher: Frontiers (Aug. 2022).
\newblock \href {https://doi.org/10.3389/fnimg.2022.948235} {\path{doi:10.3389/fnimg.2022.948235}}.
\newline\urlprefix\url{https://www.frontiersin.org/journals/neuroimaging/articles/10.3389/fnimg.2022.948235/full}

\bibitem{rondinella_boosting_2023}
A.~Rondinella, E.~Crispino, F.~Guarnera, O.~Giudice, A.~Ortis, G.~Russo, C.~Di~Lorenzo, D.~Maimone, F.~Pappalardo, S.~Battiato, \href{https://www.sciencedirect.com/science/article/pii/S0010482523004869}{Boosting multiple sclerosis lesion segmentation through attention mechanism}, Computers in Biology and Medicine 161 (2023) 107021.
\newblock \href {https://doi.org/10.1016/j.compbiomed.2023.107021} {\path{doi:10.1016/j.compbiomed.2023.107021}}.
\newline\urlprefix\url{https://www.sciencedirect.com/science/article/pii/S0010482523004869}

\bibitem{gentile_bianca-ms_2023}
G.~Gentile, M.~Jenkinson, L.~Griffanti, L.~Luchetti, M.~Leoncini, M.~Inderyas, M.~Mortilla, R.~Cortese, N.~De~Stefano, M.~Battaglini, \href{https://onlinelibrary.wiley.com/doi/abs/10.1002/hbm.26424}{{BIANCA}-{MS}: {An} optimized tool for automated multiple sclerosis lesion segmentation}, Human Brain Mapping 44~(14) (2023) 4893--4913, \_eprint: https://onlinelibrary.wiley.com/doi/pdf/10.1002/hbm.26424.
\newblock \href {https://doi.org/10.1002/hbm.26424} {\path{doi:10.1002/hbm.26424}}.
\newline\urlprefix\url{https://onlinelibrary.wiley.com/doi/abs/10.1002/hbm.26424}

\bibitem{krishnan_multi-arm_2023}
A.~P. Krishnan, Z.~Song, D.~Clayton, X.~Jia, A.~de~Crespigny, R.~A.~D. Carano, \href{https://www.nature.com/articles/s41598-023-31207-5}{Multi-arm {U}-{Net} with dense input and skip connectivity for {T2} lesion segmentation in clinical trials of multiple sclerosis}, Scientific Reports 13~(1) (2023) 4102, publisher: Nature Publishing Group.
\newblock \href {https://doi.org/10.1038/s41598-023-31207-5} {\path{doi:10.1038/s41598-023-31207-5}}.
\newline\urlprefix\url{https://www.nature.com/articles/s41598-023-31207-5}

\bibitem{cerri_open-source_2023}
S.~Cerri, D.~N. Greve, A.~Hoopes, H.~Lundell, H.~R. Siebner, M.~Mühlau, K.~Van~Leemput, \href{https://www.sciencedirect.com/science/article/pii/S2213158223000438}{An open-source tool for longitudinal whole-brain and white matter lesion segmentation}, NeuroImage: Clinical 38 (2023) 103354.
\newblock \href {https://doi.org/10.1016/j.nicl.2023.103354} {\path{doi:10.1016/j.nicl.2023.103354}}.
\newline\urlprefix\url{https://www.sciencedirect.com/science/article/pii/S2213158223000438}

\bibitem{sarica_dense_2023}
B.~Sarica, D.~Z. Seker, B.~Bayram, \href{https://www.sciencedirect.com/science/article/pii/S1386505622002799}{A dense residual {U}-net for multiple sclerosis lesions segmentation from multi-sequence {3D} {MR} images}, International Journal of Medical Informatics 170 (2023) 104965.
\newblock \href {https://doi.org/10.1016/j.ijmedinf.2022.104965} {\path{doi:10.1016/j.ijmedinf.2022.104965}}.
\newline\urlprefix\url{https://www.sciencedirect.com/science/article/pii/S1386505622002799}

\bibitem{raab_investigation_2023}
F.~Raab, W.~Malloni, S.~Wein, M.~W. Greenlee, E.~W. Lang, \href{https://www.nature.com/articles/s41598-023-48578-4}{Investigation of an efficient multi-modal convolutional neural network for multiple sclerosis lesion detection}, Scientific Reports 13~(1) (2023) 21154, publisher: Nature Publishing Group.
\newblock \href {https://doi.org/10.1038/s41598-023-48578-4} {\path{doi:10.1038/s41598-023-48578-4}}.
\newline\urlprefix\url{https://www.nature.com/articles/s41598-023-48578-4}

\bibitem{zhang_learning_2023}
L.~Zhang, R.~Tanno, M.~Xu, Y.~Huang, K.~Bronik, C.~Jin, J.~Jacob, Y.~Zheng, L.~Shao, O.~Ciccarelli, F.~Barkhof, D.~C. Alexander, \href{https://www.sciencedirect.com/science/article/pii/S0031320323001012}{Learning from multiple annotators for medical image segmentation}, Pattern Recognition 138 (2023) 109400.
\newblock \href {https://doi.org/10.1016/j.patcog.2023.109400} {\path{doi:10.1016/j.patcog.2023.109400}}.
\newline\urlprefix\url{https://www.sciencedirect.com/science/article/pii/S0031320323001012}

\bibitem{wang_energy_2023}
X.~Wang, Y.~Yang, T.~Wu, H.~Zhu, J.~Yu, J.~Tian, H.~Li, \href{https://www.frontiersin.org/journals/neuroscience/articles/10.3389/fnins.2023.1175451/full}{Energy minimization segmentation model based on {MRI} images}, Frontiers in Neuroscience 17, publisher: Frontiers (Apr. 2023).
\newblock \href {https://doi.org/10.3389/fnins.2023.1175451} {\path{doi:10.3389/fnins.2023.1175451}}.
\newline\urlprefix\url{https://www.frontiersin.org/journals/neuroscience/articles/10.3389/fnins.2023.1175451/full}

\bibitem{hindsholm_scanner_2023}
A.~M. Hindsholm, F.~L. Andersen, S.~P. Cramer, H.~J. Simonsen, M.~G. Askløf, M.~Magyari, P.~N. Madsen, A.~E. Hansen, F.~Sellebjerg, H.~B.~W. Larsson, A.~R. Langkilde, J.~L. Frederiksen, L.~Højgaard, C.~N. Ladefoged, U.~Lindberg, \href{https://www.frontiersin.org/journals/neuroscience/articles/10.3389/fnins.2023.1177540/full}{Scanner agnostic large-scale evaluation of {MS} lesion delineation tool for clinical {MRI}}, Frontiers in Neuroscience 17, publisher: Frontiers (May 2023).
\newblock \href {https://doi.org/10.3389/fnins.2023.1177540} {\path{doi:10.3389/fnins.2023.1177540}}.
\newline\urlprefix\url{https://www.frontiersin.org/journals/neuroscience/articles/10.3389/fnins.2023.1177540/full}

\bibitem{donnay_pseudo-label_2023}
C.~Donnay, H.~Dieckhaus, C.~Tsagkas, M.~I. Gaitán, E.~S. Beck, A.~Mullins, D.~S. Reich, G.~Nair, \href{https://www.frontiersin.org/journals/neuroimaging/articles/10.3389/fnimg.2023.1252261/full}{Pseudo-{Label} {Assisted} {nnU}-{Net} enables automatic segmentation of {7T} {MRI} from a single acquisition}, Frontiers in Neuroimaging 2, publisher: Frontiers (Dec. 2023).
\newblock \href {https://doi.org/10.3389/fnimg.2023.1252261} {\path{doi:10.3389/fnimg.2023.1252261}}.
\newline\urlprefix\url{https://www.frontiersin.org/journals/neuroimaging/articles/10.3389/fnimg.2023.1252261/full}

\bibitem{wahlig_3d_2023}
S.~G. Wahlig, P.~Nedelec, D.~A. Weiss, J.~D. Rudie, L.~P. Sugrue, A.~M. Rauschecker, \href{https://www.frontiersin.org/journals/neuroscience/articles/10.3389/fnins.2023.1188336/full}{{3D} {U}-{Net} for automated detection of multiple sclerosis lesions: utility of transfer learning from other pathologies}, Frontiers in Neuroscience 17, publisher: Frontiers (Oct. 2023).
\newblock \href {https://doi.org/10.3389/fnins.2023.1188336} {\path{doi:10.3389/fnins.2023.1188336}}.
\newline\urlprefix\url{https://www.frontiersin.org/journals/neuroscience/articles/10.3389/fnins.2023.1188336/full}

\bibitem{todea_multicenter_2023}
A.~R. Todea, L.~Melie-Garcia, M.~Barakovic, A.~Cagol, R.~Rahmanzadeh, R.~Galbusera, P.-J. Lu, M.~Weigel, E.~Ruberte, E.-W. Radue, S.~Schaedelin, P.~Benkert, Y.~Oezguer, T.~Sinnecker, S.~Müller, L.~Achtnichts, J.~Vehoff, G.~Disanto, O.~Findling, A.~Chan, A.~Salmen, C.~Pot, P.~Lalive, C.~Bridel, C.~Zecca, T.~Derfuss, L.~Remonda, F.~Wagner, M.~Vargas, R.~Du~Pasquier, E.~Pravata, J.~Weber, C.~Gobbi, D.~Leppert, J.~Wuerfel, T.~Kober, B.~Marechal, R.~Corredor-Jerez, M.~Psychogios, J.~Lieb, L.~Kappos, M.~B. Cuadra, J.~Kuhle, C.~Granziera, f.~t. S. M.~C. Study, \href{https://onlinelibrary.wiley.com/doi/abs/10.1002/jmri.28618}{A {Multicenter} {Longitudinal} {MRI} {Study} {Assessing} {LeMan}-{PV} {Software} {Accuracy} in the {Detection} of {White} {Matter} {Lesions} in {Multiple} {Sclerosis} {Patients}}, Journal of Magnetic Resonance Imaging 58~(3) (2023) 864--876, \_eprint: https://onlinelibrary.wiley.com/doi/pdf/10.1002/jmri.28618.
\newblock \href {https://doi.org/10.1002/jmri.28618} {\path{doi:10.1002/jmri.28618}}.
\newline\urlprefix\url{https://onlinelibrary.wiley.com/doi/abs/10.1002/jmri.28618}

\bibitem{uwaeze_automatic_2024}
J.~Uwaeze, P.~A. Narayana, A.~Kamali, V.~Braverman, M.~A. Jacobs, A.~Akhbardeh, \href{https://www.ncbi.nlm.nih.gov/pmc/articles/PMC10969435/}{Automatic {Active} {Lesion} {Tracking} in {Multiple} {Sclerosis} {Using} {Unsupervised} {Machine} {Learning}}, Diagnostics 14~(6) (2024) 632.
\newblock \href {https://doi.org/10.3390/diagnostics14060632} {\path{doi:10.3390/diagnostics14060632}}.
\newline\urlprefix\url{https://www.ncbi.nlm.nih.gov/pmc/articles/PMC10969435/}

\bibitem{de_rosa_consensus_2024}
A.~P. De~Rosa, M.~Benedetto, S.~Tagliaferri, F.~Bardozzo, A.~D’Ambrosio, A.~Bisecco, A.~Gallo, M.~Cirillo, R.~Tagliaferri, F.~Esposito, \href{https://www.nature.com/articles/s41598-024-72649-9}{Consensus of algorithms for lesion segmentation in brain {MRI} studies of multiple sclerosis}, Scientific Reports 14~(1) (2024) 21348, publisher: Nature Publishing Group.
\newblock \href {https://doi.org/10.1038/s41598-024-72649-9} {\path{doi:10.1038/s41598-024-72649-9}}.
\newline\urlprefix\url{https://www.nature.com/articles/s41598-024-72649-9}

\bibitem{bai_improving_2024}
L.~Bai, D.~Wang, H.~Wang, M.~Barnett, M.~Cabezas, W.~Cai, F.~Calamante, K.~Kyle, D.~Liu, L.~Ly, A.~Nguyen, C.-C. Shieh, R.~Sullivan, G.~Zhan, W.~Ouyang, C.~Wang, \href{https://www.sciencedirect.com/science/article/pii/S0933365724001143}{Improving multiple sclerosis lesion segmentation across clinical sites: {A} federated learning approach with noise-resilient training}, Artificial Intelligence in Medicine 152 (2024) 102872.
\newblock \href {https://doi.org/10.1016/j.artmed.2024.102872} {\path{doi:10.1016/j.artmed.2024.102872}}.
\newline\urlprefix\url{https://www.sciencedirect.com/science/article/pii/S0933365724001143}

\bibitem{huang_joint_2024}
L.~Huang, Y.~Shao, H.~Yang, C.~Guo, Y.~Wang, Z.~Zhao, Y.~Gong, A joint model for lesion segmentation and classification of {MS} and {NMOSD}, Frontiers in Neuroscience 18 (2024) 1351387.
\newblock \href {https://doi.org/10.3389/fnins.2024.1351387} {\path{doi:10.3389/fnins.2024.1351387}}.

\bibitem{wiltgen_lst-ai_2024}
T.~Wiltgen, J.~McGinnis, S.~Schlaeger, F.~Kofler, C.~Voon, A.~Berthele, D.~Bischl, L.~Grundl, N.~Will, M.~Metz, D.~Schinz, D.~Sepp, P.~Prucker, B.~Schmitz-Koep, C.~Zimmer, B.~Menze, D.~Rueckert, B.~Hemmer, J.~Kirschke, M.~Mühlau, B.~Wiestler, \href{https://www.sciencedirect.com/science/article/pii/S2213158224000500}{{LST}-{AI}: {A} deep learning ensemble for accurate {MS} lesion segmentation}, NeuroImage: Clinical 42 (2024) 103611.
\newblock \href {https://doi.org/10.1016/j.nicl.2024.103611} {\path{doi:10.1016/j.nicl.2024.103611}}.
\newline\urlprefix\url{https://www.sciencedirect.com/science/article/pii/S2213158224000500}

\bibitem{gaj_subject-based_2025}
S.~Gaj, B.~Thoomukuntla, D.~Ontaneda, K.~Nakamura, Subject-{Based} {Transfer} {Learning} in {Longitudinal} {Multiple} {Sclerosis} {Lesion} {Segmentation}, Journal of Neuroimaging: Official Journal of the American Society of Neuroimaging 35~(1) (2025) e70024.
\newblock \href {https://doi.org/10.1111/jon.70024} {\path{doi:10.1111/jon.70024}}.

\bibitem{placidi_context-dependent_2025}
G.~Placidi, L.~Cinque, G.~L. Foresti, F.~Galassi, F.~Mignosi, M.~Nappi, M.~Polsinelli, A {Context}-{Dependent} {CNN}-{Based} {Framework} for {Multiple} {Sclerosis} {Segmentation} in {MRI}, International Journal of Neural Systems 35~(3) (2025) 2550006.
\newblock \href {https://doi.org/10.1142/S0129065725500066} {\path{doi:10.1142/S0129065725500066}}.

\end{thebibliography}


\begin{thebibliography}{10}
\expandafter\ifx\csname url\endcsname\relax
  \def\url#1{\texttt{#1}}\fi
\expandafter\ifx\csname urlprefix\endcsname\relax\def\urlprefix{URL }\fi
\expandafter\ifx\csname href\endcsname\relax
  \def\href#1#2{#2} \def\path#1{#1}\fi

\bibitem{reich_multiple_2018}
D.~S. Reich, C.~F. Lucchinetti, P.~A. Calabresi, Multiple {Sclerosis}, The New England Journal of Medicine 378~(2) (2018) 169--180.
\newblock \href {https://doi.org/10.1056/NEJMra1401483} {\path{doi:10.1056/NEJMra1401483}}.

\bibitem{thompson_diagnosis_2018}
A.~J. Thompson, et~al., Diagnosis of multiple sclerosis: 2017 revisions of the {McDonald} criteria, The Lancet. Neurology 17~(2) (2018) 162--173.
\newblock \href {https://doi.org/10.1016/S1474-4422(17)30470-2} {\path{doi:10.1016/S1474-4422(17)30470-2}}.

\bibitem{on_behalf_of_the_magnims_study_group_magnims_2015}
{on behalf of the MAGNIMS study group}, \href{https://www.nature.com/articles/nrneurol.2015.157}{{MAGNIMS} consensus guidelines on the use of {MRI} in multiple sclerosis—establishing disease prognosis and monitoring patients}, Nature Reviews Neurology 11~(10) (2015) 597--606.
\newblock \href {https://doi.org/10.1038/nrneurol.2015.157} {\path{doi:10.1038/nrneurol.2015.157}}.
\newline\urlprefix\url{https://www.nature.com/articles/nrneurol.2015.157}

\bibitem{brex_longitudinal_2002}
P.~A. Brex, O.~Ciccarelli, J.~I. O'Riordan, M.~Sailer, A.~J. Thompson, D.~H. Miller, A longitudinal study of abnormalities on {MRI} and disability from multiple sclerosis, The New England Journal of Medicine 346~(3) (2002) 158--164.
\newblock \href {https://doi.org/10.1056/NEJMoa011341} {\path{doi:10.1056/NEJMoa011341}}.

\bibitem{khoury_longitudinal_1994}
S.~J. Khoury, C.~R. Guttmann, E.~J. Orav, M.~J. Hohol, S.~S. Ahn, L.~Hsu, R.~Kikinis, G.~A. Mackin, F.~A. Jolesz, H.~L. Weiner, Longitudinal {MRI} in multiple sclerosis: correlation between disability and lesion burden, Neurology 44~(11) (1994) 2120--2124.
\newblock \href {https://doi.org/10.1212/wnl.44.11.2120} {\path{doi:10.1212/wnl.44.11.2120}}.

\bibitem{rudick_significance_2006}
R.~A. Rudick, J.-C. Lee, J.~Simon, E.~Fisher, Significance of {T2} lesions in multiple sclerosis: {A} 13-year longitudinal study, Annals of Neurology 60~(2) (2006) 236--242.
\newblock \href {https://doi.org/10.1002/ana.20883} {\path{doi:10.1002/ana.20883}}.

\bibitem{montalban_revised_2024}
X.~Montalban, Revised {McDonald} criteria 2023, in: 2024 40th Congress of the {European} {Committe} for {Treatment} and {Research} in {Multiple} {Sclerosis} ({ECTRIMS} 2024), 2024, pp. 0--0.

\bibitem{maggi_central_2018}
P.~Maggi, M.~Absinta, M.~Grammatico, L.~Vuolo, G.~Emmi, G.~Carlucci, G.~Spagni, A.~Barilaro, A.~M. Repice, L.~Emmi, D.~Prisco, V.~Martinelli, R.~Scotti, N.~Sadeghi, G.~Perrotta, P.~Sati, B.~Dachy, D.~S. Reich, M.~Filippi, L.~Massacesi, Central vein sign differentiates {Multiple} {Sclerosis} from central nervous system inflammatory vasculopathies, Annals of Neurology 83~(2) (2018) 283--294.
\newblock \href {https://doi.org/10.1002/ana.25146} {\path{doi:10.1002/ana.25146}}.

\bibitem{absinta_association_2019}
M.~Absinta, P.~Sati, F.~Masuzzo, G.~Nair, V.~Sethi, H.~Kolb, J.~Ohayon, T.~Wu, I.~C.~M. Cortese, D.~S. Reich, Association of {Chronic} {Active} {Multiple} {Sclerosis} {Lesions} {With} {Disability} {In} {Vivo}, JAMA neurology 76~(12) (2019) 1474--1483.
\newblock \href {https://doi.org/10.1001/jamaneurol.2019.2399} {\path{doi:10.1001/jamaneurol.2019.2399}}.

\bibitem{elliott_slowly_2019}
C.~Elliott, J.~S. Wolinsky, S.~L. Hauser, L.~Kappos, F.~Barkhof, C.~Bernasconi, W.~Wei, S.~Belachew, D.~L. Arnold, Slowly expanding/evolving lesions as a magnetic resonance imaging marker of chronic active multiple sclerosis lesions, Multiple Sclerosis (Houndmills, Basingstoke, England) 25~(14) (2019) 1915--1925.
\newblock \href {https://doi.org/10.1177/1352458518814117} {\path{doi:10.1177/1352458518814117}}.

\bibitem{la_rosa_cortical_2022}
F.~La~Rosa, M.~Wynen, O.~Al-Louzi, E.~S. Beck, T.~Huelnhagen, P.~Maggi, J.-P. Thiran, T.~Kober, R.~T. Shinohara, P.~Sati, {others}, Cortical lesions, central vein sign, and paramagnetic rim lesions in multiple sclerosis: {Emerging} machine learning techniques and future avenues, NeuroImage: Clinical 36 (2022) 103205, publisher: Elsevier.

\bibitem{barquero_rimnet_2020}
G.~Barquero, et~al., \href{http://www.sciencedirect.com/science/article/pii/S2213158220302497}{{RimNet}: {A} deep {3D} multimodal {MRI} architecture for paramagnetic rim lesion assessment in multiple sclerosis}, NeuroImage: Clinical 28 (2020) 102412.
\newblock \href {https://doi.org/10.1016/j.nicl.2020.102412} {\path{doi:10.1016/j.nicl.2020.102412}}.
\newline\urlprefix\url{http://www.sciencedirect.com/science/article/pii/S2213158220302497}

\bibitem{maggi_cvsnet_2020}
P.~Maggi, M.~J. Fartaria, J.~Jorge, F.~L. Rosa, M.~Absinta, P.~Sati, R.~Meuli, R.~D. Pasquier, D.~S. Reich, M.~B. Cuadra, C.~Granziera, J.~Richiardi, T.~Kober, \href{https://onlinelibrary.wiley.com/doi/abs/10.1002/nbm.4283}{{CVSnet}: {A} machine learning approach for automated central vein sign assessment in multiple sclerosis}, NMR in Biomedicine 33~(5) (2020) e4283, \_eprint: https://onlinelibrary.wiley.com/doi/pdf/10.1002/nbm.4283.
\newblock \href {https://doi.org/https://doi.org/10.1002/nbm.4283} {\path{doi:https://doi.org/10.1002/nbm.4283}}.
\newline\urlprefix\url{https://onlinelibrary.wiley.com/doi/abs/10.1002/nbm.4283}

\bibitem{lou_fully_2021}
C.~Lou, P.~Sati, M.~Absinta, K.~Clark, J.~D. Dworkin, A.~M. Valcarcel, M.~K. Schindler, D.~S. Reich, E.~M. Sweeney, R.~T. Shinohara, Fully automated detection of paramagnetic rims in multiple sclerosis lesions on {3T} susceptibility-based {MR} imaging, NeuroImage. Clinical 32 (2021) 102796.
\newblock \href {https://doi.org/10.1016/j.nicl.2021.102796} {\path{doi:10.1016/j.nicl.2021.102796}}.

\bibitem{zhang_qsmrim-net_2022}
H.~Zhang, T.~D. Nguyen, J.~Zhang, M.~Marcille, P.~Spincemaille, Y.~Wang, S.~A. Gauthier, E.~M. Sweeney, \href{https://www.ncbi.nlm.nih.gov/pmc/articles/PMC8892132/}{{QSMRim}-{Net}: {Imbalance}-aware learning for identification of chronic active multiple sclerosis lesions on quantitative susceptibility maps}, NeuroImage : Clinical 34 (2022) 102979.
\newblock \href {https://doi.org/10.1016/j.nicl.2022.102979} {\path{doi:10.1016/j.nicl.2022.102979}}.
\newline\urlprefix\url{https://www.ncbi.nlm.nih.gov/pmc/articles/PMC8892132/}

\bibitem{wynen_longitudinal_2021}
M.~Wynen, F.~La~Rosa, A.~Sellimi, G.~Barquero, G.~Perrotta, V.~Lolli, V.~Van~Pesch, C.~Granziera, T.~Kober, P.~Sati, {others}, Longitudinal automated assessment of paramagnetic rim lesions in multiple sclerosis using {RimNet}, in: {ISMRM}, 2021, pp. 0--0.

\bibitem{zivadinov_effect_2008}
R.~Zivadinov, M.~Zorzon, R.~De~Masi, D.~Nasuelli, G.~Cazzato, \href{https://www.sciencedirect.com/science/article/pii/S0022510X07006557}{Effect of intravenous methylprednisolone on the number, size and confluence of plaques in relapsing–remitting multiple sclerosis}, Journal of the Neurological Sciences 267~(1) (2008) 28--35.
\newblock \href {https://doi.org/10.1016/j.jns.2007.09.025} {\path{doi:10.1016/j.jns.2007.09.025}}.
\newline\urlprefix\url{https://www.sciencedirect.com/science/article/pii/S0022510X07006557}

\bibitem{lassmann_multiple_2014}
H.~Lassmann, \href{https://www.sciencedirect.com/science/article/pii/S0014488613003610}{Multiple sclerosis: {Lessons} from molecular neuropathology}, Experimental Neurology 262 (2014) 2--7.
\newblock \href {https://doi.org/10.1016/j.expneurol.2013.12.003} {\path{doi:10.1016/j.expneurol.2013.12.003}}.
\newline\urlprefix\url{https://www.sciencedirect.com/science/article/pii/S0014488613003610}

\bibitem{harris_serial_1991}
J.~O. Harris, J.~A. Frank, N.~Patronas, D.~E. McFarlin, H.~F. McFarland, Serial gadolinium-enhanced magnetic resonance imaging scans in patients with early, relapsing-remitting multiple sclerosis: implications for clinical trials and natural history, Annals of Neurology 29~(5) (1991) 548--555.
\newblock \href {https://doi.org/10.1002/ana.410290515} {\path{doi:10.1002/ana.410290515}}.

\bibitem{dworkin_automated_2018}
J.~D. Dworkin, et~al., \href{http://www.ajnr.org/content/39/4/626}{An {Automated} {Statistical} {Technique} for {Counting} {Distinct} {Multiple} {Sclerosis} {Lesions}}, American Journal of Neuroradiology 39~(4) (2018) 626--633, publisher: American Journal of Neuroradiology Section: ADULT BRAIN.
\newblock \href {https://doi.org/10.3174/ajnr.A5556} {\path{doi:10.3174/ajnr.A5556}}.
\newline\urlprefix\url{http://www.ajnr.org/content/39/4/626}

\bibitem{wynen_conflunet_2024}
M.~Wynen, M.~Istasse, P.~M. Gordaliza, A.~Stölting, P.~Maggi, B.~Macq, M.~B. Cuadra, \href{https://ieeexplore.ieee.org/document/10635707}{Conflunet: {Improving} {Confluent} {Lesion} {Identification} {In} {Multiple} {Sclerosis} {With} {Instance} {Segmentation}}, in: 2024 {IEEE} {International} {Symposium} on {Biomedical} {Imaging} ({ISBI}), 2024, pp. 1--5, iSSN: 1945-8452.
\newblock \href {https://doi.org/10.1109/ISBI56570.2024.10635707} {\path{doi:10.1109/ISBI56570.2024.10635707}}.
\newline\urlprefix\url{https://ieeexplore.ieee.org/document/10635707}

\bibitem{hafiz_survey_2020}
A.~M. Hafiz, G.~M. Bhat, \href{https://doi.org/10.1007/s13735-020-00195-x}{A survey on instance segmentation: state of the art}, International Journal of Multimedia Information Retrieval 9~(3) (2020) 171--189.
\newblock \href {https://doi.org/10.1007/s13735-020-00195-x} {\path{doi:10.1007/s13735-020-00195-x}}.
\newline\urlprefix\url{https://doi.org/10.1007/s13735-020-00195-x}

\bibitem{gu_review_2022}
W.~Gu, S.~Bai, L.~Kong, \href{https://www.sciencedirect.com/science/article/pii/S0262885622000300}{A review on {2D} instance segmentation based on deep neural networks}, Image and Vision Computing 120 (2022) 104401.
\newblock \href {https://doi.org/10.1016/j.imavis.2022.104401} {\path{doi:10.1016/j.imavis.2022.104401}}.
\newline\urlprefix\url{https://www.sciencedirect.com/science/article/pii/S0262885622000300}

\bibitem{cheng_panoptic-deeplab_2020}
B.~Cheng, \href{http://arxiv.org/abs/1911.10194}{Panoptic-{DeepLab}: {A} {Simple}, {Strong}, and {Fast} {Baseline} for {Bottom}-{Up} {Panoptic} {Segmentation}}, arXiv:1911.10194 [cs] (Mar. 2020).
\newline\urlprefix\url{http://arxiv.org/abs/1911.10194}

\bibitem{he_masked_2021}
K.~He, X.~Chen, S.~Xie, Y.~Li, P.~Dollár, R.~Girshick, \href{http://arxiv.org/abs/2111.06377}{Masked {Autoencoders} {Are} {Scalable} {Vision} {Learners}}, arXiv:2111.06377 [cs]ArXiv: 2111.06377 version: 1 (Nov. 2021).
\newline\urlprefix\url{http://arxiv.org/abs/2111.06377}

\bibitem{nasir_nuclei_2023}
E.~S. Nasir, A.~Parvaiz, M.~M. Fraz, \href{https://doi.org/10.1007/s10462-022-10372-5}{Nuclei and glands instance segmentation in histology images: a narrative review}, Artificial Intelligence Review 56~(8) (2023) 7909--7964.
\newblock \href {https://doi.org/10.1007/s10462-022-10372-5} {\path{doi:10.1007/s10462-022-10372-5}}.
\newline\urlprefix\url{https://doi.org/10.1007/s10462-022-10372-5}

\bibitem{isensee_nnu-net_2021}
F.~Isensee, P.~F. Jaeger, S.~A.~A. Kohl, J.~Petersen, K.~H. Maier-Hein, \href{https://www.nature.com/articles/s41592-020-01008-z}{{nnU}-{Net}: a self-configuring method for deep learning-based biomedical image segmentation}, Nature Methods 18~(2) (2021) 203--211, number: 2 Publisher: Nature Publishing Group.
\newblock \href {https://doi.org/10.1038/s41592-020-01008-z} {\path{doi:10.1038/s41592-020-01008-z}}.
\newline\urlprefix\url{https://www.nature.com/articles/s41592-020-01008-z}

\bibitem{bilic_liver_2023}
P.~Bilic, P.~Christ, H.~B. Li, E.~Vorontsov, A.~Ben-Cohen, G.~Kaissis, A.~Szeskin, C.~Jacobs, G.~E.~H. Mamani, G.~Chartrand, F.~Lohöfer, J.~W. Holch, W.~Sommer, F.~Hofmann, A.~Hostettler, N.~Lev-Cohain, M.~Drozdzal, M.~M. Amitai, R.~Vivanti, J.~Sosna, I.~Ezhov, A.~Sekuboyina, F.~Navarro, F.~Kofler, J.~C. Paetzold, S.~Shit, X.~Hu, J.~Lipková, M.~Rempfler, M.~Piraud, J.~Kirschke, B.~Wiestler, Z.~Zhang, C.~Hülsemeyer, M.~Beetz, F.~Ettlinger, M.~Antonelli, W.~Bae, M.~Bellver, L.~Bi, H.~Chen, G.~Chlebus, E.~B. Dam, Q.~Dou, C.-W. Fu, B.~Georgescu, X.~Gir\'o-i Nieto, F.~Gruen, X.~Han, P.-A. Heng, J.~Hesser, J.~H. Moltz, C.~Igel, F.~Isensee, P.~Jäger, F.~Jia, K.~C. Kaluva, M.~Khened, I.~Kim, J.-H. Kim, S.~Kim, S.~Kohl, T.~Konopczynski, A.~Kori, G.~Krishnamurthi, F.~Li, H.~Li, J.~Li, X.~Li, J.~Lowengrub, J.~Ma, K.~Maier-Hein, K.-K. Maninis, H.~Meine, D.~Merhof, A.~Pai, M.~Perslev, J.~Petersen, J.~Pont-Tuset, J.~Qi, X.~Qi, O.~Rippel, K.~Roth, I.~Sarasua, A.~Schenk, Z.~Shen, J.~Torres, C.~Wachinger, C.~Wang,
  L.~Weninger, J.~Wu, D.~Xu, X.~Yang, S.~C.-H. Yu, Y.~Yuan, M.~Yue, L.~Zhang, J.~Cardoso, S.~Bakas, R.~Braren, V.~Heinemann, C.~Pal, A.~Tang, S.~Kadoury, L.~Soler, B.~van Ginneken, H.~Greenspan, L.~Joskowicz, B.~Menze, \href{https://www.sciencedirect.com/science/article/pii/S1361841522003085}{The {Liver} {Tumor} {Segmentation} {Benchmark} ({LiTS})}, Medical Image Analysis 84 (2023) 102680.
\newblock \href {https://doi.org/10.1016/j.media.2022.102680} {\path{doi:10.1016/j.media.2022.102680}}.
\newline\urlprefix\url{https://www.sciencedirect.com/science/article/pii/S1361841522003085}

\bibitem{wynen_lesion_2024}
M.~Wynen, P.~M. Gordaliza, A.~Stölting, P.~Maggi, M.~Bach~Cuadra, \href{https://dial.uclouvain.be/pr/boreal/object/boreal:285623}{Lesion {Instance} {Segmentation} in {Multiple} {Sclerosis}: {Assessing} the {Efficacy} of {Statistical} {Lesion} {Splitting}}, in: {ISMRM}, ISMRM, 2024, pp. 0--0.
\newline\urlprefix\url{https://dial.uclouvain.be/pr/boreal/object/boreal:285623}

\bibitem{malinin_shifts_2022}
A.~Malinin, et~al., \href{http://arxiv.org/abs/2206.15407}{Shifts 2.0: {Extending} {The} {Dataset} of {Real} {Distributional} {Shifts}}, arXiv:2206.15407 [cs, stat] (Sep. 2022).
\newline\urlprefix\url{http://arxiv.org/abs/2206.15407}

\bibitem{kingma_adam_2017}
D.~P. Kingma, J.~Ba, \href{http://arxiv.org/abs/1412.6980}{Adam: {A} {Method} for {Stochastic} {Optimization}}, arXiv:1412.6980 [cs] (Jan. 2017).
\newblock \href {https://doi.org/10.48550/arXiv.1412.6980} {\path{doi:10.48550/arXiv.1412.6980}}.
\newline\urlprefix\url{http://arxiv.org/abs/1412.6980}

\bibitem{grahl_evidence_2019}
S.~Grahl, V.~Pongratz, P.~Schmidt, C.~Engl, M.~Bussas, A.~Radetz, G.~Gonzalez-Escamilla, S.~Groppa, F.~Zipp, C.~Lukas, J.~Kirschke, C.~Zimmer, M.~Hoshi, A.~Berthele, B.~Hemmer, M.~Mühlau, Evidence for a white matter lesion size threshold to support the diagnosis of relapsing remitting multiple sclerosis, Multiple Sclerosis and Related Disorders 29 (2019) 124--129.
\newblock \href {https://doi.org/10.1016/j.msard.2019.01.042} {\path{doi:10.1016/j.msard.2019.01.042}}.

\bibitem{avants_symmetric_2008}
B.~B. Avants, C.~L. Epstein, M.~Grossman, J.~C. Gee, Symmetric diffeomorphic image registration with cross-correlation: evaluating automated labeling of elderly and neurodegenerative brain, Medical Image Analysis 12~(1) (2008) 26--41.
\newblock \href {https://doi.org/10.1016/j.media.2007.06.004} {\path{doi:10.1016/j.media.2007.06.004}}.

\bibitem{hoopes_synthstrip_2022}
A.~Hoopes, J.~S. Mora, A.~V. Dalca, B.~Fischl, M.~Hoffmann, \href{https://www.sciencedirect.com/science/article/pii/S1053811922005900}{{SynthStrip}: skull-stripping for any brain image}, NeuroImage 260 (2022) 119474.
\newblock \href {https://doi.org/10.1016/j.neuroimage.2022.119474} {\path{doi:10.1016/j.neuroimage.2022.119474}}.
\newline\urlprefix\url{https://www.sciencedirect.com/science/article/pii/S1053811922005900}

\bibitem{vanden_bulcke_bmat_2022}
C.~Vanden~Bulcke, M.~Wynen, J.~Detobel, F.~La~Rosa, M.~Absinta, L.~Dricot, B.~Macq, M.~B. Cuadra, P.~Maggi, {BMAT}: {An} open-source {BIDS} managing and analysis tool, NeuroImage: Clinical 36 (2022) 103252, publisher: Elsevier.

\bibitem{yushkevich_itk-snap_2016}
P.~A. Yushkevich, Y.~Gao, G.~Gerig, \href{https://www.ncbi.nlm.nih.gov/pmc/articles/PMC5493443/}{{ITK}-{SNAP}: an interactive tool for semi-automatic segmentation of multi-modality biomedical images}, Conference proceedings : ... Annual International Conference of the IEEE Engineering in Medicine and Biology Society. IEEE Engineering in Medicine and Biology Society. Annual Conference 2016 (2016) 3342--3345.
\newblock \href {https://doi.org/10.1109/EMBC.2016.7591443} {\path{doi:10.1109/EMBC.2016.7591443}}.
\newline\urlprefix\url{https://www.ncbi.nlm.nih.gov/pmc/articles/PMC5493443/}

\bibitem{maier-hein_metrics_2024}
L.~Maier-Hein, A.~Reinke, P.~Godau, M.~D. Tizabi, F.~Buettner, E.~Christodoulou, B.~Glocker, F.~Isensee, J.~Kleesiek, M.~Kozubek, M.~Reyes, M.~A. Riegler, M.~Wiesenfarth, A.~E. Kavur, C.~H. Sudre, M.~Baumgartner, M.~Eisenmann, D.~Heckmann-Nötzel, T.~Rädsch, L.~Acion, M.~Antonelli, T.~Arbel, S.~Bakas, A.~Benis, M.~B. Blaschko, M.~J. Cardoso, V.~Cheplygina, B.~A. Cimini, G.~S. Collins, K.~Farahani, L.~Ferrer, A.~Galdran, B.~van Ginneken, R.~Haase, D.~A. Hashimoto, M.~M. Hoffman, M.~Huisman, P.~Jannin, C.~E. Kahn, D.~Kainmueller, B.~Kainz, A.~Karargyris, A.~Karthikesalingam, F.~Kofler, A.~Kopp-Schneider, A.~Kreshuk, T.~Kurc, B.~A. Landman, G.~Litjens, A.~Madani, K.~Maier-Hein, A.~L. Martel, P.~Mattson, E.~Meijering, B.~Menze, K.~G.~M. Moons, H.~Müller, B.~Nichyporuk, F.~Nickel, J.~Petersen, N.~Rajpoot, N.~Rieke, J.~Saez-Rodriguez, C.~I. Sánchez, S.~Shetty, M.~van Smeden, R.~M. Summers, A.~A. Taha, A.~Tiulpin, S.~A. Tsaftaris, B.~Van~Calster, G.~Varoquaux, P.~F. Jäger,
  \href{https://www.nature.com/articles/s41592-023-02151-z}{Metrics reloaded: recommendations for image analysis validation}, Nature Methods 21~(2) (2024) 195--212, publisher: Nature Publishing Group.
\newblock \href {https://doi.org/10.1038/s41592-023-02151-z} {\path{doi:10.1038/s41592-023-02151-z}}.
\newline\urlprefix\url{https://www.nature.com/articles/s41592-023-02151-z}

\bibitem{raina_tackling_2023}
V.~Raina, N.~Molchanova, M.~Graziani, A.~Malinin, H.~Muller, M.~B. Cuadra, M.~Gales, \href{https://ieeexplore.ieee.org/document/10230755/}{Tackling {Bias} in the {Dice} {Similarity} {Coefficient}: {Introducing} {NDSC} for {White} {Matter} {Lesion} {Segmentation}}, in: 2023 {IEEE} 20th {International} {Symposium} on {Biomedical} {Imaging} ({ISBI}), 2023, pp. 1--5, iSSN: 1945-8452.
\newblock \href {https://doi.org/10.1109/ISBI53787.2023.10230755} {\path{doi:10.1109/ISBI53787.2023.10230755}}.
\newline\urlprefix\url{https://ieeexplore.ieee.org/document/10230755/}

\bibitem{schmidt_bayesian_2017}
P.~Schmidt, \href{https://edoc.ub.uni-muenchen.de/id/eprint/20373}{Bayesian inference for structured additive regression models for large-scale problems with applications to medical imaging}, Ph.D. thesis, Ludwig-Maximilians-Universität München, medium: application/pdf (2017).
\newblock \href {https://doi.org/10.5282/EDOC.20373} {\path{doi:10.5282/EDOC.20373}}.
\newline\urlprefix\url{https://edoc.ub.uni-muenchen.de/id/eprint/20373}

\bibitem{dereskewicz_flames_2025}
E.~Dereskewicz, F.~L. Rosa, J.~d.~S. Silva, E.~Sizer, A.~Kohli, M.~Wynen, W.~A. Mullins, P.~Maggi, S.~Levy, K.~Onyemeh, B.~Ayci, A.~J. Solomon, J.~Assländer, O.~Al-Louzi, D.~S. Reich, J.~Sumowski, E.~S. Beck, \href{https://www.medrxiv.org/content/10.1101/2025.05.19.25327707v1}{{FLAMeS}: {A} {Robust} {Deep} {Learning} {Model} for {Automated} {Multiple} {Sclerosis} {Lesion} {Segmentation}}, pages: 2025.05.19.25327707 (May 2025).
\newblock \href {https://doi.org/10.1101/2025.05.19.25327707} {\path{doi:10.1101/2025.05.19.25327707}}.
\newline\urlprefix\url{https://www.medrxiv.org/content/10.1101/2025.05.19.25327707v1}

\bibitem{la_rosa_flames_2023}
F.~La~Rosa, J.~Dos Santos~Silva, W.~A. Mullins, H.~Greenspan, J.~Sumowski, D.~S. Reich, E.~S. Beck, \href{https://zenodo.org/records/7626121}{{FLAMeS}: {FLAIR} {Lesion} {Analysis} in {Multiple} {Sclerosis}} (Feb. 2023).
\newblock \href {https://doi.org/10.5281/zenodo.7626121} {\path{doi:10.5281/zenodo.7626121}}.
\newline\urlprefix\url{https://zenodo.org/records/7626121}

\bibitem{cerri_contrast-adaptive_2021}
S.~Cerri, O.~Puonti, D.~S. Meier, J.~Wuerfel, M.~Mühlau, H.~R. Siebner, K.~Van~Leemput, \href{https://www.sciencedirect.com/science/article/pii/S1053811920309563}{A contrast-adaptive method for simultaneous whole-brain and lesion segmentation in multiple sclerosis}, NeuroImage 225 (2021) 117471.
\newblock \href {https://doi.org/10.1016/j.neuroimage.2020.117471} {\path{doi:10.1016/j.neuroimage.2020.117471}}.
\newline\urlprefix\url{https://www.sciencedirect.com/science/article/pii/S1053811920309563}

\bibitem{noauthor_scipyorg_nodate}
\href{https://www.scipy.org/}{{SciPy}.org — {SciPy}.org} (2025).
\newline\urlprefix\url{https://www.scipy.org/}

\bibitem{wynen_multi-modal_2024}
M.~Wynen, P.~M. Gordaliza, A.~Stölting, P.~Maggi, M.~B. Cuadra, B.~Macq, Multi-modal segmentation for paramagnetic rim lesion detection in multiple sclerosis, in: Medical {Imaging} 2024: {Imaging} {Informatics} for {Healthcare}, {Research}, and {Applications}, Vol. 12931, SPIE, 2024, pp. 241--246.

\bibitem{maggi_b_2023}
P.~Maggi, C.~Vanden~Bulcke, E.~Pedrini, C.~Bugli, A.~Sellimi, M.~Wynen, A.~Stölting, W.~A. Mullins, G.~Kalaitzidis, V.~Lolli, {others}, B cell depletion therapy does not resolve chronic active multiple sclerosis lesions, EBioMedicine 94, publisher: Elsevier (2023).

\bibitem{wynen_machine_2025}
M.~Wynen, C.~Vanden~Bulcke, S.~Borrelli, P.~M. Gordaliza, A.~Stölting, F.~Guisset, C.~Cordier, M.~S. Martire, A.~Tamanti, B.~Macq, {others}, Machine {Learning}-{Based} {Combination} of the {Central} {Vein} {Sign}, {Cortical} {Lesions} and {Paramagnetic} {Rim} {Lesions}: {A} {Web}-{Based} {Tool} for the {Diagnosis} of {Multiple} {Sclerosis} (2025).
\newblock \href {https://doi.org/https://dx.doi.org/10.2139/ssrn.5192817} {\path{doi:https://dx.doi.org/10.2139/ssrn.5192817}}.

\bibitem{gerin_exploring_2024}
B.~Gérin, M.~Zanella, M.~Wynen, S.~Mahmoudi, B.~Macq, C.~De~Vleeschouwer, Exploring viability of test-time training: {Application} to {3D} segmentation in multiple sclerosis, in: 2024 {IEEE} {Conference} on {Artificial} {Intelligence} ({CAI}), IEEE, 2024, pp. 557--562.

\end{thebibliography}
\end{document}